\documentclass[final,3p,times]{elsarticle}

\usepackage{amssymb}
\usepackage{amsthm}
\usepackage{amsmath}
\usepackage{libertine}
\usepackage{libertinust1math}
\usepackage[backref=page]{hyperref}
\usepackage{paralist}
\usepackage{graphicx}
\usepackage{algorithm}
\usepackage{algpseudocode}
\usepackage{xcolor}
\usepackage{subcaption}
\usepackage{makecell}
\usepackage{xspace}
\graphicspath{{./figures/}}

\makeatletter
\newcommand\newtag[2]{#1\def\@currentlabel{#1}\label{#2}}
\makeatother

\algnewcommand{\Inputs}[1]{%
  \State \textbf{Inputs:}
  \Statex \hspace*{\algorithmicindent}\parbox[t]{.8\linewidth}{\raggedright #1}
}
\algnewcommand{\Initialize}[1]{%
  \State \textbf{Initialize:}
  \Statex \hspace*{\algorithmicindent}\parbox[t]{.8\linewidth}{\raggedright #1}
}

\newcommand{\etal}{\emph{et al.}\xspace}

\journal{Physica A}

\begin{document}

\begin{frontmatter}



\title{Discerning media bias within a network of political allies and opponents: Disruption by partisans}

\author[inst1]{Yutong Bu\corref{cor1}}
\ead{buy1@student.unimelb.edu.au}
\cortext[cor1]{Corresponding author}

\affiliation[inst1]{organization={School of Physics},
            addressline={University of Melbourne}, 
            city={Parkville},
            state={VIC 3010},
            country={Australia}}

\author[inst1,inst2]{Andrew Melatos}

\affiliation[inst2]{organization={Australian Research Council Centre of Excellence for Gravitational Wave Discovery (OzGrav)},
            addressline={University of Melbourne}, 
            city={Parkville},
            state={VIC 3010},
            country={Australia}}

\begin{abstract}

An individual's opinions about media bias derive from their own independent assessment of media outputs combined with peer pressure from networked political allies and opponents. 
Here we generalize previous idealized, probabilistic models of the perception formation process, based on a network of Bayesian learners inferring the bias of a coin, by introducing obdurate agents (partisans), whose opinions stay fixed. 
It is found that even one partisan destabilizes an allies-only network, stopping it from achieving asymptotic learning and forcing persuadable agents to vacillate indefinitely (turbulent nonconvergence) between the true coin bias $\theta_0$ and the partisan's belief $\theta_{\rm p}$. 
The dwell time $t_{\rm d}$ at the partisan's belief increases, as the partisan fraction $f$ increases, and decreases, when multiple partisans disagree amongst themselves. 
In opponents-only networks, asymptotic learning occurs, whether or not partisans are present. 
However, the counterintuitive tendency to reach wrong conclusions first, identified in previous work with zero partisans, does not persist in general for $\theta_0 \neq \theta_{\rm p}$ in complete networks; it is a property of sparsely connected systems (e.g.\  Barab\'{a}si-Albert networks with attachment parameter $\lesssim 10$). 
In mixed networks containing allies and opponents, partisans drive counterintuitive outcomes, which depend sensitively, on where they reside. 
A strongly balanced triad exhibits intermittency with a partisan (sudden transitions between long intervals of static beliefs and turbulent nonconvergence) and asymptotic learning without a partisan. 
Counterintuitively, in an unbalanced triad, one of the persuadable agents achieves asymptotic learning at $\theta_0$, when the partisan is located favorably, but is driven away from $\theta_0$, when there is no partisan. The above results are interpreted briefly in terms of the social science theory of structural balance.

\end{abstract}




\end{frontmatter}

\section{Introduction}
\label{sec:intro}
Media bias influences society through its impact on the political voting system, because the enfranchised public learns about political parties and public policy through print, broadcast, and social media \cite{eberl_one_2017}. Media bias can be direct, e.g.\ favoritism towards a party \cite{alizadeh_effect_2015}, or indirect, e.g.\ filtering positive or negative information \cite{barzilai-nahon_gatekeeping_2009,dalessio_media_2000}, or arguing false equivalence \cite{boykoff_balance_2004}.  In the classic definition by Williams, media bias is ``volitional", ``influential" and ``threatening to widely held conventions" \cite{williams_unbiased_1975},  so its capacity to affect voters' opinions is substantial.

The dynamics of people's perceptions of media bias can be modeled through a network approach.  An individual agent's opinion about media bias is shaped directly by consuming media outputs and indirectly by observing the opinions of networked acquaintances about the same media outputs.  Many opinion dynamics models in the literature are deterministic, in the sense that every agent holds a single belief with complete certainty at an instant in time, although of course the belief changes from one instant to the next \cite{degroot_reaching_1974,vaz_martins_mass_2010,shi_evolution_2016,mcquade_social_2019}.  Such studies often focus on belief convergence or asymptotic learning between like-minded agents (``allies'') \cite{degroot_reaching_1974,hegselmann_opinion_2002} although there are exceptions that focus on antagonistic interactions between agents  (``opponents'') \cite{shi_evolution_2016}. In comparison, there is a lesser but growing focus in the media bias literature on probabilistic Bayesian learners, i.e.\ agents who hold a spectrum of uncertain beliefs at an instant in time \cite{low_discerning_2022,low_vacillating_2022,fang_opinion_2020,fang_social_2019}, similar to the real world.  Within the Bayesian framework, an agent's beliefs are represented by a probability distribution, which evolves deterministically in response to stochastic external stimuli (e.g.\ daily newspaper editorials) and peer pressure within the network, leading to drift and diffusion of beliefs.

Many opinion dynamics models treat agents as {\em persuadable}, that is, open to modifying their beliefs in response to their own observations as well as peer pressure from other agents in the network. In contrast, in this paper, we consider the role of partisans: {\em obdurate} agents who refuse to change their opinion, regardless of external inputs or peer pressure. Their obduracy may be subconscious and psychological or conscious and cynical (e.g.\ a deliberately disruptive strategy by politically motivated ``trolls'' on social media).  The idea of partisans was first proposed by Mobilia \& Mauro \cite{mobilia_does_2003}, who introduced the equivalent term ``zealots'' in a deterministic, non-Bayesian framework. Mobilia \& Mauro \cite{mobilia_does_2003} analyzed the inhomogeneous voter model, in which the network is a $d$-dimensional hypercubic lattice. They and other authors found that a single zealot can persuade the rest of the network to converge on the zealot's belief for $d < 3$, and that a relatively small number of zealots can hinder the formation of consensus or even a clear-cut majority opinion for arbitrary $d$ \cite{mobilia_role_2007,liggett_stochastic_nodate, belitsky_mixture_2001}.  Moreover, the long-term disposition of opinions within the network depends on the network graph structure (e.g.\ connectivity or scale invariance) and the opinions and locations of the partisans \cite{yildiz_binary_2013}. 

In this paper, we examine how partisans shape perceptions of media bias in a Bayesian framework for the first time.
The paper generalizes previous opinion dynamics models of media bias among Bayesian learners, in which all the agents are persuadable \cite{low_discerning_2022,low_vacillating_2022,fang_opinion_2020}. 
It also generalizes previous deterministic models involving partisans, such as the voter model \cite{mobilia_does_2003,yildiz_binary_2013}, to the new application of media bias.  The goal is to investigate the disruptive influence of partisans on perceptions of media bias as a function of their number within allies-only, opponents-only, and mixed networks.  The paper is structured as follows. Section \ref{sec:partisans} introduces the opinion update rule for the idealized model of a biased coin introduced in Ref.\ \cite{low_discerning_2022} and explains how partisans are implemented. Monte Carlo simulations are performed on complete networks with allies only, opponents only, and a mixture of allies and opponents in Sections \ref{sec:alliesonly}, \ref{sec:opponentonly}, and \ref{sec:mixed} respectively.  The goal of the simulations is to identify how partisans modify the phenomena observed in networks of Bayesian learners without partisans, such as the time-scale to reach asymptotic learning, the emergence of turbulent nonconvergence and intermittency, and the tendency of agents in opponents-only networks to reach wrong conclusions first \cite{low_discerning_2022,low_vacillating_2022,fang_opinion_2020}.  
A brief, preliminary study of how the above phenomena translate to partly connected networks is included in Sections \ref{subsec:wrong_conclusion_first} and \ref{sec:BA}, as a prelude to future work.
The results are interpreted briefly in terms of the social science theory of structural balance in Section \ref{sec:conclusion}. 

\section{An idealized model of media bias: inferring the bias of a coin}
\label{sec:partisans}
In this section, we extend the media bias model formulated by Low \& Melatos \cite{low_discerning_2022} to include obdurate partisans.  Section \ref{subsec:modelintro} summarizes how the evolution of perceptions about media bias maps onto an idealized model, in which a network of political allies and opponents infers the bias of a coin. A two-step update rule is presented, which updates the belief of each networked agent in response to independent observations of the coin and peer pressure from the network \cite{low_discerning_2022,low_vacillating_2022}.  
The implementation of partisans is described in Section \ref{subsec:intropartisans}, and the pseudocode for the resulting, iteratively updated automaton is presented in Section \ref{subsec:automaton}.  
The paper focuses on complete networks for technical reasons justified in Section \ref{subsec:network}. Complete networks are adequate to demonstrate the central points of the paper, but the study of other network topologies is an important issue which is deferred to future work (except for a brief discussion in Sections \ref{sec:opponentonly} and \ref{sec:BA}). 
The extended idealized model is compared with other models in the literature in Section \ref{subsec:relationtoothers}.

\subsection{Updating beliefs in two steps: independent observations and peer pressure}
\label{subsec:modelintro}
In Ref.\ \cite{low_discerning_2022}, a network of $n$ agents attempts to learn the true bias $0 \leq \theta_0 \leq 1 $ of a coin though a sequence of $T$ coin tosses, where $\theta_0$ denotes the probability of heads for a single coin toss.  The coin toss is an analog for a piece of politically relevant information (e.g.\ editorials, articles), which the agents consume from a media outlet, and $\theta_0$ is an analog for the political bias of the media outlet (e.g.\ on a simplified, left-right spectrum).  The $i$-th agent harbors probabilistic beliefs about the political bias of the media outlet, or analogously the bias of the coin, which are described by the probability density function (PDF) $x_i(t,\theta)$ at time $t$. The beliefs can be multi-valued and hence uncertain in general; the agent may be equally confident about two distinct values of the coin's bias, for example, with $x_i(t,\theta_1)=0.5=x_i(t,\theta_2 \neq \theta_1)$. 
\footnote{Multimodal belief PDFs are realistic in various settings.  In the media bias context, for example, a reader of a newspaper's editorials may be unsure as to whether the newspaper leans right or left politically, if the editorials lean right on economic issues and left on social issues.  The reader may develop a bimodal belief PDF in this scenario (i.e., the newspaper actually leans right or left but not both, and the reader is unsure which option to prefer) instead of a unimodal belief PDF (i.e., the newspaper occupies the middle of the road politically, and the reader believes its bias lies between the right and left extremes).  Multimodality is more likely, when there are multiple ways to map the underlying bias variable to the observed output signal  (e.g., editorials), and readers (and maybe even the newspaper's editors themselves) do not know consciously which mapping applies in practice. The point is especially pertinent, when a reader overlays an oversimplified mental model (e.g., left-right dichotomy) on a more complicated bias structure (e.g., left-right economic-social matrix). }
Starting from initial PDFs $x_1(t=0,\theta), \dots, x_n(t=0,\theta)$, the beliefs of the $i$-th agent in the network are updated in two steps, every time a coin toss occurs. 
\begin{enumerate}
    \item The $i$-th agent observes the coin toss at time $t$ and invokes Bayes's rule to combine their prior beliefs at time $t$ with the result of the coin toss to generate an intermediate, provisional PDF $x_i'(t+1/2,\theta)$.
    \item The $i$-th agent is influenced by positive and negative peer pressure from their allies and opponents respectively and adjusts $x_i'(t+1/2,\theta)$ accordingly to obtain the fully updated PDF at the conclusion of that time step, $x_i(t+1,\theta)$.
\end{enumerate}
The outcome of each coin toss is a public external signal, which is simultaneous and unfiltered for all agents, i.e.\ all agents observe the outcome and accept it without demur.

The first half of the update rule above, based on an independent observation of the coin, follows immediately from Bayes's rule. We write
\begin{equation}\label{eq:updatefirsthalf}
    x'_i(t+1/2, \theta) = \frac{P[S(t) | \theta]}{P[S(t)]}x_i (t, \theta),
\end{equation}
where $S(t)$ is the outcome of the coin toss (heads or tails), $P[S(t)| \theta]$ is the likelihood function, and $P[S(t)] = \sum_{\theta}P[S(t)| \theta]x_i (t, \theta)$ is the normalizing constant. The likelihood function takes the form
\begin{equation}\label{eq:likelihood}
    P[S(t)| \theta] = \begin{cases} \theta, & \mbox{if } S(t)\mbox{ is heads} \\
    1-\theta,  & \mbox{if } S(t)\mbox{ is tails.} \end{cases}
\end{equation}

The second half of the update rule shapes the belief of each agent under the influence of political allegiances.  In much of the literature, agents are typically conditioned to copy, or at least move towards, the beliefs of their neighbors in the network; that is, they implicitly regard their neighbors as allies \cite{degroot_reaching_1974,hegselmann_opinion_2002,fang_social_2019,jadbabaie_non-bayesian_2012,deffuant_mixing_2000}.  In this paper, we generalize the network to include allies and opponents, such that antagonistic  interactions involving negative feedback between agents are taken into account.  The network is modelled by a graph, where nodes represent agents, and edges are tagged with the political allegiance between two agents.  A weighted graph with $n \times n$ adjacency matrix $A$ is used with entries 
\begin{equation}
    A_{ij} = 
    \begin{cases}
        + 1 , \quad & \text{agents $i$ and $j$ are allies} \\
        0 , \quad & \text{agents $i$ and $j$ are disconnected} \\
        - 1 , \quad & \text{agents $i$ and $j$ are opponents} \\
    \end{cases}
\end{equation}
We note that agents $i$ and $j$ can be connected (and therefore influence each other) indirectly through one or more third parties (e.g. $A_{ik}\neq0$ and $A_{kj} \neq 0$ for $k \neq i,j$), even if they are not connected directly ($A_{ij}=0$).  In this paper, we assume for simplicity that $A$ is symmetric (i.e.\ $A_{ij} = A_{ji}$),  and that allies and opponents exert equal and opposite peer pressure.

Under peer pressure from allies and opponents, the provisional belief PDF at $t+1/2$ of the $i$-th agent is updated according to
\begin{equation}\label{eq:undatesecondhalf}
    x_i(t+1, \theta) \propto \max\left[0, x_i '(t+1/2, \theta) + \mu \Delta x_i '(t + 1/2, \theta)\right]
\end{equation} 
to give the PDF at time $t+1$, concluding the two-step update rule. The proportionality constant is set by normalization.  The learning rate $0.0 < \mu \leq 0.5$ quantifies the susceptibility of an agent to their neighbors' beliefs and is covariant with the duration of each time step; halving $\mu$ is equivalent to doubling the time step without loss of generality.  The inequality $\mu \leq 0.5$ prevents the beliefs from overshooting and is borrowed from the Deffuant-Weisbuch model \cite{deffuant_mixing_2000}.  The displacement 
\begin{equation}\label{eq:xiprimed}
    \Delta x_i '(t + 1/2, \theta) = a_i^{-1} \sum_{i\neq j} A_{ij}[x_j '(t+1/2, \theta) - x_i '(t+1/2, \theta)]
\end{equation}
with $a_i = \sum_{j \neq i} |A_{ij} |$, equals the average difference in belief (at each $\theta$) between agent $i$ and all its neighbors. We only consider connected graphs, where every agent has at least one ally or opponent, i.e.\ $a_i \neq 0$ for all $i$. Equations \eqref{eq:undatesecondhalf} and \eqref{eq:xiprimed} drive an agent's belief towards its allies ($A_{ij} > 0$) and away from its opponents ($A_{ij} < 0$).  Without the maximization operation in equation \eqref{eq:undatesecondhalf}, the antagonistic interaction can produce $x_i(t+1, \theta) < 0$ via equation \eqref{eq:xiprimed}, which is invalid for a probability density; see Section 2.3 in Ref.\ \cite{low_discerning_2022} for more details.  All agents observe the coin toss and combine their prior belief with the result of the coin toss synchronously via equation \eqref{eq:updatefirsthalf} before agents are influenced by positive or negative peer pressure from the network via equation \eqref{eq:undatesecondhalf}, which completes the full time-step from $t$ to $t+1$. 

\subsection{Implementation of obdurate partisans}
\label{subsec:intropartisans}

A partisan is obdurate, in the sense that their opinion does not change with time in response to independent observations or peer pressure. 
In general, the PDF of a partisan can take any time-independent form, viz. $x_i(t,\theta)=x_i(\theta)$.  
In this paper, we  specialize mainly to the situation where partisans hold a single belief, with $ x_i(t,\theta) = \delta(\theta-\theta_{\rm p})$ for all $t$, except in Section \ref{subsec:partisandifferentopinion} where we treat dueling partisans.  
One key goal of the analysis is to test how the behavior of the network depends on $\theta_{\rm p}$. 
For example, does the impact of the partisan increase, as $|\theta_{\rm p} - \theta_0|$ increases?  
In general, when $x_i(\theta)$ is not simply a delta function, one can pose a more sophisticated question: can a partisan increase or decrease their impact on persuadable agents by adjusting the shape of $x_i(\theta)$? 
This question is relevant, when a partisan's beliefs are shaped by a deliberate strategy to manipulate the opinions of other agents in the network, instead of subconscious, psychological factors. Its study is postponed to future work.

In this paper, a partisan's PDF is implemented as a narrow Gaussian distribution with mean $\theta_{\rm p}$ and standard deviation $\sigma_{\rm p} = 10^{-3}$, truncated to the domain $0 \leq \theta \leq 1$.  Although obdurate partisans do not change their opinions, other agents still interact with them, learn from the partisans, and evolve their opinions accordingly.

In this paper, for numerical convenience only, we apply Eqs.\ \eqref{eq:updatefirsthalf} and \eqref{eq:undatesecondhalf} to all agents synchronously, both persuadable and obdurate. We then reset the partisans' PDFs to their initial forms after applying Eq.\ \eqref{eq:updatefirsthalf} and do so again after applying Eq.\ \eqref{eq:undatesecondhalf}, to ensure that the partisans' beliefs are unchanged by observing the coin toss or interacting with other agents. In our implementation, the network graph is undirected, and the update steps in Eqs.\ \eqref{eq:updatefirsthalf} and \eqref{eq:undatesecondhalf} involve matrix multiplications (see \ref{sec:zoomzoom} for a more detailed discussion). Hence it is simpler to reset beliefs instead of checking for partisanship when applying Eqs.\ \eqref{eq:updatefirsthalf} and \eqref{eq:undatesecondhalf} via matrix multiplication.  Alternatively, one can employ a mask matrix to locate partisans in the network, and multiply by the mask matrix when evaluating Eqs.\ \eqref{eq:updatefirsthalf} and \eqref{eq:undatesecondhalf}, which costs additional space and runtime.  If we extend the model, such that $A$ is no longer symmetric, i.e.\ $A_{ij} \neq A_{ji}$, and use a directed graph to represent the network, partisans can be implemented as nodes with zero out-degree.
\footnote{That is, one sets $A_{i{\rm p}} = 0$ and $A_{{\rm p }i} \neq 0$ for all persuadable agents $i$ who are adjacent to the partisan.}

\subsection{Automaton}
\label{subsec:automaton}

We present a discrete-time automaton in Algorithm \ref{alg:algo} to implement the model in Sections \ref{subsec:modelintro} and \ref{subsec:intropartisans} and show clearly how the numerical simulation is set up and run.

\begin{algorithm}
    \caption{Probabilistic discrete-time automaton for the idealized coin bias application}\label{alg:algo}
    \begin{algorithmic}
        \Inputs {
            network topology $A_{ij}$\\
            \mbox{true coin bias $\theta_0$, partisan coin bias $\theta_{\rm p}$, learning rate $\mu$,
            maximum time-step $T$}
        }
        \Initialize {
            $x_p(t=0, \theta) \gets$ truncated Gaussian with mean $\theta_{\rm p}$, standard deviation $10^{-3}$\\  
            \mbox{$x_{i\neq \rm p}(t=0, \theta) \gets$ truncated Gaussian with mean $\in [0, 1]$, standard deviation $\in [0.2, 0.8]$}
        }\\
        \State $t \gets 0$
        \Repeat { (for each time-step $t$)}
            \State Select $S(t) \sim$ Bernoulli $(\theta_0)$ \Comment{Simulate coin toss}
            \State Update beliefs of all agents in response to one coin toss at each time-step (Eq.\ \eqref{eq:updatefirsthalf})
            \State Reset beliefs of partisan agents to $x_p(t=0,\theta)$
            \State Blend beliefs of all agents with network neighbors (Eq.\ \eqref{eq:undatesecondhalf})            
            \State Reset beliefs of partisan agents to $x_p(t=0,\theta)$
            \State $t \gets t + 1$
        \Until{$t = T$}
    \end{algorithmic}
\end{algorithm}

The simulation is initialized with values of $\theta_0, \theta_{\rm p}, \mu, T$ and a network topology $A_{ij}$, which determines the number, connectivity, and political allegiances of the agents. The beliefs of persuadable agents are initialized as truncated Gaussian distributions, with randomly chosen mean in the range $[0.0, 1.0]$ and standard deviation in the range $[0.2,0.8]$.  
That is, one obtains $x_{i\neq {\rm p}} (t=0,\theta) \neq x_{j\neq {\rm p}} (t=0,\theta)$ for all $i \neq j$ in general.  We adopt the Gaussian instead of a uniform distribution, because the right-hand side of Eq.\ \eqref{eq:xiprimed} equals zero for uniform distributions, whereupon none of the agents would subsequently change their beliefs, which is not illuminating.
The beliefs of the partisans are initialized as discussed in Section \ref{subsec:intropartisans}.  For the purpose of numerical simulation, the continuous variable $\theta$ is discretized into 21 regularly-spaced values, $\theta \in \{0.00, 0.05, \ldots, 1.00\}$, following Ref.\ \cite{low_discerning_2022,low_vacillating_2022}\footnote{Interestingly, Ref.\ \cite{tee_quantized_2019} finds that the human brain quantizes probability into $\approx 16$ bins, based on observing the in-game decisions and profits of human gamblers. }.
This reflects the practical reality that human beliefs about media bias are coarse-grained.
\begin{figure}[h]
    \includegraphics[scale=0.7]{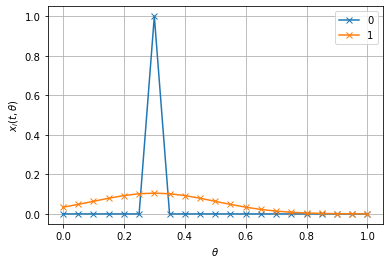}
    \centering
    \caption{Example of the initial beliefs of two agents in a network.  Agent 0 (blue curve) is a partisan with $\theta_{\rm p} = 0.3$.  Agent 1 (orange curve) is a persuadable agent whose initial belief PDF has mean  0.3 and standard deviation 0.2 (Gaussian truncated to $0 \leq \theta \leq 1$).}
    \label{fig:init_belief}
\end{figure}

At each time step in Algorithm \ref{alg:algo}, a coin toss is simulated by a Bernoulli trial with success probability $\theta_0$.  

We are interested in the long-term behavior of the persuadable agents in the system.  For example, does the PDF of a particular persuadable agent tend to a constant function of $\theta$ or does it fluctuate indefinitely, as in the phenomenon of turbulent nonconvergence observed in Ref.\ \cite{low_vacillating_2022,mobilia_role_2007}?  If the belief of an agent remains unchanged within some tolerance (e.g. 1\%) for a user-selected amount of time $\tau_{\rm max}$, viz.\
\begin{equation} \label{eq:asymlearncondition}
    \max_{\theta} |x_i(t+\tau, \theta) - x_i(t, \theta)| < 0.01 \max_{\theta} |x_i(t, \theta)| \quad \text{for } 1 \leq \tau \leq \tau_{\rm max} , 
\end{equation}
then we say that the agent achieves asymptotic learning. If all the agents achieve asymptotic learning, viz.\
\begin{equation} \label{eq:sysasymlearncondition}
    \max_{\theta} |x_i(t+\tau, \theta) - x_i(t, \theta)| < 0.01 \max_{\theta} |x_i(t, \theta)| \quad \text{for } 1 \leq \tau \leq \tau_{\rm max} \text{ and all persuadable } i, 
\end{equation}
then we say that the system achieves asymptotic learning.  In this paper, we take $\tau_{\rm max} = 99$ typically, an arbitrary choice also made in Ref.\ \cite{low_discerning_2022}.  
If Eq.\ \eqref{eq:sysasymlearncondition} holds for $t \geq t_{\rm a}$, we call $t_{\rm a}$ the asymptotic learning time.  
The automaton iterates until the system achieves asymptotic learning, or the maximum run-time $T$ is reached.  
In this paper, we adopt the convention that the partisans are not part of the conditions \eqref{eq:asymlearncondition} and \eqref{eq:sysasymlearncondition} for asymptotic learning; they are obdurate, so they do not learn anything.

We aim to discover
\begin{inparaenum}[(i)]
    \item if partisans change $t_{\rm a}$, or even prevent the system from achieving asymptotic learning at all; and 
    \item if partisans mislead the persuadable agents to infer the bias of the coin incorrectly. 
\end{inparaenum}
Both questions (i) and (ii) have been asked in the context of deterministic models previously \cite{mobilia_does_2003,mobilia_role_2007,yildiz_binary_2013,mobilia_voting_2005,yildiz_discrete_2011,yildiz_opinion_2021,abrahamsson_opinion_2019,galam_role_2007,ghaderi_opinion_2014,klamser_zealotry_2017}, but they are raised here in the context of media bias and Bayesian learners for the first time. 
First, we consider allies-only networks in Section \ref{sec:alliesonly} and investigate how different numbers of partisans disrupt the beliefs of persuadable agents.  
We then investigate how partisans disrupt opponents-only networks in Section \ref{sec:opponentonly}.
Later, in Section \ref{sec:mixed}, we explore networks with both allies and opponents.


\subsection{Mathematical model of networks: complete versus Barab\'{a}si-Albert}
\label{subsec:network}
We use mathematical graphs to model the sociopolitical connections between agents. In this paper we study complete graphs, which ensure that all agents (partisan and persuadable) are connected to each other.   This is not realistic in most actual social contexts, where everybody does not know everybody else.
However, the central goal of this paper is to quantify how partisans disrupt opinion formation among persuadable agents.  
This is hard to test in a controlled fashion, when the network is partly connected.  
For example, some persuadable agents may connect mostly with partisans, while others may connect mostly with fellow persuadable agents, and the two groups exchange members unpredictably, when partly connected networks are generated randomly.  
In contrast, it is easier to perform controlled tests on a complete network, where everybody knows the same number of partisans or persuadable agents.  
In this paper, we investigate systematically how the connectivity of partisans affects opinion formation by adjusting the fraction of partisans in a complete network in Sections \ref{subsec:dwelltime_and_frac} and \ref{subsec:switching_between_belief}, secure in the knowledge that every persuadable agent knows every partisan in every random realization of the system. 
In future work, we intend to grapple with the challenge of partly connected networks and generalize the results to Barab\'{a}si-Albert networks, for instance, following Low \& Melatos \cite{low_discerning_2022,low_vacillating_2022}.
(Some preliminary simulations involving Barab\'{a}si-Albert networks are presented in Sections \ref{subsec:wrong_conclusion_first} and \ref{sec:BA}.)  
Barab\'{a}si-Albert networks are scale-free; their degree distribution follows a power law, which is a good approximation to many real-world networks \cite{barabasi_emergence_1999,tang_survey_2016,kumar_structure_2016,maniu_building_2011}.
In future work, for example, one can investigate the behavior of persuadable agents indirectly connected to a partisan via a chain of allies and/or opponents, and compare the influence of highly versus sparsely connected partisans.

In the studies by Mobilia \etal on zealotry \cite{mobilia_does_2003,mobilia_voting_2005,mobilia_role_2007,mobilia_nonlinear_2015}, the network is implemented on a $d$-dimensional cubic lattice of size $(2L + 1)^d$, where each agent (voter) is labelled by a vector $r$ with components $-L \leq r_i \leq L$ and $1 \leq i \leq d$. 
The lattice model used by Mobilia \etal is not a complete network as each node is only connected to its nearest neighbors.  Nor is it scale-free, as the degree distribution does not follow a power law; the degree of each node is constant and equals $2d$ for a $d$-dimensional cubic lattice.
The system is evolved by picking a random agent and checking whether or not they are partisans (zealots).  If they are partisans, nothing happens, as a partisan's opinion never changes. If not, the agent adopts the opinion of a random nearest neighbor.
This approach differs from the current paper, where every agent interrogates the PDF of every other agent to whom they are connected at every time step (as well as the public coin toss, of course).

\subsection{Relation to other models}
\label{subsec:relationtoothers}
The model introduced in previous sections is distinct from but related to models in the literature, particularly the DeGroot model \cite{degroot_reaching_1974} and the Deffuant-Weisbuch model \cite{deffuant_mixing_2000}.  
In the Deffuant-Weisbuch model, the adjacency matrix is modified according to  $A_{ij} \neq 0 $ for $|x_j(t) - x_i(t)| < \epsilon$, i.e.\ when the opinions of two agents differ sufficiently, they stop communicating with each other.  In contrast, the model in Section \ref{subsec:modelintro} holds $A_{ij}$ constant in time and therefore neglects the time-dependent ``echo chamber'' or ``silo'' effect captured by the Deffuant-Weisbuch model \cite{deffuant_mixing_2000}.
The belief displacement $\Delta x_i'$ in Eq.\ \eqref{eq:undatesecondhalf} takes inspiration from the DeGroot model, which averages the opinion of an agent's allies. 
The DeGroot and Deffuant-Weisbuch models are deterministic.  
A few models do exist that describe the opinion of an agent as a PDF like in this paper, including \cite{fang_opinion_2020,fang_social_2019,jadbabaie_non-bayesian_2012} for example.  
In particular, Fang \etal's \cite{fang_social_2019} model allows the external signal to be interpreted differently by agents, unlike in this paper, where the external signal is broadcast to and accepted by every agent.
Moreover, Fang \etal \cite{fang_social_2019} implemented a dynamic network, in which $A_{ij}$ depends on $x_i(t,\theta) - x_j(t,\theta)$, whereas $A_{ij}$ is static in this paper.
A more detailed comparison between the model in this paper and others is presented in Section 2.4 in Ref.\ \cite{low_discerning_2022}.  

Most existing work investigating zealots treats the opinions of persuadable agents deterministically, i.e.\ defined by a single number rather than a PDF.  Often, the opinions are also binary, e.g.\ voting for one party or the other. The voter model \cite{liggett_stochastic_nodate,castellano_statistical_2009}, for example, represents the opinion of agents by a discrete spin value ($+1$ or $-1$).  
Some papers about zealots test for the long-term emergence of a consensus, implying general agreement among all agents, whereas the notion of asymptotic learning in Eq.\ \eqref{eq:asymlearncondition} allows agents to believe steadily in a mixture of views, e.g. the $i$-th agent may hold two opinions $\theta_1$ and $\theta_2 \neq \theta_1$ with equal confidence for $\tau_{\rm max}$ consecutive time steps.

We are not aware of any model that considers antagonistic interactions between partisans and agents in the context of the media bias problem involving Bayesian learners. However, extensions of the DeGroot and Deffuant-Weibuch models certainly exist, that consider antagonistic interactions without partisans \cite{vaz_martins_mass_2010,shi_evolution_2016,chen_opinion_2019,he_discrete-time_2021}. The latter papers focus on the divergence of opinion, when antagonism is introduced. 

\section{Allies only}
\label{sec:alliesonly}

We prepare to study mixed networks containing allies and opponents by investigating first the impact of partisans in simpler networks containing allies only. A representative example of the evolution of an allies-only network is presented in Section \ref{subsec:samethetap}. It exemplifies the general result, that even a single partisan prevents the system from achieving equilibrium (asymptotic learning), because the coin and the partisan ``pull'' the persuadable agents in opposite directions. 
The distribution of dwell times in specific states, and the role played by the fraction of partisans in the network, are explored in Sections \ref{subsec:dwelltime_and_frac} and \ref{subsec:switching_between_belief} respectively. 
Section \ref{subsec:partisandifferentopinion} explores the opinion dynamics when there exist partisans with different beliefs.  The special cases $\theta_0=0$ and $\theta_0 = 1$ are discussed in \ref{sec:twoallies}. 

For the rest of the paper, except where stated otherwise, we take $\theta_0=0.6$, $\theta_{\rm p} = 0.3$, and $\mu = 0.25$. These values are arbitrary but representative.

\subsection{Disruption by a single partisan}
\label{subsec:samethetap}

\begin{figure}[h!]
    \centering
    \begin{subfigure}{0.45\textwidth}
        \centering
        \includegraphics[width=\linewidth]{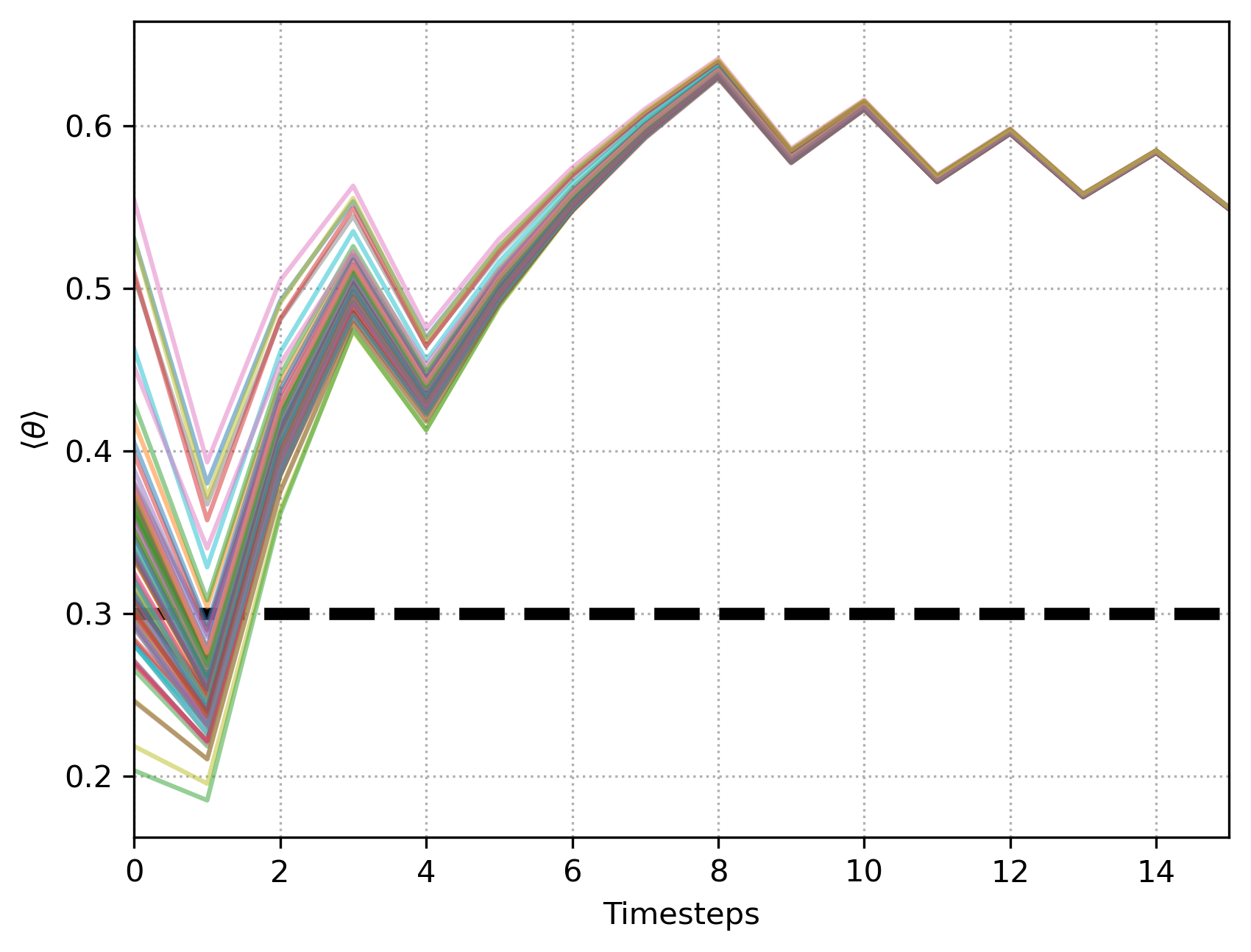}
        \caption{}\label{fig:meanconvergent}
    \end{subfigure}%
    \hspace*{\fill}  
        \centering
        \begin{subfigure}{0.45\textwidth}
        \includegraphics[width=\linewidth]{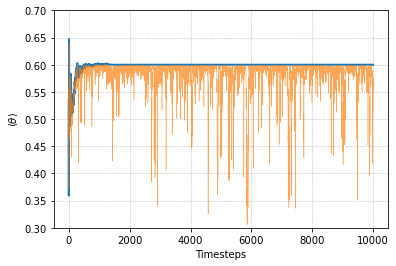}
        \caption{}\label{fig:meannonpartisan}
    \end{subfigure} 
    \begin{subfigure}{0.45\textwidth}
        \includegraphics[width=\linewidth]{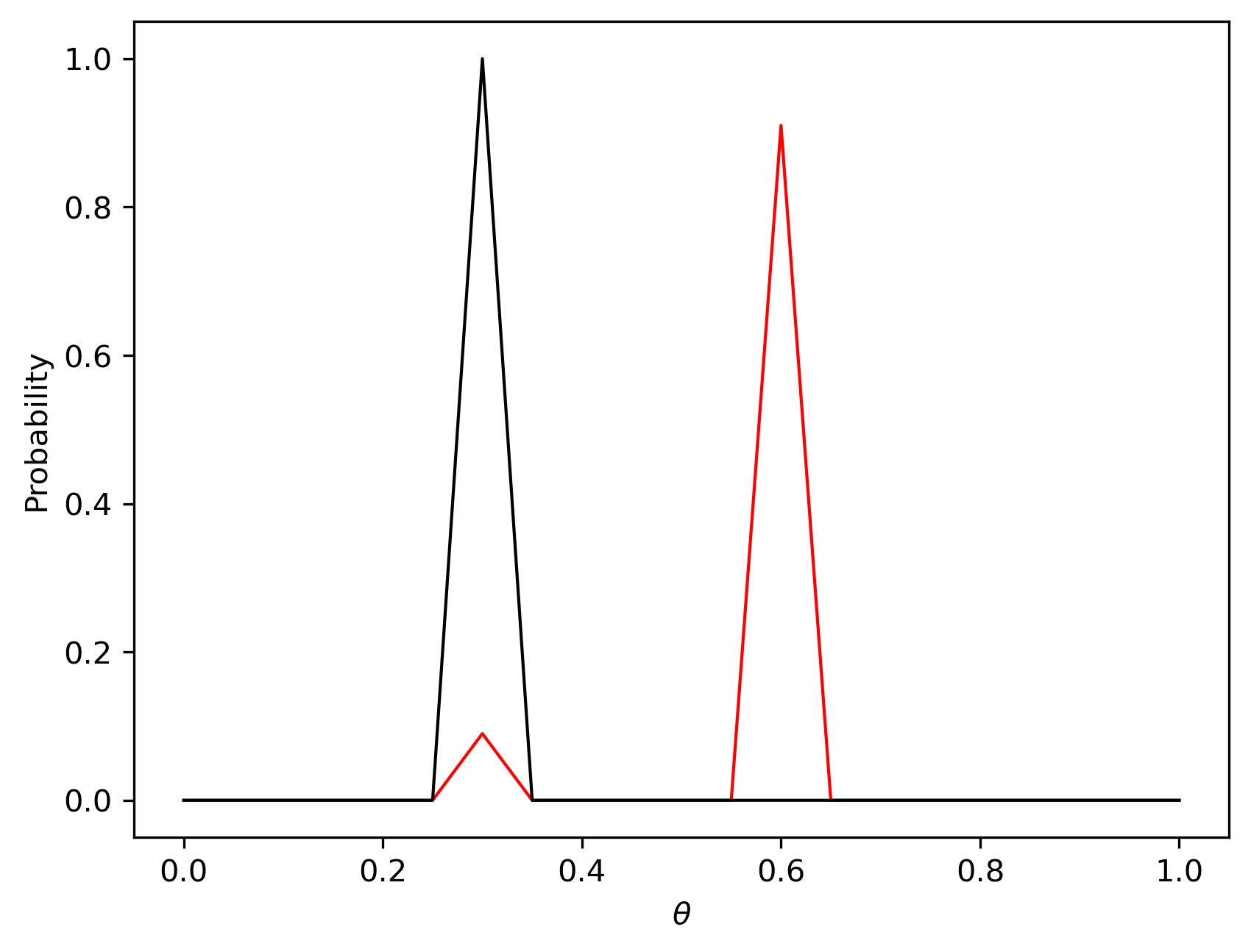}
        \caption{}\label{fig:0.6truebias}
    \end{subfigure}
    \hspace*{\fill}
    \begin{subfigure}{0.45\textwidth}
        \centering
        \includegraphics[width=\linewidth]{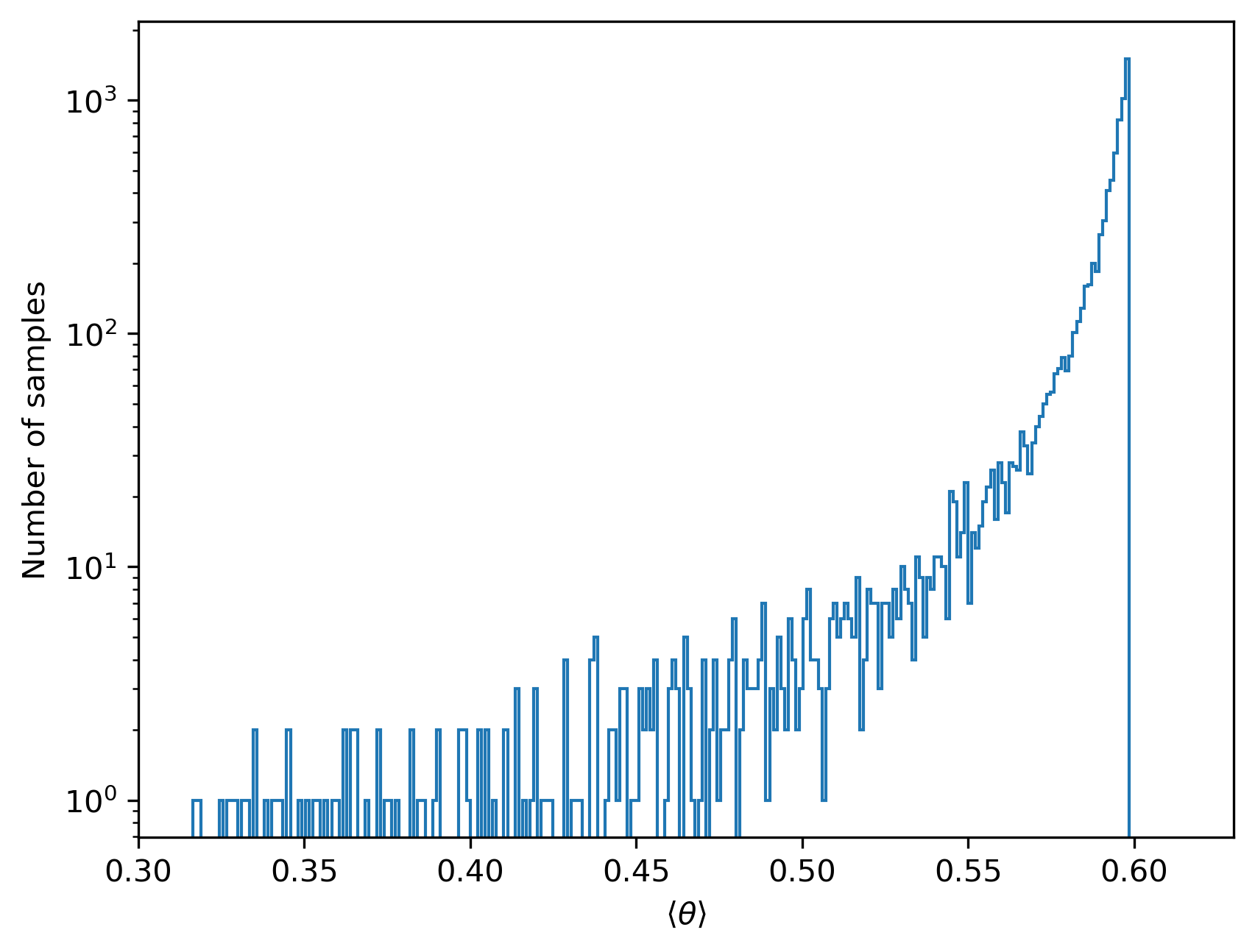}
        \caption{}\label{fig:meanhist}
    \end{subfigure}
    \caption{Disruption of equilibrium in an allies-only network with 99 persuadable agents and one partisan. 
    (a) Rapid formation of consensus among persuadable agents: mean belief $\langle \theta \rangle$ of every agent versus time $t$ near the start of the simulation ($0 \leq t \leq 15$). The colored and black dashed curves correspond to the persuadable agents and partisan respectively. 
    (b) Inhibition of asymptotic learning among persuadable agents: $\langle\theta\rangle$ versus $t$ for an arbitrary persuadable agent (representative of the persuadable consensus; orange curve) throughout the full simulation ($0\leq t \leq 10^4$), demonstrating that $\langle \theta \rangle$ never asymptotes to a fixed value, compared to the situation without a partisan (blue curve), where $\langle\theta\rangle$ asymptotes to $\theta_0$. 
    (c) Bimodal belief PDF snapshot at $t = 2400$ with $x_i(2400,\theta_0) = 0.09$ and $x_i(2400,\theta_{\rm p}) = 0.91$ for an arbitrary persuadable agent (red curve) and the partisan (black curve).  
    (d) Histogram of $\langle \theta \rangle$ sampled at every time step in the interval $2 \times 10^3\leq t \leq 1\times 10^4$, yielding $\langle\theta\rangle \geq 0.56$ for 90\% of the time. Persuadable agents usually believe in $\theta_0$ and rarely believe in $\theta_{\rm p}$. The jaggedness at $\langle\theta\rangle \lesssim 0.55$ is caused by small number statistics per bin. }
    \label{fig:meanconverge}
\end{figure}

Consider an allies-only network with $n=100$, containing 99 persuadable agents, and one partisan labelled `agent p'. We run a number of simulations with randomized priors and coin toss sequences. In this section, we focus on a representative simulation, whose output is plotted in Fig.\ \ref{fig:meanconverge}.
We find that even a single partisan disrupts the perceptions of the persuadable agents, by dragging them away from believing in the true coin bias. 
\ref{sec:coinvsprior} discusses how the priors and coin toss sequences influence the long-term behavior of the system. 
In short, the priors are largely inconsequential, but a specific subsequence of coin tosses (say, five heads in a row) can have an appreciable impact on persuadable agents.


In a network containing only allies, persuadable agents' beliefs converge on a relatively short timescale of order 10 time steps, before vacillating between $\theta_0$ and $\theta_{\rm p}$ over the longer term.  Fig.\  \ref{fig:meanconvergent} and Fig.\ \ref{fig:meannonpartisan} display the evolution of the mean belief $\langle\theta\rangle$ in one particular simulation.  
The bold black dashed line represents the partisan, whose mean belief is constant.  
We run an ensemble of simulations with randomized prior and coin toss sequences, choosing different values for $\theta_0$ and $\theta_{\rm p}$ but excluding the special case of $\theta_0 = 0$ or 1 (i.e.\ the coin returns heads or tails only).
For brevity, we present the simulation described in Fig.\ \ref{fig:meanconverge} to demonstrate how a single partisan disrupts the belief of persuadable agents.
Similar behavior is observed throughout this ensemble for all different $\theta_0 \neq 0, 1, \theta_{\rm p}$.  

Over the short term, the persuadable agents reach a consensus, in the sense that they agree among themselves, viz.\
$\text{max}_{\theta} |x_i(t,\theta) -x_j(t,\theta) | < \epsilon \text{ max}_{\theta} \left[ x_i(t, \theta), x_j(t, \theta) \right]$, 
for any pair of persuadable agents $i$ and $j$, where $\epsilon = 0.01$ is a user-selected tolerance.  
This is observed in Fig.\ \ref{fig:meanconvergent}, where $\langle \theta \rangle$ converges rapidly to approximately the same value for all 99 persuadable agents over $\sim 10$ time steps. 
Visually this is observed as a reduction in the spread of the 99 colored, zig-zag curves, as time passes. 
The zig-zag dynamics are a typical response to the coin: if it comes up heads, $\langle\theta\rangle$ increases, otherwise $\langle\theta\rangle$ decreases. For example, $\langle\theta\rangle$ increases monotonically for $4 \leq t \leq 8$, because the coin returns four heads in a row. 
At $t=14$, one has $\langle \theta \rangle \approx 0.58 \neq \theta_{\rm p}$ for all 99 persuadable agents; that is, the partisan is yet to exert their influence fully.

Consensus is not the same as asymptotic learning. Over the long term, the persuadable agents vacillate as a group between believing in $\theta=\theta_0$ and $\theta=\theta_{\rm p}$, maintaining consensus among themselves. This is observed in Fig.\ \ref{fig:meannonpartisan} where $\langle \theta \rangle$ is plotted versus time for an arbitrary, representative, persuadable agent.  
Without any partisans, namely the situation shown by the blue curve in Fig.\ \ref{fig:meannonpartisan}, one finds $\theta \rightarrow \theta_0$ monotonically for large $t$, irrespective of the coin toss sequence, as in Fig.\ 2 in Ref.\ \cite{low_discerning_2022}. 
In contrast, the orange curve in Fig.\ 2b fluctuates in the range $0.32 \leq \langle\theta\rangle \leq 0.60$. The partisan disrupts the beliefs of the persuadable agents, causing them to vacillate indefinitely, even when there is only one partisan among 99 persuadable agents. 
The mean belief $\langle\theta\rangle$ spends $\approx 86 \%$ and $0.04 \%$ of the $8\times 10^3$ timesteps (after the initial transient $0\leq t \leq 2\times 10^3$) in Fig.\ \ref{fig:meanhist} near $\theta_0$ and $\theta_{\rm p}$ respectively, with $ | \langle\theta\rangle - \theta_0 | \leq 0.025 \theta_0$ and $ | \langle\theta\rangle - \theta_{\rm p} | \leq 0.025 \theta_{\rm p}$, and spends $\approx 14\%$ of the time near neither value. The dwell time in each state is quantified further in Section \ref{subsec:switching_between_belief}.

Once the persuadable agents gather enough information to move beyond their initial priors, at $t \gtrsim 2\times 10^3$, their belief PDF vanishes uniformly except at $\theta_0$ and $\theta_{\rm p}$, as shown in  Fig.\ \ref{fig:0.6truebias}.  
The heights of the peaks at $\theta_0$ and $\theta_{\rm p}$ fluctuate continuously. Fig.\ \ref{fig:meanhist} shows a histogram of $\langle \theta \rangle$ for $2 \times 10^3\leq t \leq 1\times 10^4$, with $\langle\theta\rangle$ sampled at each of the $8\times 10^3$ intervening time steps.
With only one partisan, we find $\langle\theta\rangle \geq 0.56$ during 90\% of the time. That is, the persuadable agents prefer $\theta_0$ without rejecting $\theta_{\rm p}$ completely. These preferences reverse, as the fraction of partisans increases, as discussed in Section \ref{subsec:switching_between_belief}. 
When the persuadable agents form a consensus around the true coin bias, with $x_j(t,\theta) \approx \delta(\theta - \theta_0)$ for most $j$, the displacement $\Delta x_i'$ in Eq.\ \eqref{eq:xiprimed} is dominated by the ``pull'' of the disagreeing partisan, who has $x_{\rm p}(t,\theta) = \delta(\theta - \theta_{\rm p})$ with $\theta_{\rm p} \neq \theta_0$. All the persuadable agents feel the partisan's ``pull'' simultaneously, as well as the influence from the coin toss described by Eqs.\ \eqref{eq:updatefirsthalf} and \eqref{eq:likelihood}. 
Hence the characteristic timescale to move away from $\langle \theta \rangle \approx \theta_0$ depends on $\mu^{-1}$ and the coin toss sequence. 
On the other hand, when the persuadable agents form a consensus around the partisan's belief, with $x_j(t,\theta)=\delta(\theta-\theta_{\rm p})$ for most $j$, the internal signals and hence the displacement $\Delta x_i'$ are small, and the coin toss dominates the update rule via Eq.\ \eqref{eq:updatefirsthalf} and \eqref{eq:likelihood}. The characteristic timescale to move away from $\langle\theta\rangle \approx \theta_{\rm p}$ depends only on the coin toss sequence.  
Hence, on balance, the persuadable agents feel a more durable ``pull'' when the consensus is formed around $\theta_0$ instead of $\theta_{\rm p}$. 

The behavior of a network containing one partisan differs from Fig.\ \ref{fig:init_belief}, if the agents' beliefs are deterministic (single value per agent), or the update rule does not involve an external signal. 
For example, in the voter model analyzed by Mobilia et al.\ \cite{abrahamsson_opinion_2019,mobilia_does_2003,mobilia_voting_2005}, every agent's opinion converges to the partisan and subsequently remains unchanged. Once consensus is reached, the system enters a steady state, because the interaction between agents involves randomly adopting the belief of a neighbor, and all neighbors hold the same belief.
In Ref.\ \cite{abrahamsson_opinion_2019,mobilia_does_2003,mobilia_voting_2005}, the persuadable agents only change their belief in response to their neighbors, whereas the update rule in Section \ref{subsec:modelintro} responds to both the coin toss and the neighbors.  
When persuadable agents are exposed to two or more sources of contradictory information about $\theta$ (e.g.\ the coin tosses and the partisan), their beliefs do not settle to a steady state, as Fig.\ \ref{fig:meannonpartisan} illustrates. The disruption occurs even if there is only one partisan in a large network with $n \gg 1$ agents.

\subsection{Dwell times and the fraction of partisans in a network}
\label{subsec:dwelltime_and_frac}

\begin{figure}[h]
    \begin{subfigure}{0.5\textwidth}
        \centering
        \includegraphics[width=\linewidth]{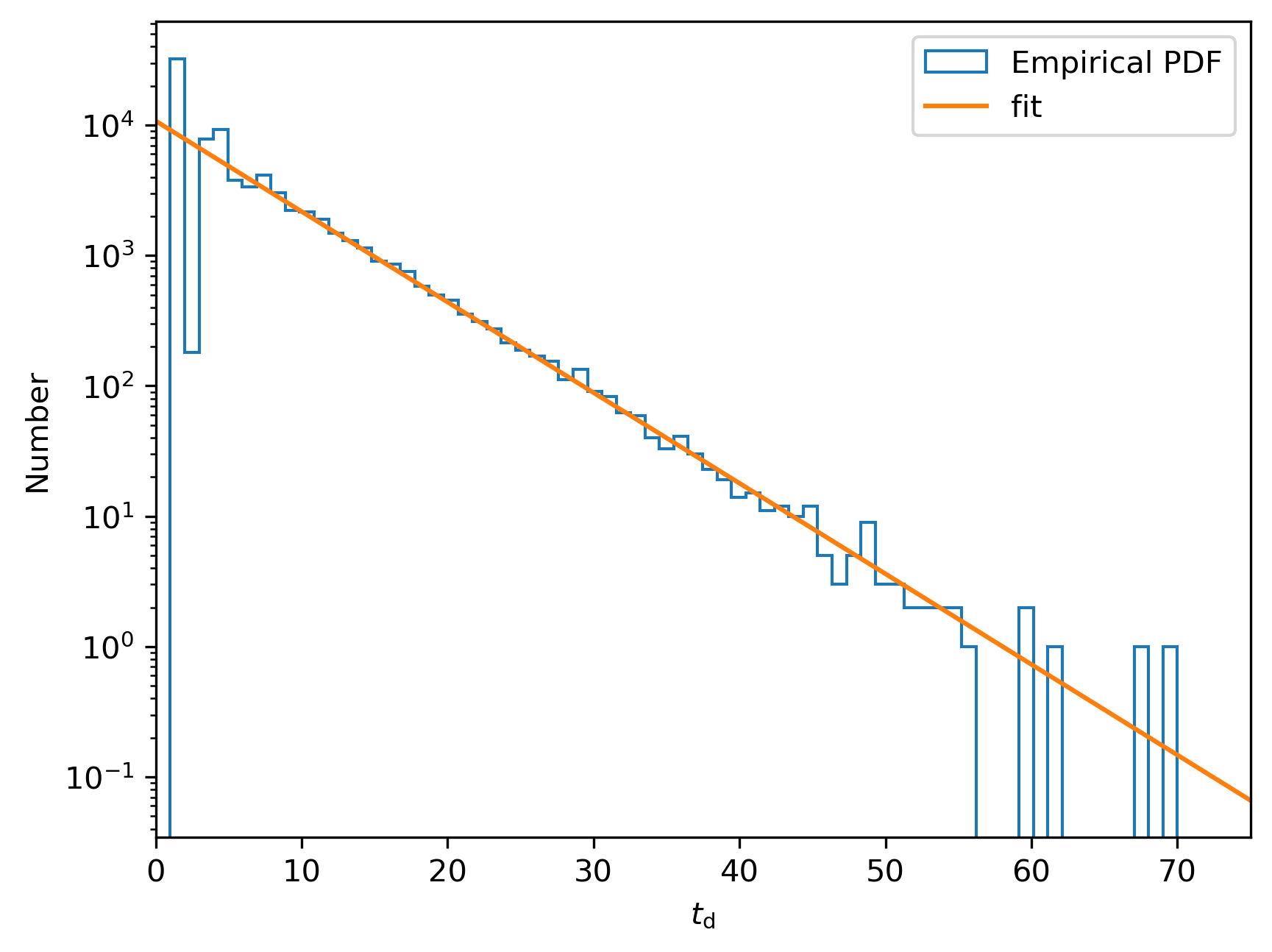}
    \caption{}\label{fig:dwell_log_fit}
    \end{subfigure} 
    \hspace*{\fill}
    \begin{subfigure}{0.5\textwidth}
        \centering
        \includegraphics[width=\linewidth]{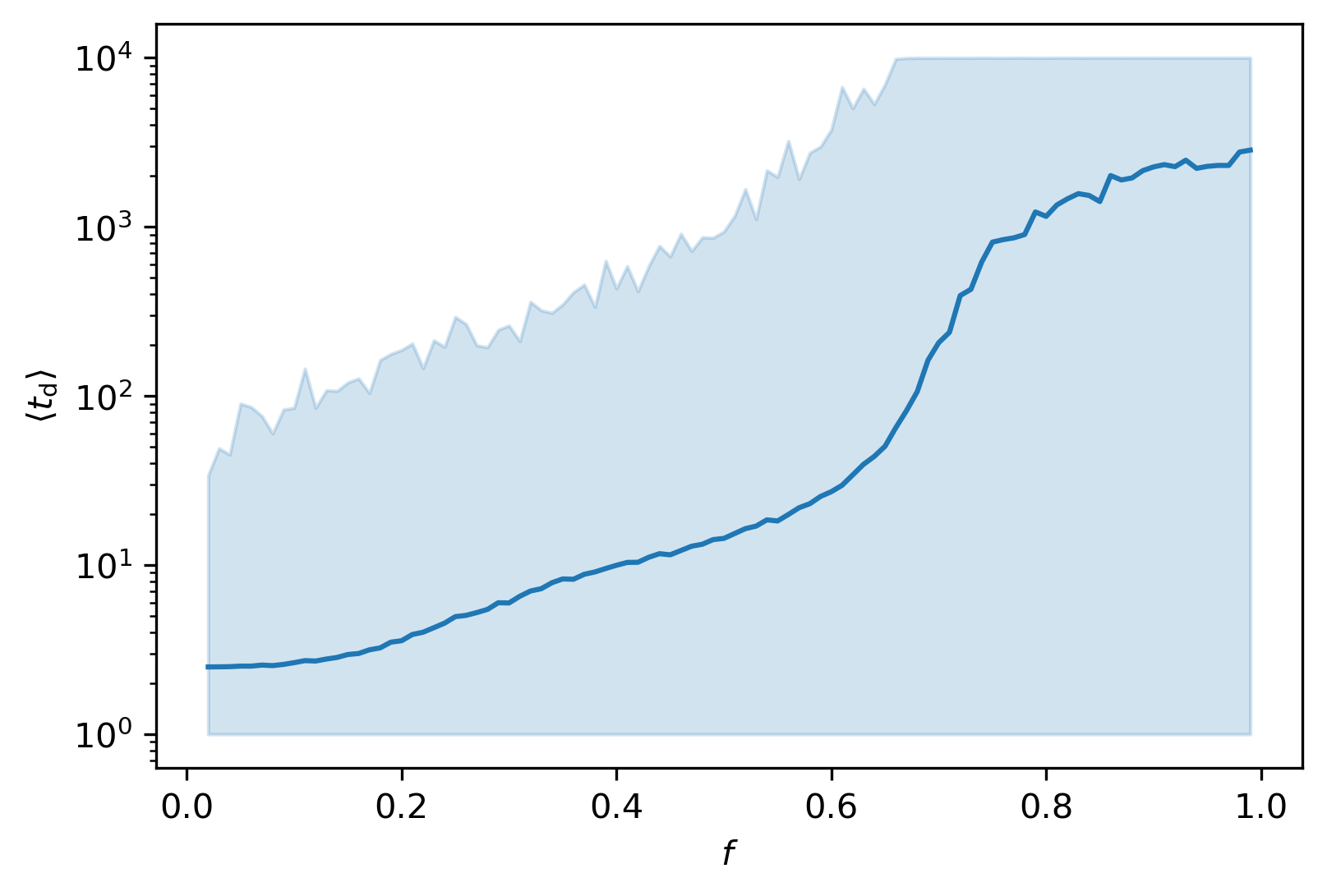}
    \caption{}\label{fig:dwell_vs_frac_partisan}
    \end{subfigure} 
    \caption{ 
        Statistics of dwell times $t_{\rm d}$ defined by Eq.\ \eqref{eq:dwelltime} in an allies-only network with one or more partisans.
        (a) Dwell time PDF $p(t_{\rm d})$ plotted as a histogram (blue curve) for a network containing 99 persuadable agents and one partisan and a single realization of the coin toss sequence, which produces 80,698 dwell time samples across $T=10^6$ total time steps. The orange curve indicates an exponential best fit of the form $p(t_{\rm d}) \propto \exp(-0.16 t_{\rm d})$ with coefficient of determination statistic $R^2 > 0.9$ \cite{javed_probability_nodate}.
        (b) Mean dwell time $\langle t_{\rm d} \rangle$ versus the partisan fraction $f$ in allies-only networks with $n=100$. For each value of $f$ in the range $0.01 \leq f \leq 0.99$, we run an ensemble of 100 simulations for $T=10^4$, where the coin tosses are independent and identically distributed. The solid blue curve indicates the ensemble mean, and the light blue shading spans the ensemble minimum and maximum. }
\end{figure}

In the presence of a partisan, persuadable agents never settle in their beliefs, and the asymptotic learning time defined by Eq.\ \eqref{eq:asymlearncondition} is unsuitable to quantify the long-term behavior of the system.  
Eq.\ \eqref{eq:sysasymlearncondition} holds temporarily, but in the long run we observe turbulent nonconvergence, as observed in Section 4.3 of Ref.\ \cite{low_discerning_2022} for different reasons. 
To quantify the behavior of the persuadable agents, we define the dwell time $t_{\rm d}$ such that 
\begin{equation} \label{eq:dwelltime}
    \max_{\theta} |x_i(t', \theta) - x_i(t, \theta)| < \delta \max_{\theta} |x_i(t, \theta)|
\end{equation}
is true for $t < t' \leq t + t_{\rm d}$ and $\delta = 0.01$ is a user-selected tolerance.
Eq.\ \eqref{eq:dwelltime} defines blocks of time, where the system dwells in the same state to a good approximation, e.g.\ the system may spend $t_1 \leq t' \leq t_1+t_{\rm d1}$ with $\langle\theta\rangle \approx \theta_0$, then $t_2 \leq t' \leq t_2+t_{\rm d2}$ with $\langle\theta\rangle \approx \theta_{\rm p}$, and so on. The blocks of time do not overlap, nor are they necessarily contiguous; in the above examples, one may have $t_2 > t_1 + t_{\rm d1}$, if the two blocks are separated by an interval of turbulent nonconvergence.
Fig.\ \ref{fig:dwell_log_fit} shows a histogram of the 80,698 dwell times observed in an allies-only network with 99 persuadable agents, one partisan, and $T = 10^6$. For this network, dwelling rarely lasts for more than 50 time steps. 
The dwell time PDF $p(t_{\rm d})$ is exponential to a good approximation, with $p(t_{\rm d}) \propto \exp(-0.16 t_{\rm d})$ and hence $\langle t_{\rm d} \rangle \approx 5.4$\footnote{Note that $ \langle t_{\rm d} \rangle ^{-1}= (5.4)^{-1}$ does not equal 0.16 exactly, as expected for an exponential, because we cannot fit reliably to long dwell intervals ($t_{\rm d} >55$) due to small number statistics per bin.}.
We also find that 90\% of the dwell times satisfy $t_{\rm d} \leq 13$.
Similar behavior is observed in networks with no partisans.
For example, Fig.\ 5 in Ref.\ \cite{low_vacillating_2022} shows a simulation with $T = 10^6$ coin tosses, with the network generated using the Barab\'{a}si-Albert model with $n = 100$ and attachment parameter $m = 3$ \cite{hagberg_exploring_2008}. 
An exponential trend is also observed with best fit $p(t_{\rm d}) \propto \exp(-0.0069 t_d)$\footnote{
    Strictly speaking, Ref.\ \cite{low_vacillating_2022} distinguishes between two versions of the dwell time, termed turbulent ($t_{\rm t}$) and stable ($t_{\rm s}$), which are defined by Eq.\ \eqref{eq:dwelltime} with the conditions $t_{\rm d} < 100$ and $t_{\rm d} \geq 100$ respectively. 
    We cannot draw this distinction usefully in this paper, as we find ${\rm max}(t_{\rm d})< 100$ for most realizations of a complete network. 
    Nevertheless, it is interesting that all approaches lead to exponential PDFs for $p(t_{\rm d})$ (this paper), $p(t_{\rm t})$ (Ref.\ \cite{low_vacillating_2022}), and usually $p(t_{\rm s})$ (Ref.\ \cite{low_vacillating_2022}).
}.

Fig.\ \ref{fig:dwell_vs_frac_partisan} shows how the dwell time depends on the fraction of partisans, $f$, with $0.01 \leq f \leq 0.99$ in steps of $0.01$.
When $f$ is low, the mean dwell time $\langle t_{\rm d} \rangle$ is short, and the persuadable agents frequently change their beliefs.  
As $f$ increases, $\langle t_{\rm d} \rangle$ and $\max(t_{\rm d})$ increase. 
More partisans produce a stronger internal signal between the persuadable agents and partisans, strengthening the ``pull'' towards $\theta_{\rm p}$ and away from $\theta_0$. 
For $f \gtrsim 0.6$, $\langle t_{\rm d} \rangle$ and $\max(t_{\rm d})$ rise sharply (note the logarithmic scale), while $\min(t_{\rm d}) = 1$ remains unchanged. 
Short dwell times occur early in the simulations, when the persuadable agents start to gather information about the coin sequence and partisans and are still influenced strongly by their initial priors. 
Long dwell times occur later in the simulations; for example, one obtains $\max(t_{\rm d}) \approx T = 10^4$, when the persuadable agents maintain their beliefs until the end of the simulation, once the initial transient dies out.
The flatter trend in $\langle t_{\rm d} \rangle$ for $f \gtrsim 0.8$ is caused by the ceiling $t_{\rm d} \leq T$.

The histogram in Fig.\ \ref{fig:dwell_log_fit} contains 80,698 dwell times covering a total of 435,080 out of $T=10^6$ time steps. Hence, for $f=10^{-2}$, the system spends 43.5\% of its time dwelling near $\theta_0$ or $\theta_{\rm p}$ and the rest of its time fluctuating in a state of turbulent nonconvergence similar to the one identified in Ref.\ \cite{low_vacillating_2022} (in the latter paper, the turbulence is caused by different dynamics unrelated to the presence of partisans). 
The fraction of time spent in turbulent nonconvergence depends on $f$. 
For example, at $f=0.5$, one finds 208 dwell time intervals covering a total of 2997 out of $T=10^4$ time steps, and the system spends 70\% of its time in a state of turbulent nonconvergence, whereas at $f=0.9$, one finds typically four dwell time intervals covering a total of 9952 out of $T=10^4$ time steps, and the system spends 0.4\% of its time in a state of turbulent nonconvergence. 
When $f$ is high, the ``pull'' from $fn$ partisans ${\rm p}_1,\dots,{\rm p}_{fn}$ on an arbitrary persuadable agent $i$ scales as $A_{i{\rm p}_1}+\dots+A_{i{\rm p}_{fn}}$ and is relatively large, which prevents the persuadable agents from changing their mind. 
Hence, we only observe turbulent nonconvergence early in the simulation, before persuadable agents dwell for a long time at $\theta_{\rm p}$. 

\subsection{Switching between beliefs}
\label{subsec:switching_between_belief}
What does $x_i(t,\theta)$ look like, when an agent dwells near some belief, and how does switching between beliefs correlate with $t_{\rm d}$ statistically? 
The picture is complicated, because $x_i(t,\theta)$ may peak at $\theta_0$, $\theta_{\rm p}$, or both $\theta_0$ and $\theta_{\rm p}$ at the same time, as shown in Fig.\ \ref{fig:0.6truebias}. Theoretically, to categorize the belief at each time step in a dwell interval, we should define an orthogonal set of belief templates and match $x_i(t,\theta)$ against all of them. 
However, this approach is challenging numerically, when $x_i(t,\theta)$ is continuously valued, and the number of belief templates is large. 
Instead, we assume that each dwell interval is characterized approximately by an average belief, defined by $\langle\theta\rangle$ at the final time step in the interval. We can then compare the belief and dwell statistics efficiently.


\begin{figure}[h]
    \centering
    \begin{subfigure}{0.5\textwidth}
        \centering
        \includegraphics[width=\linewidth]{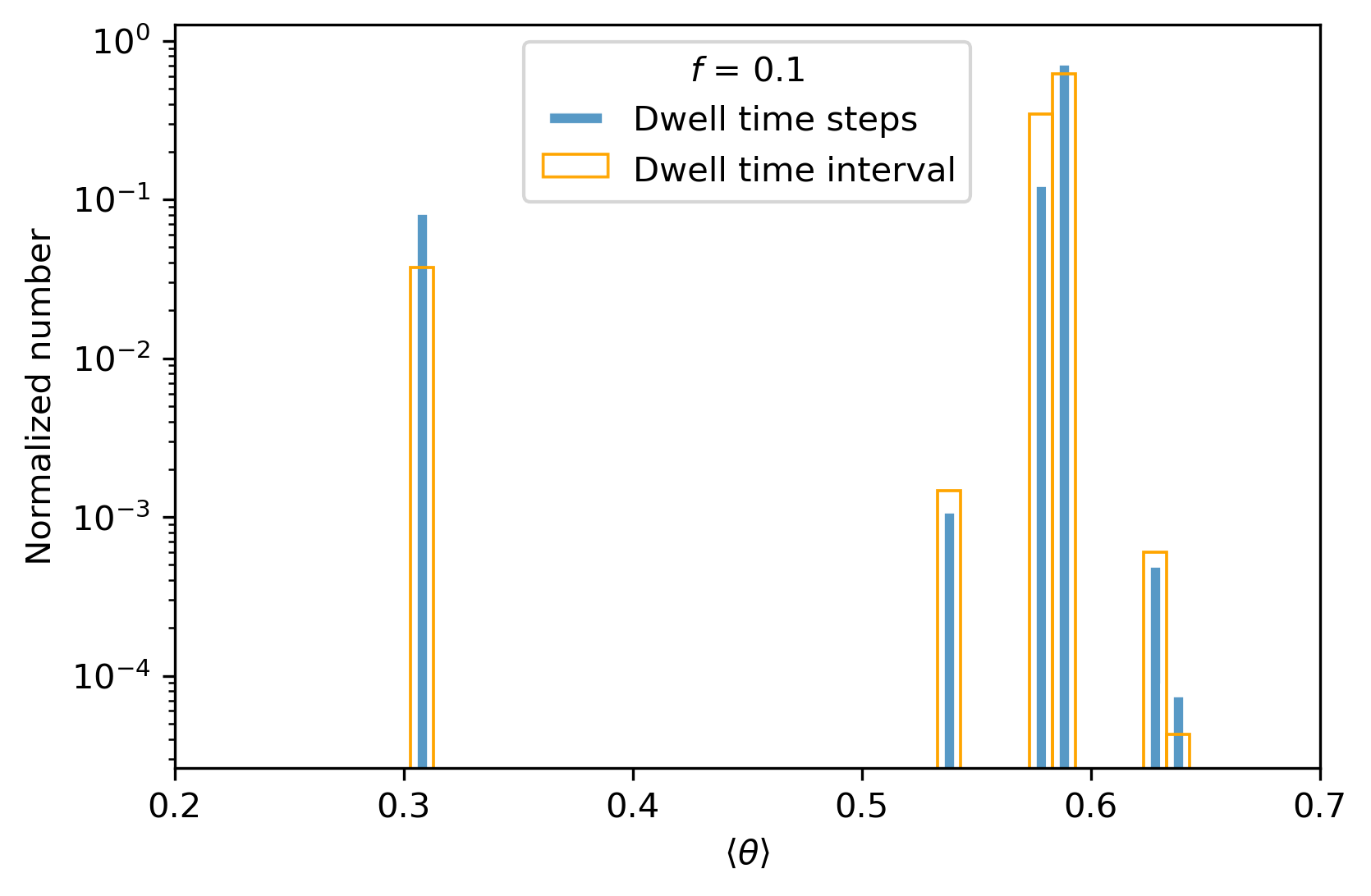}
        \caption{}\label{fig:dt_dti_0.1}
    \end{subfigure}%
    \begin{subfigure}{0.5\textwidth}
        \centering
        \includegraphics[width=\linewidth]{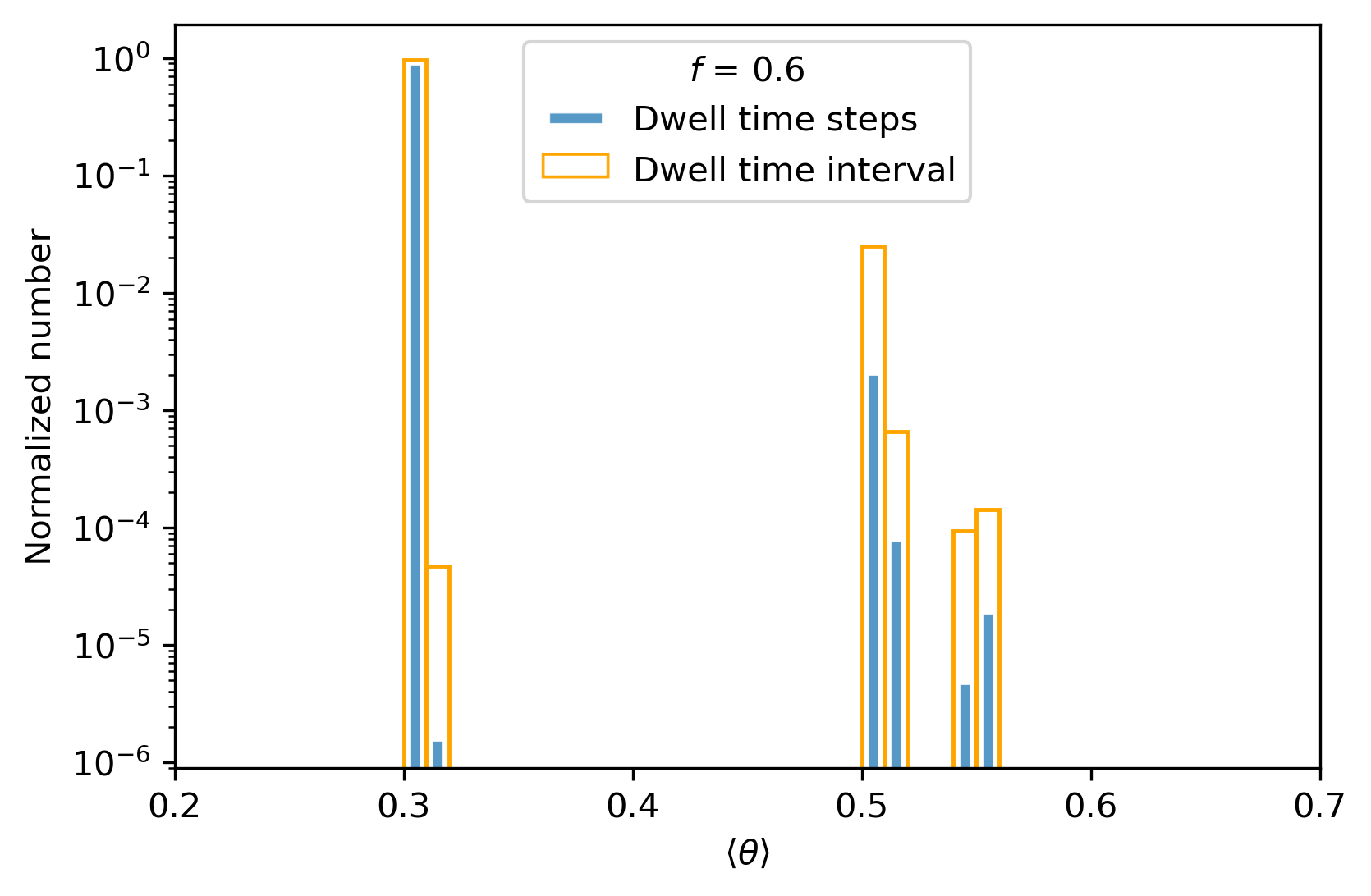}
    \caption{}\label{fig:dt_dti_0.6}
    \end{subfigure} 
    \begin{subfigure}{0.5\textwidth}
        \centering
        \includegraphics[width=\linewidth]{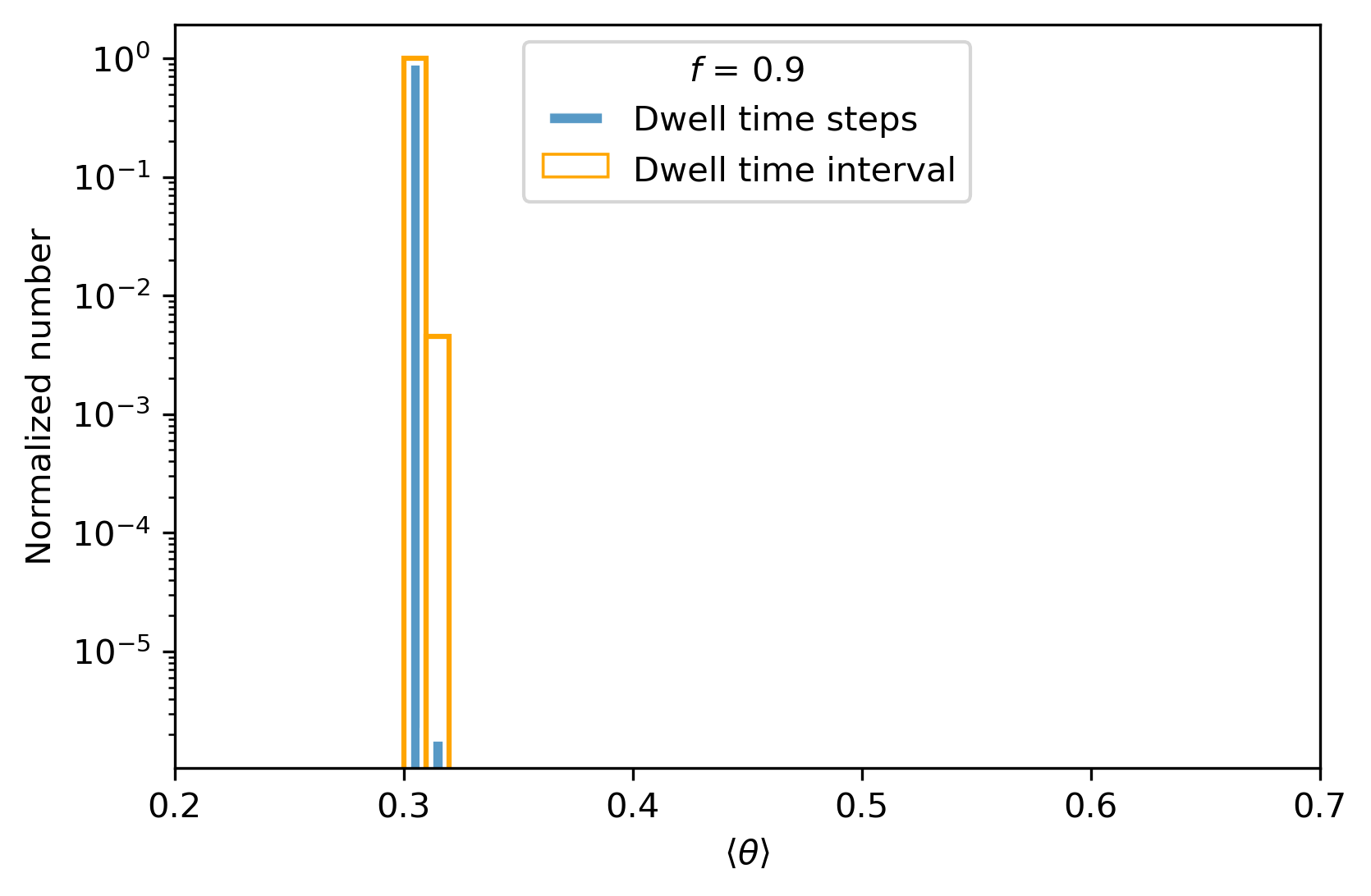}
    \caption{}\label{fig:dt_dti_0.9}
    \end{subfigure} 
    \caption{Enduring versus transitory beliefs. Histograms of the number of dwell time steps (blue bars) and dwell intervals (orange bars) during which the average belief is measured to be $\langle \theta \rangle$, plotted as functions of $\langle\theta\rangle$ for (a) $f=0.1$, (b) $f=0.6$, and (c) $f=0.9$. Each histogram is constructed from (a) $\sim 2\times 10^2$, (b) $\sim 2\times 10^2$, and (c) $\sim 4$ samples per simulation for an ensemble of 100 simulations, i.e. 100 independent, identically distributed coin toss sequences.  The vertical axis is normalized by the total number of orange and blue samples in every panel.}
    \label{fig:dt_dti}
\end{figure}

Fig.\ \ref{fig:dt_dti} displays histograms counting the number of dwell time steps (blue bars) and dwell intervals (orange bars) as functions of the mean belief $\langle\theta\rangle$ during the time steps and intervals respectively. The aim is to understand what beliefs are enduring or transitory at low, moderate, and high $f$. 
At low $f=0.1$ (Fig.\ \ref{fig:dt_dti_0.1}), the persuadable agents spend $\sim 10$ times longer at $\theta_0$ than at $\theta_{\rm p}$ (blue bars) but switch between $\theta_0$ and $\theta_{\rm p}$ frequently, as shown by the short dwell time $\langle t_{\rm d} \rangle = 2.65$ in Fig.\ \ref{fig:dwell_vs_frac_partisan}. 
The agents also dwell sometimes at $\langle\theta\rangle \neq \theta_0, \theta_{\rm p}$ but only rarely ($0.2\%$ of the time) and predominantly early in the simulation. 
At moderate $f=0.6$ (Fig.\ \ref{fig:dt_dti_0.6}), near the sharp low-to-high-$f$ transition observed in Fig.\ \ref{fig:dwell_vs_frac_partisan}, the persuadable agents dwell at $\langle\theta\rangle \approx 0.50$ and $0.55$, corresponding to the belief PDF peaking simultaneously at both $\theta_0$ and $\theta_{\rm p}$ with $x_i(t,\theta_{\rm p}) \gtrsim x_i(t,\theta_0)$ i.e.\ the persuadable agents are more confident in $\theta_{\rm p}$ while not completely discounting $\theta_0$. 
As $f$ increases, the persuadable agents grow their confidence in $\theta_{\rm p}$. 
At $f = 0.6$, 97\% of the dwell intervals have $x_i(t,\theta_{\rm p}) = 1$, while only 4\% have this property for $f = 0.1$. 
At high $f=0.9$ (Fig.\ \ref{fig:dt_dti_0.9}), agents never dwell at $\theta_0$ despite observing the coin continuously.  
The internal interaction from the partisans produces $x_i(t,\theta) = \delta(\theta-\theta_{\rm p})$ for all persuadable agents, i.e.\ the belief is zeroed out everywhere except at $\theta_{\rm p}$.  
This happens because Bayes's rule in Eqs.\ \eqref{eq:updatefirsthalf} and \eqref{eq:likelihood} is multiplicative.
Once the $i$-th persuadable agent achieves $x_i(t,\theta_0)=0$ at some $t$, they stop feeling the  ``pull'' from the coin at every subsequent $t' > t$, resulting in $x_i(t',\theta_0) = 0$ for all $t' > t$ and hence a long  dwell time $t_{\rm d} \lesssim T = 10^4$ at $\theta_{\rm p}$.  
As partisans become the supermajority ($f \gtrsim 0.8$) in the network, the persuadable agents are misled by the partisans and reject the true coin bias $\theta_0$ in favor of $\theta_{\rm p}$.

\subsection{Dueling partisans}
\label{subsec:partisandifferentopinion}

\begin{figure}[h!]
    \centering
    \begin{subfigure}{0.47\textwidth}
        \centering
        \includegraphics[width=\linewidth]{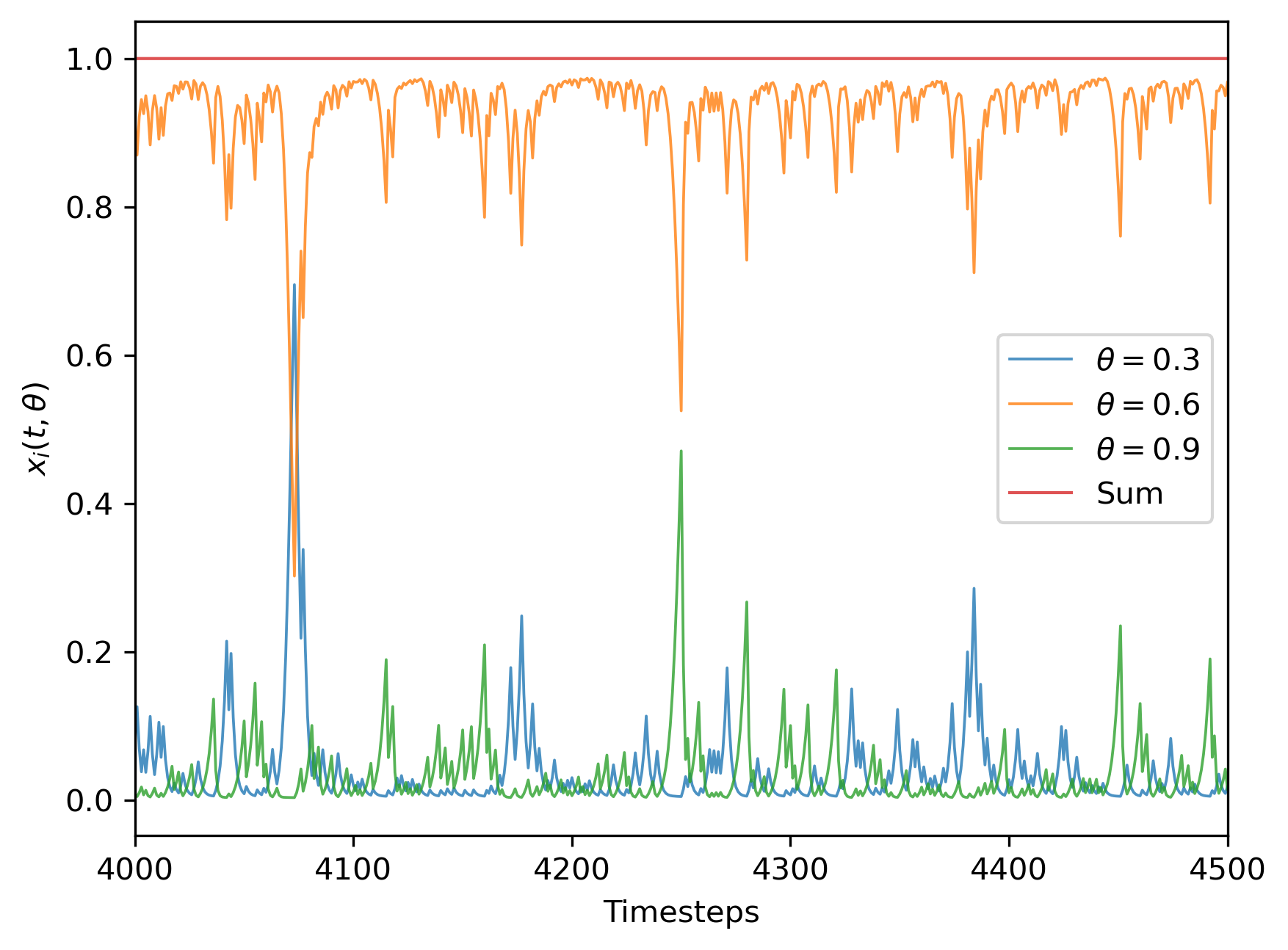}
        \caption{}\label{fig:dueling_belief_turb}
    \end{subfigure}%
    \hspace*{\fill}  
        \centering
        \begin{subfigure}{0.47\textwidth}
        \includegraphics[width=\linewidth]{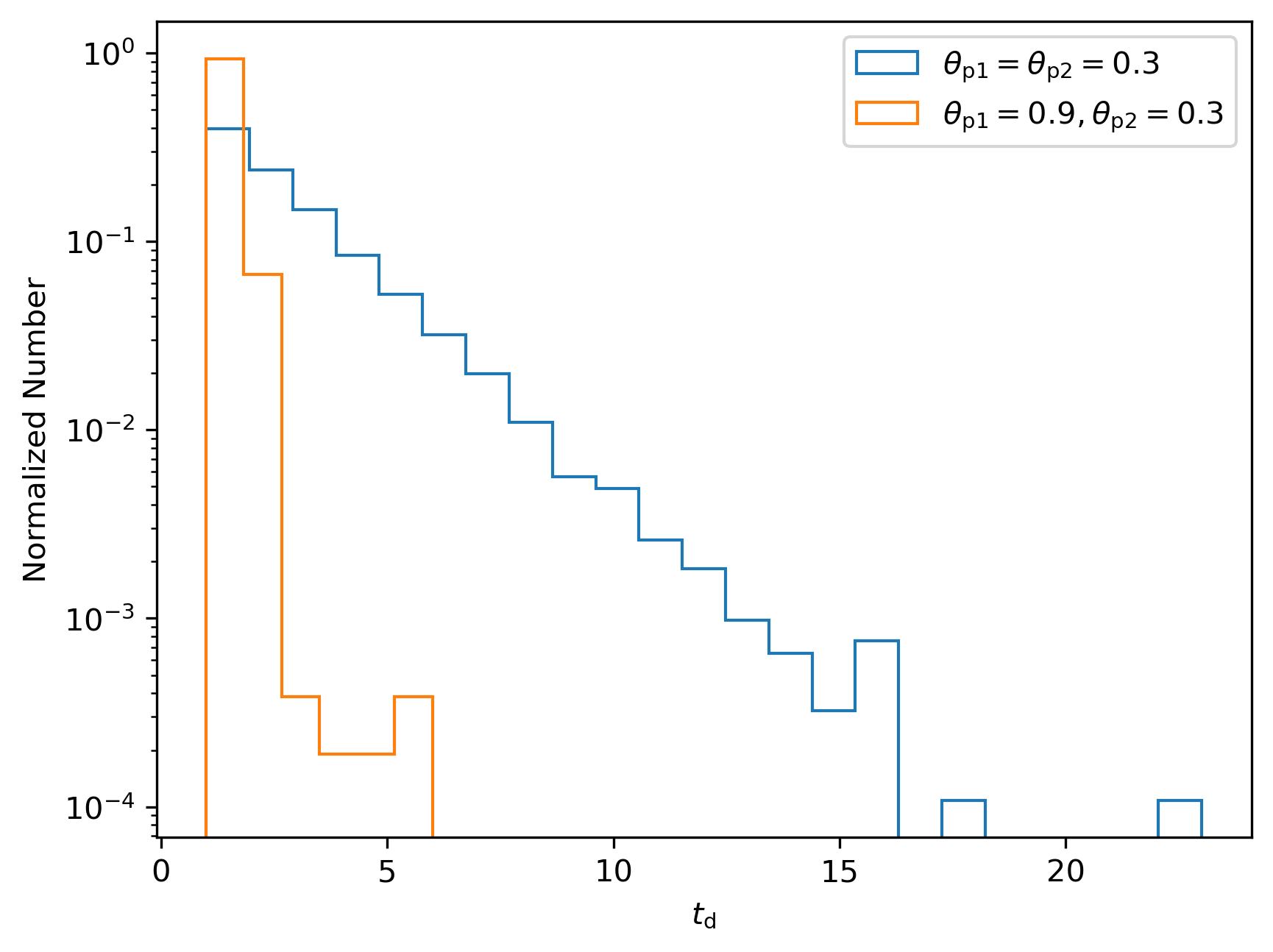}
        \caption{}\label{fig:dueling_dt_f0.02}
    \end{subfigure} 
    \begin{subfigure}{0.47\textwidth}
        \includegraphics[width=\linewidth]{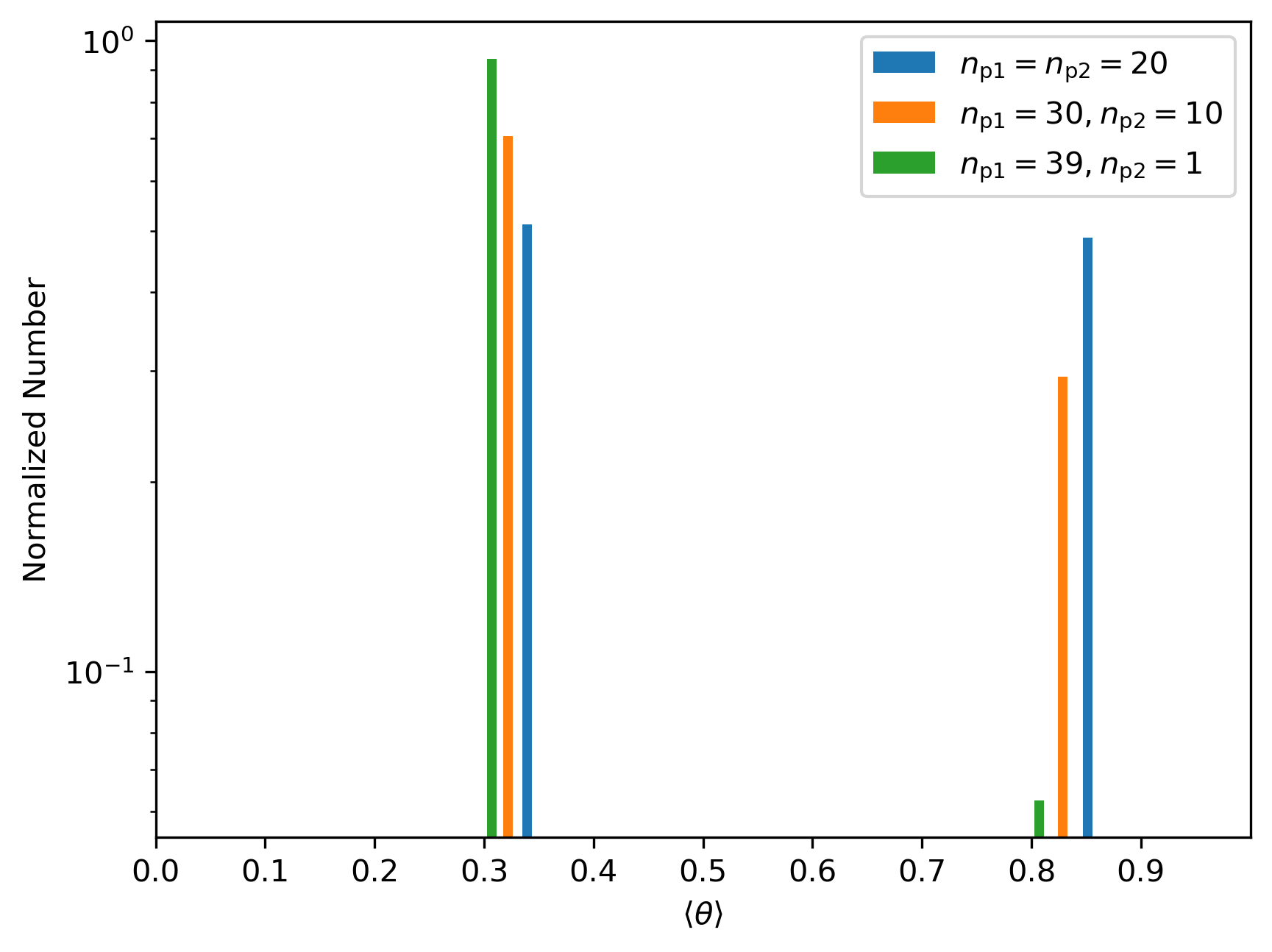}
        \caption{}\label{fig:uneven_belief}
    \end{subfigure}
    \hspace*{\fill}
    \begin{subfigure}{0.47\textwidth}
        \centering
        \includegraphics[width=\linewidth]{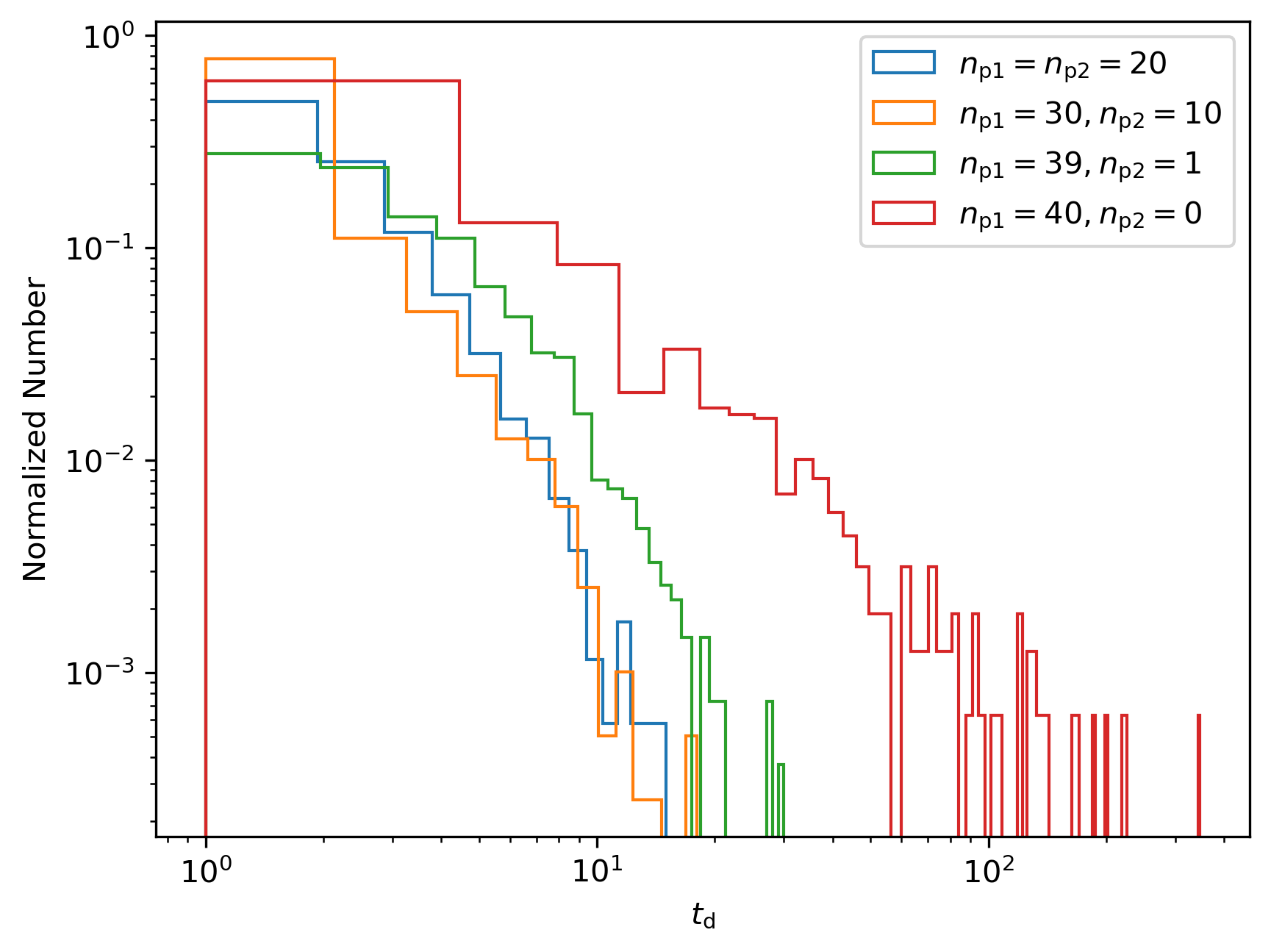}
        \caption{}\label{fig:uneven_intervals}
    \end{subfigure}
    \caption{ 
        Dueling (i.e.\ disagreeing) partisans in allies-only networks with $n = 100$, $\theta_{\rm p1} = 0.3$, $\theta_{\rm p2} = 0.9$ and $\theta_0 = 0.6$.
        (a) Turbulent nonconvergence of $x_i(t,\theta)$ in a network with $f=0.02$ for the $i$-th arbitrary agent, whose beliefs switch continuously during the representative interval $4.0 \times 10^3 \leq t \leq 4.5 \times 10^3$ between favoring $\theta = \theta_{\rm p1}$ (blue curve), $\theta=\theta_0$ (orange curve) and $\theta = \theta_{\rm p2}$ (green curve). The sum $x_i(t,\theta_0)+x_i(t,\theta_{\rm p1}) + x_i(t,\theta_{\rm p2})$ is plotted as a red curve.
        (b) Dwell time histogram for agreeing partisans (blue curve) and disagreeing partisans (orange curve; full simulation from panel (a)) with $f = 0.02$ for both networks, showing that disagreement among partisans destabilizes the beliefs of persuadable agents. 
        (c) and (d) Histograms of (c) number of dwell intervals and (d) number of dwell time steps during which the average belief is measured to be $\langle \theta \rangle$, plotted as functions of $\langle\theta\rangle$, color-coded as in the legend and corresponding to cases (i), (ii), and (iii) defined in the third paragraph of Section \ref{subsec:partisandifferentopinion}. The red curve in (d) shows a network containing agreeing partisans for comparison. The vertical axis is normalized by the total number of orange, blue, green and red samples in every panel. All networks have $f = 0.4$, which is a representative example.
        Each simulation runs to $T = 10^5$ with randomized priors and coin tosses. }
    \label{fig:dueling_partisan}
\end{figure}

Now suppose that the network contains two groups of partisans ``p1'' and ``p2'' who disagree, with $\theta_{\rm p1}=0.3$, $\theta_{\rm p2}=0.9$, and $\theta_0 = 0.6$ (illustrative values chosen arbitrarily). We say that the partisans are ``dueling'' to convey that they disagree and compete in their influence, even though they may disagree accidentally or cordially rather than deliberately or aggressively.
The results resemble those described in Section \ref{subsec:samethetap}, except that the belief PDF becomes trimodal, with $x_i(t, \theta) \neq 0$ at $\theta = \theta_0, \theta_{\rm p1}, $ and $\theta_{\rm p2}$. 
Fig.\ \ref{fig:dueling_belief_turb} shows the time evolution of $x_i(t, \theta)$ at $\theta = \theta_0, \theta_{\rm p1}, $ and $\theta_{\rm p2} $ for the representative interval $ 4\times 10^3 \leq t \leq 4.5\times 10^3$ and the $i$-th arbitrary agent.
The red line shows $x_i(t,\theta_0)+x_i(t,\theta_{\rm p1}) + x_i(t,\theta_{\rm p2})$, which constantly equals one, indicating zero belief in $\theta \neq \theta_0, \theta_{\rm p1}, \theta_{\rm p2}$. 
The $i$-th agent favors $\theta_0$ without discounting $\theta_{\rm p1}$ and $\theta_{\rm p2}$ completely. During the $5\times 10^2$ plotted time steps, the orange curve for $x_i(t,\theta_0)$ never settles down, dropping eight times to $x_i(t,\theta_0) < 0.8$. It also drops twice to $x_i(t,\theta_0) < 0.6$, once each in favor of $\theta_{\rm p1}$ and $\theta_{\rm p2}$. 

Dueling partisans destabilize the beliefs of the persuadable agents more than a single partisan. Fig.\ \ref{fig:dueling_dt_f0.02} shows the dwell time histogram of two networks, both with $f = 0.02$,  where one network contains two partisans who agree (blue curve), and the other contains two partisans who disagree (orange curve).
The shorter times marked by the orange curve indicate that the persuadable agents change their belief more frequently in the company of dueling partisans.  
This makes sense intuitively; the persuadable agents are ``pulled'' in three directions rather than two. 
For example, when the persuadable agents dwell near $\theta_{\rm p1}$, where the internal signal between partisan p1 and the persuadable agents is negligible, the internal signal between the partisan p2 and the persuadable agents remains substantial.

What happens when uneven numbers of partisans disagree? Let $n_{\rm p1}$ and $n_{\rm p2}$ be the numbers of partisans with belief PDFs $\delta(\theta-\theta_{\rm p1})$ and $\delta(\theta-\theta_{\rm p2})$ respectively. 
We consider three situations, all with $f=0.4$ for the sake of illustration: 
(i) $n_{\rm p1}= 20 = n_{\rm p2}$, (ii) $n_{\rm p1} = 30, n_{\rm p2} = 10$, and (iii)  $n_{\rm p1} = 39, n_{\rm p2} = 1$. 
Fig.\ \ref{fig:uneven_belief} displays histograms counting the number of dwell intervals for cases (i), (ii), and (iii), presented as functions of $\langle\theta\rangle$ during the dwell interval like in Fig.\ \ref{fig:dt_dti}.
When the number of disagreeing partisans is equal, as in case (i), we observe roughly equal numbers of dwell intervals centered on $\theta_{\rm p1}$ (1774 times) and $\theta_{\rm p2}$ (1690 times) (blue bars in Fig.\ \ref{fig:uneven_belief}). 
In case (ii), we observe 2796 dwell intervals centered on $\theta_{\rm p1}$ and 1162 dwell intervals centered on $\theta_{\rm p2}$ (orange bars in Fig.\ \ref{fig:uneven_belief}), while in case (iii), we observe 2547 dwell intervals centered on $\theta_{\rm p1}$ and only 170 dwell intervals centered on $\theta_{\rm p2}$ (green bars in Fig.\ \ref{fig:uneven_belief}). 
The persuadable agents dwell more frequently at the belief held by most of the partisans. 

The number of dwell intervals at each belief is roughly proportional to the partisan population. For $ n_{\rm p1} / (n_{\rm p1}+n_{\rm p2}) = 0.75$, we find that 70\% of the dwell intervals happen at $\langle \theta \rangle \approx \theta_{\rm p1}$; for $ n_{\rm p1} / (n_{\rm p1}+n_{\rm p2}) = 0.98$, we find that 94\% of the dwell intervals happen at $\langle \theta \rangle \approx \theta_{\rm p1}$.  
Note that $\langle \theta \rangle$ does not exactly equal $\theta_{\rm p1}$ or $\theta_{\rm p2}$, because the belief PDF is bimodal, peaking at both $\theta_{\rm p1}$ and $\theta_{\rm p2}$.
The persuadable agents no longer believe in $\theta_0$ for the reason described at the end of Section \ref{subsec:switching_between_belief}. 

Fig.\ \ref{fig:uneven_intervals} shows the histograms of the number of dwell time steps as a function of the mean belief $\langle \theta \rangle$ for cases (i), (ii), and (iii), as well as a control network containing agreeing partisans for comparison.
We observe longer dwell times for $n_{\rm p1} \gg n_{\rm p2}$ (Fig.\ \ref{fig:uneven_intervals}) compared with $n_{\rm p1} = n_{\rm p2}$ or $n_{\rm p1} \gtrsim n_{\rm p2}$, due to the stronger ``pull'' from the dominant group, but $\langle t_{\rm d} \rangle$ is still shorter by a factor of $\approx 3$ than for networks in which all the partisans agree (red curve in Fig.\ \ref{fig:uneven_intervals}). 
Interestingly, even one disagreeing partisan is enough to shorten $\langle t_{\rm d} \rangle$ by a factor $\approx 3$; compare the green and red curves in Fig.\ \ref{fig:uneven_intervals}, for example. In contrast, there is not much difference between $n_{\rm p1}/n_{\rm p2} =39$ and $n_{\rm p1}/n_{\rm p2} = 1$, which yield $\langle t_{\rm d} \rangle = 3.4$ and $2.1$ respectively; see the green and blue curves in Fig.\ \ref{fig:uneven_intervals}. 

\newpage

\section{Opponents only}
\label{sec:opponentonly}
We now change our focus to opponents-only networks to investigate the impact of partisans under antagonistic interactions. 
Section \ref{subsec:asym_learning} examines a network with only one partisan to build intuition through a baseline system. 
We find that a partisan does not prevent the system from achieving equilibrium; unlike in an allies-only network, opposing partisans do not ``pull'' the persuadable agents towards a unique belief $\theta_{\rm p} \neq \theta_0$, because everybody is in mutual opposition.  
The results concerning asymptotic learning resemble those without partisans in Ref.\ \cite{low_discerning_2022}.
Section \ref{subsec:wrong_conclusion_first} examines if the wrong conclusion is reached first in this simulation, as observed in Ref.\ \cite{low_discerning_2022}, for various network structures. 
The special case $\theta_{\rm p} = \theta_0$ is investigated, where indeed asymptotic learning does occur more slowly at $\theta_0$ than at other (wrong) values of $\theta$, just as in Ref.\ \cite{low_discerning_2022}. 
More generally, however, for $\theta_{\rm p} \neq \theta_0$, the connectivity of the network controls whether or not the wrong conclusion is reached first, unlike the system without partisans in Ref.\ \cite{low_discerning_2022}.
The asymptotic learning time $t_{\rm a}$ is computed as a function of the partisans fraction $f$ in Section \ref{subsec:opponent_frac_partisan}. 

\subsection{Asymptotic learning}
\label{subsec:asym_learning}
Opponents-only networks with and without partisans exhibit similar long-term outcomes: individual persuadable agents, and the system as a whole, achieve asymptotic learning as defined by Eqs.\ \eqref{eq:asymlearncondition} and \eqref{eq:sysasymlearncondition}, and as observed in Ref.\ \cite{low_discerning_2022}.

\begin{figure}[h!]
    \centering
    \begin{subfigure}{0.45\textwidth}
        \centering
        \includegraphics[width=\linewidth]{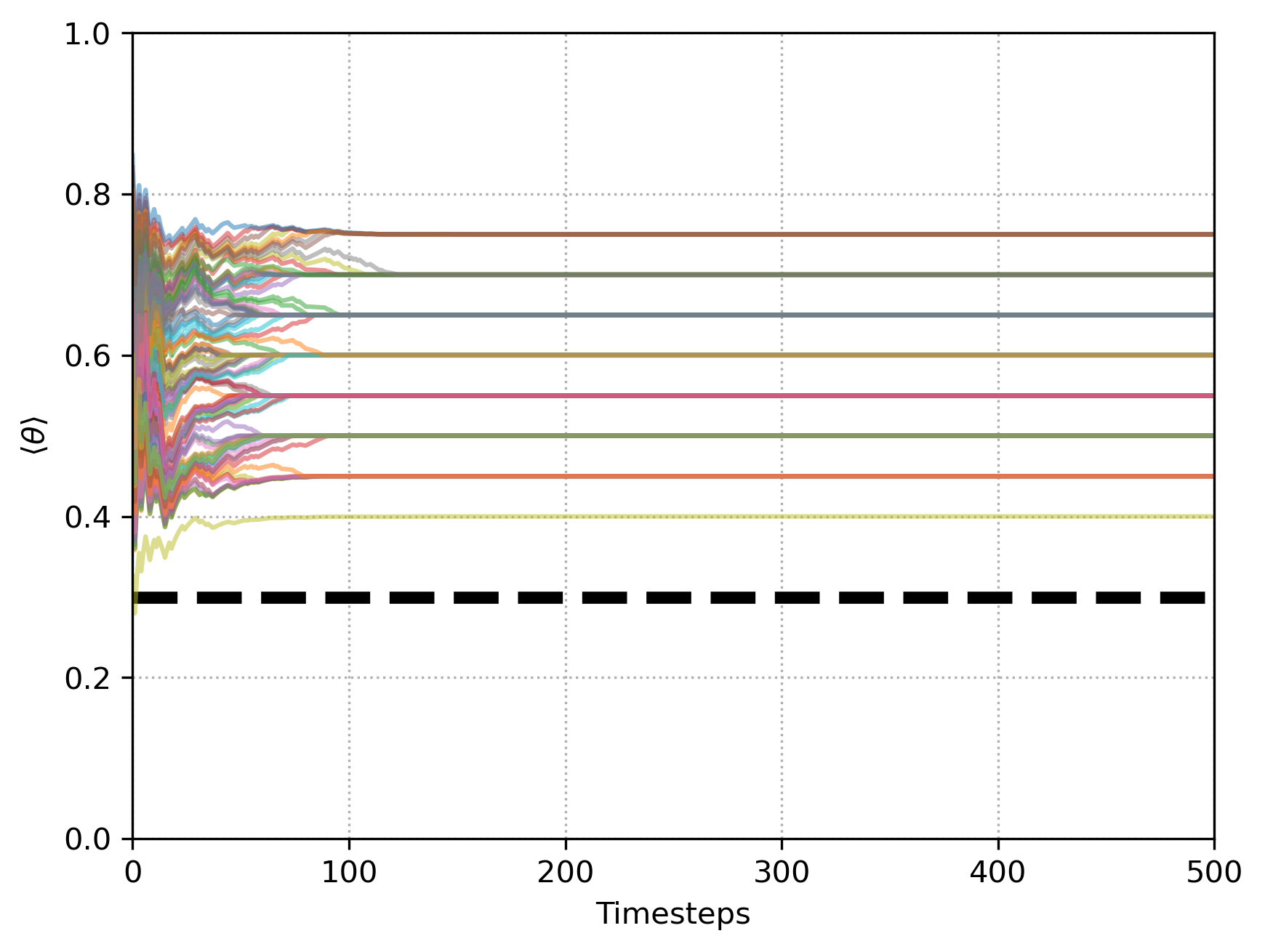}
        \caption{}\label{fig:mean_1_o}
    \end{subfigure}%
    \hspace*{\fill}  
    \begin{subfigure}{0.55\textwidth}
        \centering
        \includegraphics[width=\linewidth]{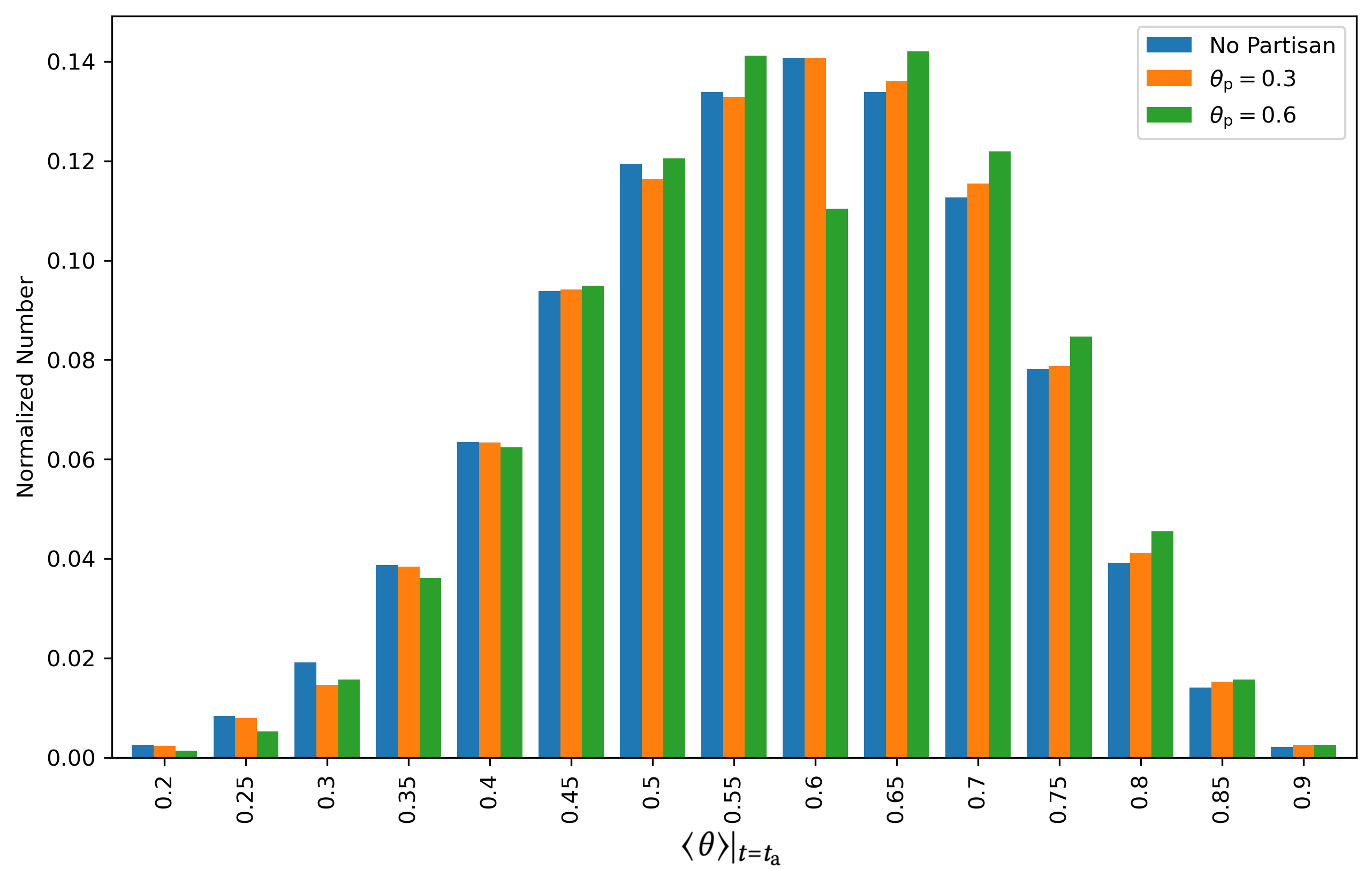}
        \caption{}\label{fig:diff_theta_p_o}
    \end{subfigure} \\
    \begin{subfigure}{0.55\textwidth}
        \centering
        \includegraphics[width=\linewidth]{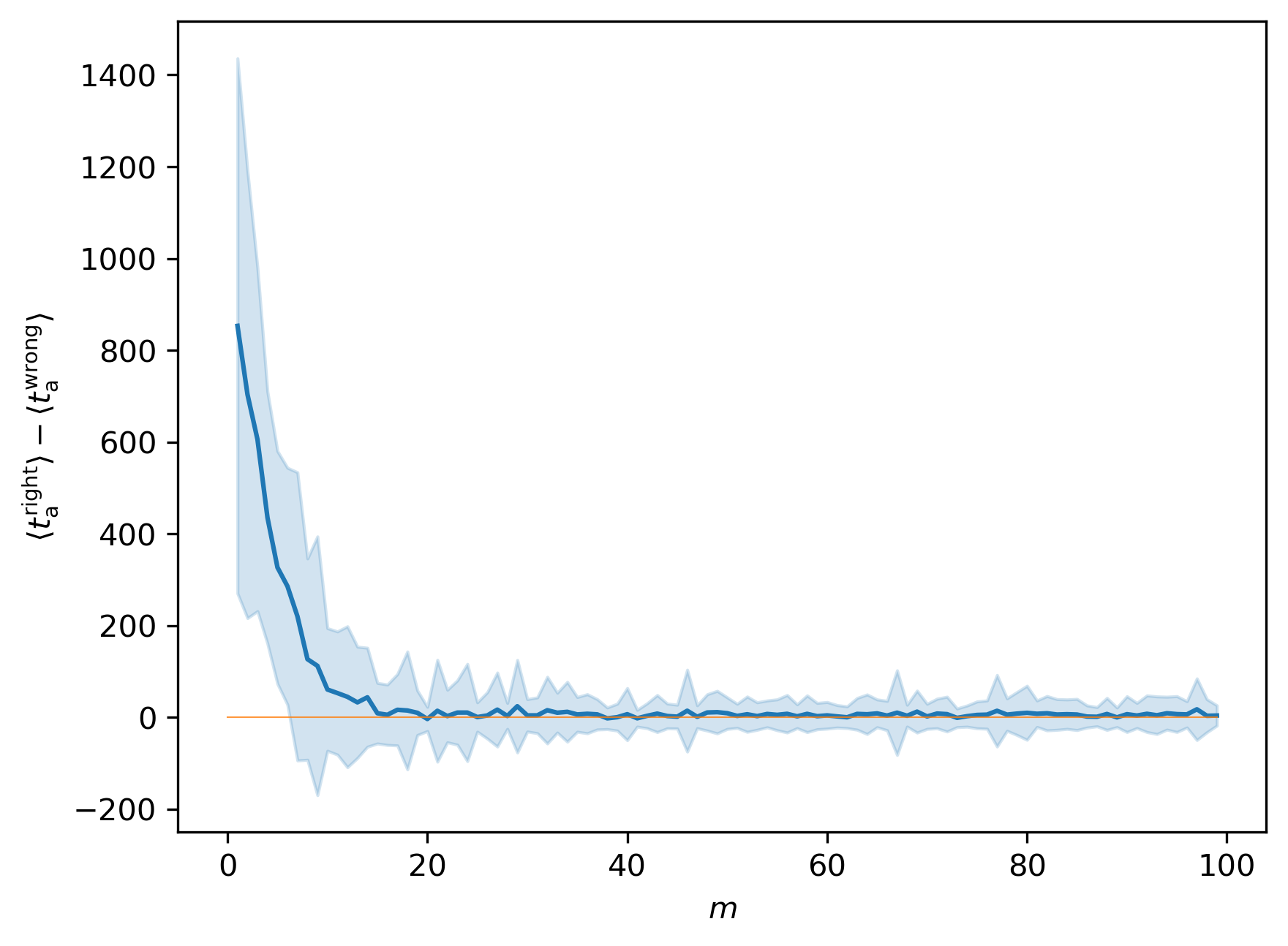}
        \caption{}\label{fig:bba_sweep}
    \end{subfigure}%
    \caption{
        Asymptotic learning in opponents-only networks. 
        (a) Mean belief $\langle \theta \rangle$ of every agent versus time $t$ for a particular simulation with $n=100$ and one partisan with $\theta_{\rm p} = 0.3$. The simulation runs to $T = 10^5$ but only the interval $0 \leq t\leq 500$ is plotted for clarity. The system achieves asymptotic learning at $t_{\rm a} = 221$ as defined by Eq.\ \eqref{eq:sysasymlearncondition}. The final beliefs of persuadable agents cluster around $\theta_0$, spaced by the $\theta$ bin width (0.05).
        The solid colored and black dashed curves correspond to the persuadable agents and partisan respectively. 
        (b) Histograms of the number of agents with average belief $\langle \theta \rangle$ measured at $t_{\rm a}$, plotted as functions of $\langle\theta\rangle|_{t=t_{\rm a}}$, for a complete network with $n=100$ and zero partisan (blue bars), one partisan with $\theta_{\rm p} = 0.3 \neq \theta_0$ (orange bars), and one partisan with $\theta_{\rm p} = 0.6 = \theta_0$ (green bars). 
        Each network is analyzed via an ensemble of $10^3$ simulations, where the coin tosses are independent and identically distributed, and the priors are randomized. 
        (c) $\langle t_{\rm a}^{\rm right} \rangle - \langle t_{\rm a}^{\rm wrong} \rangle$ for a Barab\'{a}si-Albert network with $n=100$ plotted against the attachment parameter $1 \leq m \leq 99$. 
        Each network is run for 100 simulations with independent priors and coin tosses. 
        The heavy blue curve shows $\langle t_{\rm a}^{\rm right} \rangle - \langle t_{\rm a}^{\rm wrong} \rangle$ for each $m$, while the light blue shading shows the range corresponding to two standard deviations. The orange line indicates $\langle t_{\rm a}^{\rm right} \rangle = \langle t_{\rm a}^{\rm wrong} \rangle$. 
    }
    \label{fig:single_opp_only}
\end{figure}

Consider a simulation like the one conducted in Section \ref{subsec:samethetap}, but with $A_{ij} = -1$ for all $i\neq j$.
We no longer observe turbulent nonconvergence as in Fig.\ \ref{fig:meannonpartisan}.
Fig.\ \ref{fig:mean_1_o} shows how $\langle\theta\rangle$ evolves in a particular simulation for $0 \leq t \leq 500$. 
Only the first 500 timesteps are shown in Fig.\ \ref{fig:mean_1_o} for brevity, but the simulation runs for $T=10^5$, and $\langle \theta \rangle$ tends to a constant for $t >221$ for every agent. 
That is, the system achieves asymptotic learning at $t_{\rm a} = 221$.

The persuadable agents do not reach consensus, in contrast to Section \ref{subsec:samethetap}, because antagonistic interactions drive their beliefs apart.
The beliefs of persuadable agents cluster around the true bias $\theta_0$, as shown in Fig.\ \ref{fig:mean_1_o}, specifically in the range $0.40 \leq \theta \leq 0.75$ in steps of 0.05 due to the discretization discussed in Section \ref{subsec:automaton}. 
This behaviour is also observed in a Barab\'{a}si-Albert network, as shown in Fig.\ 10a in Ref.\ \cite{low_discerning_2022}. 
Fig.\ \ref{fig:diff_theta_p_o} presents a histogram of the number of persuadable agents with average belief $\langle\theta\rangle|_{t=t_{\rm a}}$. The distribution has the shape of a bell curve, peaking at $\langle\theta\rangle|_{t=t_{\rm a}} = \theta_0$. 
The dependence of the histogram on $|\theta_0 - \theta_{\rm p}| $ is studied in Section \ref{subsec:wrong_conclusion_first}.
Multiple persuadable opponents can occupy the same $\langle\theta\rangle$ bin, because the internal peer pressure is zero if two opponents agree with each other according to Eq.\ \eqref{eq:xiprimed}, i.e.\ there is no belief repulsion between opponents who agree, allowing opponents to settle on the same belief. 
Eq.\ \eqref{eq:xiprimed} implies that such ``grudging agreement'' is unstable, when the network contains only two persuadable opponents, because a small disagreement grows with time via \eqref{eq:xiprimed}. 
However, when there are many persuadable opponents who hold adjacent beliefs, the situation stabilizes.  For example, agents with $x_i(t,\theta)=\delta(\theta-0.50)$ and $x_j(t,\theta)=\delta(\theta-0.60)$ exert opposing repulsive peer pressure on, and hence stabilize, multiple agents labeled $k$ with $x_k(t,\theta)=\delta(\theta-0.55)$.
This resembles the scenario, when two opponents begrudgingly agree because they are both forced into the same opinion by other opponents.  

Asymptotic learning in opponents-only networks containing more than one partisan occurs similarly to Fig.\ \ref{fig:mean_1_o}. It is investigated in Section \ref{subsec:opponent_frac_partisan} as part of a study of $t_{\rm a}$ versus $f$.

\subsection{Reaching the wrong conclusion first}
\label{subsec:wrong_conclusion_first}
A key observation in opponents-only Barab\'{a}si-Albert networks without partisans is that agents who reach the wrong conclusion do so before their opponents reach the right conclusion. 
This counterintuitive tendency is quantified in detail in Section 4.2 and 5.1 in Ref.\ \cite{low_discerning_2022}. 
Here we test if the tendency still holds for complete networks with a single partisan. In this section, unlike in the rest of Sections \ref{sec:alliesonly} -- \ref{sec:mixed}, we consider scale-free, Barab\'{a}si-Albert networks as well as complete networks, because it turns out that connectivity plays an important role in reaching the wrong conclusion first.

We start by investigating the special situation where the partisan believes in the true bias, with $\theta_{\rm p} = \theta_0 = 0.6$.
Table \ref{tab:right_and_wrong} compares the statistics of $ t_{\rm a}^{\rm right}$ and $ t_{\rm a}^{\rm wrong}$ for four different systems, where the superscripts ``right'' and ``wrong'' label $t_{\rm a}$ for the right and wrong conclusions respectively. 
When a partisan is present in a complete network with $\theta_{\rm p} = \theta_0$, the wrong conclusion does tend to be reached first, but the difference between $\langle t_{\rm a}^{\rm right} \rangle$ and $\langle t_{\rm a}^{\rm wrong} \rangle$ is smaller than in a Barab\'{a}si-Albert network, as evidenced by columns \ref{col:ab_no_partisan} and \ref{col:complete_06} in Table \ref{tab:right_and_wrong}. 
This is because, at the beginning of the simulation, the persuadable agents feel persistent repulsion from the partisan at $\theta = \theta_0$.
Persistent repulsion also affects how many agents converge on a particular final belief. 
Fig.\ \ref{fig:diff_theta_p_o} presents a histogram of the number of agents as a function of $\langle \theta \rangle|_{t=t_{\rm a}}$; as $x_i(t, \theta)$ peaks narrowly for all $i$ and $t \geq t_{\rm a}$, it is accurate to approximate agents' final beliefs by $\langle \theta\rangle|_{t=t_{\rm a}}$. 
For $\theta_{\rm p} =\theta_0 = 0.6$, i.e.\ the green bars in the histogram, $\langle\theta\rangle|_{t=t_{\rm a}} = 0.6$ is not the modal value, unlike for the control experiment with zero partisans (blue bars). Quantitatively, 11\% of the agents tend to $\langle\theta\rangle|_{t=t_{\rm a}} = 0.6$, compared to 14\% in the control experiment. This is the same trend identified in Ref.\ \cite{low_discerning_2022}, namely reaching the wrong conclusion first, except that $t_{\rm a}^{\rm right} - t_{\rm a}^{\rm wrong}$ is smaller. 


\begin{table}[h]
    \centering\begin{tabular}{|c||c|c||c|c||c|c||c|c|}
    \hline
     & 
    \multicolumn{2}{c||}{\makecell{\newtag{(a)}{col:ab_no_partisan} \\ No partisan \\ Barab\'{a}si-Albert}} &
    \multicolumn{2}{c||}{\makecell{\newtag{(b)}{col:complete_no_partisan} \\ No partisan \\ Complete}} &
    \multicolumn{2}{c||}{\makecell{\newtag{(c)}{col:complete_03} \\ $\theta_{\rm p} = 0.3 \neq \theta_0$ \\ Complete}} &
    \multicolumn{2}{c|}{\makecell{\newtag{(d)}{col:complete_06} \\ $\theta_{\rm p} = 0.6 = \theta_0$ \\ Complete}} \\
    \hline
    Property of $t_{\rm a}$
    &$t_{\rm a} ^{\rm right}$ & $t_{\rm a}^{\rm wrong}$ 
    & $t_{\rm a} ^{\rm right}$ & $t_{\rm a}^{\rm wrong}$
    & $t_{\rm a} ^{\rm right}$ & $t_{\rm a}^{\rm wrong}$
    & $t_{\rm a} ^{\rm right}$ & $t_{\rm a}^{\rm wrong}$\\
    \hline
    First quartile &337&60&50&47&50&47&51&45 \\
    Median &715&92&60&57&60&58&63&56 \\
    Third quartile &1093&174&74&73&76&74&82&73 \\
    \hline
    \makecell{Total number of \\ asymptotic learning agents} &26110&33220&14181&85819&13919&85081&10926&88074 \\
    \hline
    \end{tabular}
    \caption{
        Summary statistics of $t_{\rm a}^{\rm right}$ and $t_{\rm a}^{\rm wrong}$, the asymptotic learning times for right and wrong beliefs respectively, accumulated over $10^3$ simulations, for $n=100$, $T=10^5$, and four different network structures.  
        The first, second, and third quartiles are listed for each network.}
    \label{tab:right_and_wrong}
\end{table}

Interestingly, the tendency to reach the wrong conclusion first does not occur in complete networks in the general case $\theta_{\rm p} \neq \theta_0$. 
Columns \ref{col:complete_no_partisan} and \ref{col:complete_03} in Table \ref{tab:right_and_wrong} show that one has $t_{\rm a}^{\rm right} \approx t_{\rm a}^{\rm wrong}$ within statistical fluctuations for complete networks with $\theta_{\rm p}=0.3 \neq \theta_0$ (column \ref{col:complete_03}) and without a partisan (column \ref{col:complete_no_partisan}). 
This differs from the behavior of Barab\'{a}si-Albert networks studied in Ref.\ \cite{low_discerning_2022}. 
The Barab\'{a}si-Albert attachment parameter $m$ is often small in applications ($m = 3$ in Ref.\ \cite{low_discerning_2022}), meaning that each Barab\'{a}si-Albert agent has only a few opponents, while each agent has 99 opponents in the complete network with $n=100$. 
Hence the influence of one particular opponent, and the ``lock-out effect'' identified in Section 4.2 in Ref.\ \cite{low_discerning_2022}, are less significant in a complete network than in a Barab\'{a}si-Albert network when calculating the sum in Eq.\ \eqref{eq:xiprimed}.

We test this effect by considering different values of $m$ for Barab\'{a}si-Albert networks with $n=100$. 
Fig.\ \ref{fig:bba_sweep} shows the difference in mean asymptotic learning time between the right and wrong beliefs for $1\leq m \leq 99$. 
We observe that for networks with fewer connections, the wrong conclusion is reached first.  
Quantitatively, we find $\langle t_{\rm a}^{\rm right} \rangle - \langle t_{\rm a}^{\rm wrong} \rangle \leq 0.01 \langle t_{\rm a}^{\rm wrong} \rangle $ for $m \geq 15$. 
This agrees with the findings of Low \& Melatos \cite{low_discerning_2022} on Barab\'{a}si-Albert networks with $m=3$.  
More generally, as $m$ increases, the connectivity within the network increases, and each agent is connected to more opponents, suppressing the influence of individual agents.
The light blue shading shows the range corresponding to two standard deviations.
The dispersion in $\langle t_{\rm a}^{\rm right} \rangle - \langle t_{\rm a}^{\rm wrong} \rangle$ decreases with $m$. 


\subsection{Fraction of partisans: effect on $t_{\rm a}$}
\label{subsec:opponent_frac_partisan}
We now increase the number of partisans in opponents-only networks and examine the behavior of persuadable agents. 
All networks achieve asymptotic learning. 
Turbulent nonconvergence is not observed, because partisans do not ``pull'' other agents (whom they oppose) towards their own partisan beliefs, even when the partisans agree among themselves.

\begin{figure}[h]
    \centering
    \includegraphics[scale=0.58]{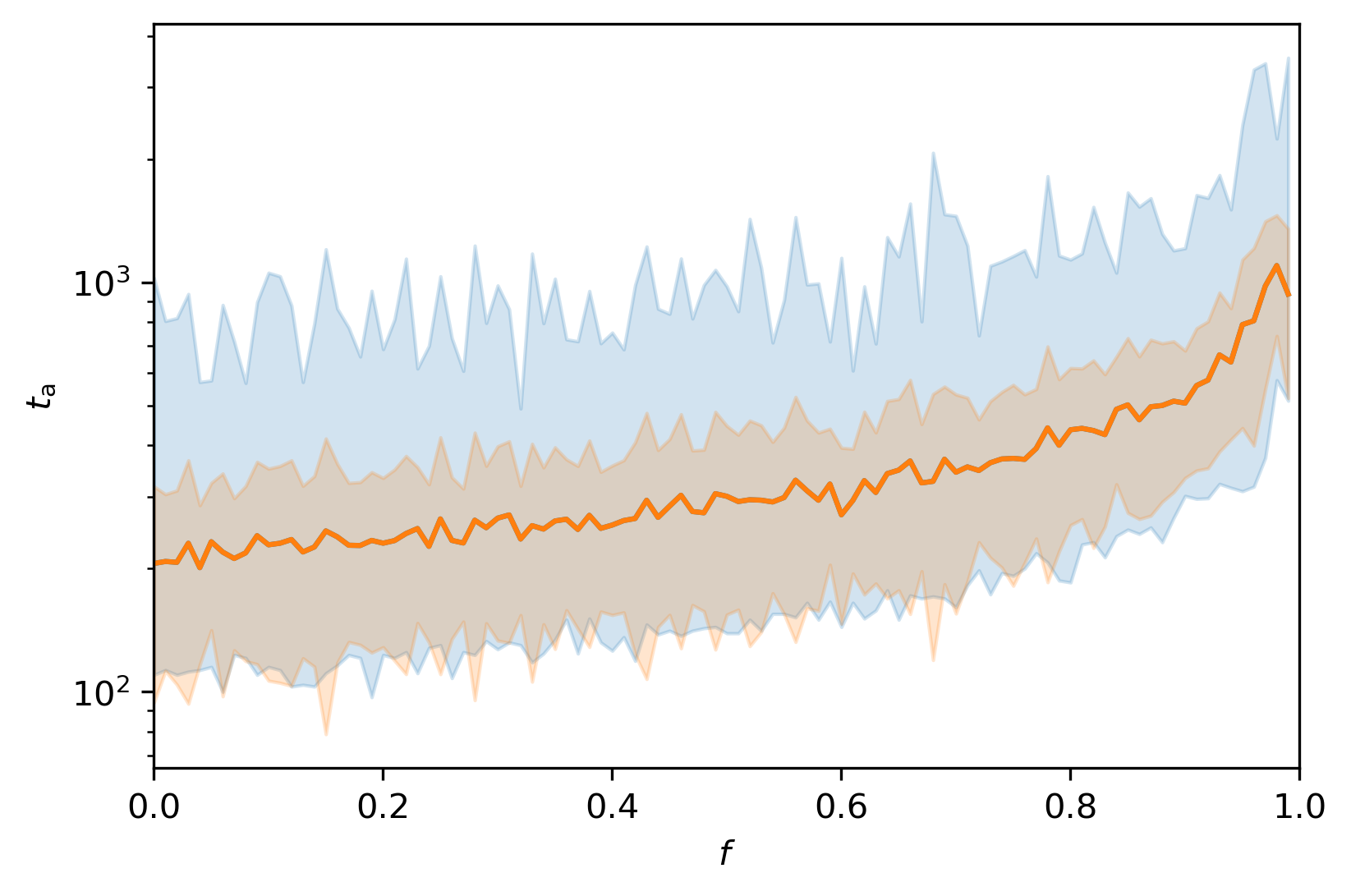}
    \caption{
        Asymptotic learning time $t_{\rm a}$ versus partisan fraction $f$ in opponents-only networks with $n =100$.  For each value of $f$ in the range $0.01 \leq f \leq 0.99$, we run an ensemble of 100 simulations until $T=t_{\rm a}$, where the coin tosses are independent and identically distributed. The solid orange curve indicates the ensemble mean.
        The light blue shading spans the ensemble minimum and maximum, and the orange shading spans two standard deviations.
    }
    \label{fig:ta_vs_f}
\end{figure}

Fig.\ \ref{fig:ta_vs_f} shows how the asymptotic learning time depends on the fraction of partisans $f$, all for $\theta_{\rm p} = 0.3$, with $0.01 \leq f \leq 0.99$ in steps of 0.01.
The persuadable agents achieve asymptotic learning more slowly as $f$ increases.
The reason is related to the ``lock-out'' mechanism identified in Section 4.2 of Ref.\ \cite{low_discerning_2022}. 
Consider opponents $i$ and $j$, whose beliefs do not overlap at some $\theta$. 
For $x'_i(t+1/2,\theta)=0$ and $x'_j(t+1/2,\theta) \geq 0$ without loss of generality, we have $A_{ij} [ x'_j(t+1/2,\theta) - x'_i(t+1/2,\theta) ] \leq 0$. 
Summing over all such opponents $j$ for fixed $i$, we obtain $x_i(t+1,\theta)=0$ from Eq.\ \eqref{eq:undatesecondhalf} at the value of $\theta$ being considered. 
Therefore agents cannot respond to the full likelihood of the coin, because part of the $\theta$ domain is zeroed out by opponents, i.e. $x_i(t,\theta') = 0$ implies $x'_i(t+1/2,\theta') = 0$ for locked-out $\theta'$ values, because Eq.\ \eqref{eq:updatefirsthalf} is multiplicative. 
For low $f$, there are more persuadable agents forming beliefs at various $\theta$ values, so more of the $\theta$ domain for every agent is locked out by opponents, causing every agent to ``see'' a narrow likelihood when observing the coin tosses.  
On the other hand, when networks contain more partisans (i.e.\ higher $f$), who all agree on $\theta_{\rm p}$, fewer $\theta$ values are occupied by persuadable agents, and all persuadable agents ``see'' more of the likelihood. 
As  $x_i(t,\theta)$ is normalized, agents who see a narrower likelihood achieve asymptotic learning faster than agents who see a wider likelihood. 
Therefore, $t_{\rm a}$ increases with $f$ in an opponents-only network.
The light blue shading in Fig.\ \ref{fig:ta_vs_f} brackets the ensemble minimum and maximum, and the orange shading brackets the range out to two standard deviations.

\section{Mixed allegiances}
\label{sec:mixed}
We now turn our focus to more realistic networks containing mixed political allegiances, where some agents interact simultaneously with allies and opponents.
An exhaustive analysis of mixed networks lies outside the scope of this paper. 
A preliminary step towards this challenging problem without partisans was taken in Refs.\ \cite{low_discerning_2022} and \cite{low_vacillating_2022}, mainly (but not exclusively) in the context of Barab\'{a}si-Albert networks. 
Here we focus instead on the special (and simpler) case of complete networks to take advantage of the baseline studies in Sections \ref{sec:alliesonly} and \ref{sec:opponentonly}, trading off some richness in network topology in favor of simplifying and therefore clarifying the novel effects introduced by partisans. 
In Section \ref{subsec:unbalanced_triad}, we consider the dynamics of an unbalanced triad, which is the smallest nontrivial subunit of a mixed network and the driver of much (although not all) of the counterintuitive behavior reported in Refs.\ \cite{low_discerning_2022} and \cite{low_vacillating_2022}. 
In Section \ref{subsec:larger_network}, we consider a representative example with $n=100$, to gain a preliminary sense of how the behavior of larger mixed networks compares with allies- and opponents-only networks. 
The results in Section \ref{subsec:larger_network} point to some productive avenues for future work but are not intended to be exhaustive.

\subsection{Triads}
\label{subsec:unbalanced_triad}

\begin{figure}[h!]
    \centering
    \includegraphics[scale = 0.25]{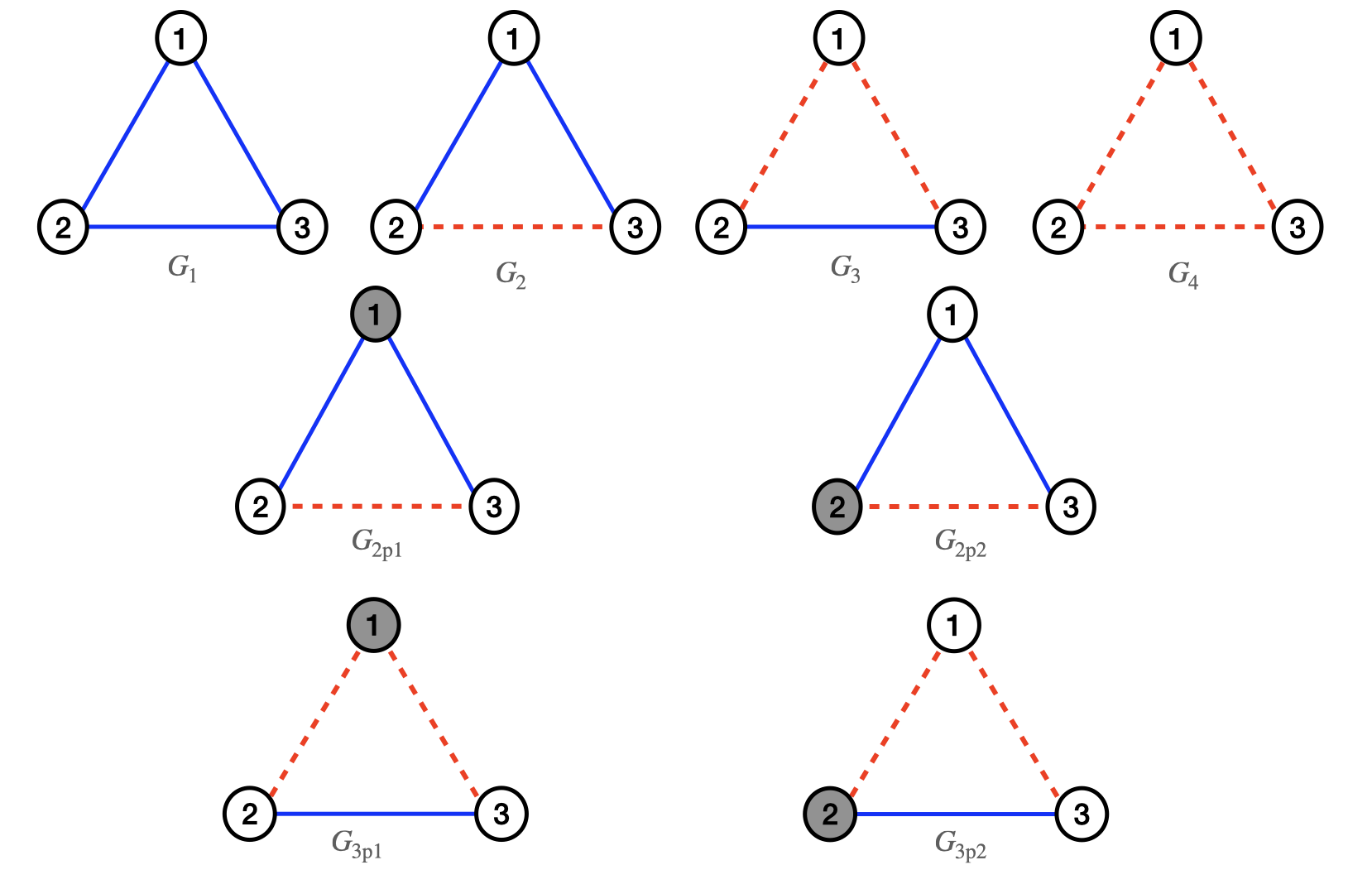}
    \caption{
        (Top Row) The four unique triads without partisans. Blue solid lines represent allies, and red dashed lines represent opponents.  In the absence of partisans, permutation of the vertex labels generates identical graphs.
        (Middle row) Two distinguishable versions of the unbalanced triad $G_2$ with one partisan (grey shading) assuming the role of agent 1 (left; $G_{2{\rm p}1}$) and agent 2 (right; $G_{2{\rm p}2}$). 
        (Bottom row) Two distinguishable versions of $G_3$ with one partisan (grey shading).
    }
    \label{fig:triad}
\end{figure}

Four unique triads with $n=3$ can be constructed. They are labeled $G_1, \dots, G_4$ in the top row of  Fig.\ \ref{fig:triad}.
$G_1$ and $G_4$ are allies-only and opponents-only networks, studied in Sections \ref{sec:alliesonly} and \ref{sec:opponentonly} respectively.
$G_3$ is nominally a mixed network but it behaves the same as an opponents-only network with $n=2$, when there are no partisans, as is clear visually. 
Interestingly, though, it exhibits counterintuitive behavior, when a partisan is introduced, as discussed below in this section. 
The unbalanced triad $G_2$ leads to important, counterintuitive, new behavior even without partisans, as demonstrated in Ref.\ \cite{low_discerning_2022}. 
It exhibits internal tension: agent 1 is attracted to the beliefs of its allies, agents 2 and 3, but the beliefs of agents 2 and 3 tend to diverge, because agents 2 and 3 are opponents. 
In other words, agent 1 is in the invidious position of striving to agree with two individuals, who are predisposed to disagree. In general, this leads to unsteady dynamics, such as turbulent nonconvergence \cite{low_discerning_2022}. 
In an unbalanced triad, the impact of a partisan depends sensitively on where the partisan is inserted; they may assume the role of agent 1 or agent 2 (equivalent to agent 3), as depicted in the bottom row of Fig.\ \ref{fig:triad}. 
We consider both scenarios in this section.

\begin{figure}[h!]
    \centering
    \begin{subfigure}{0.5\textwidth}
        \centering
        \includegraphics[width=\linewidth]{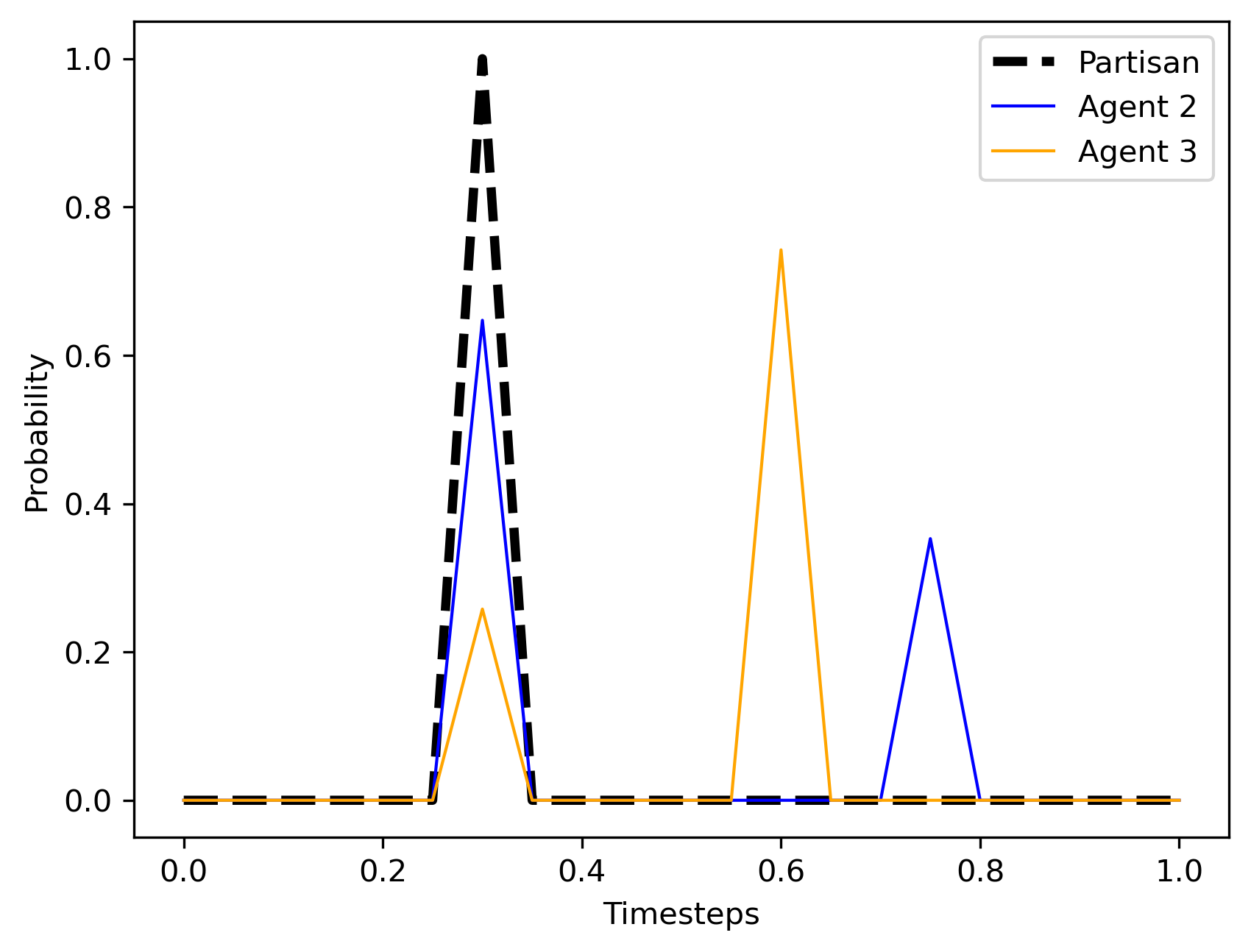}
        \caption{}\label{fig:G_2_prior}
    \end{subfigure}%
    \hspace*{\fill}  
    \begin{subfigure}{0.5\textwidth}
        \centering
        \includegraphics[width=\linewidth]{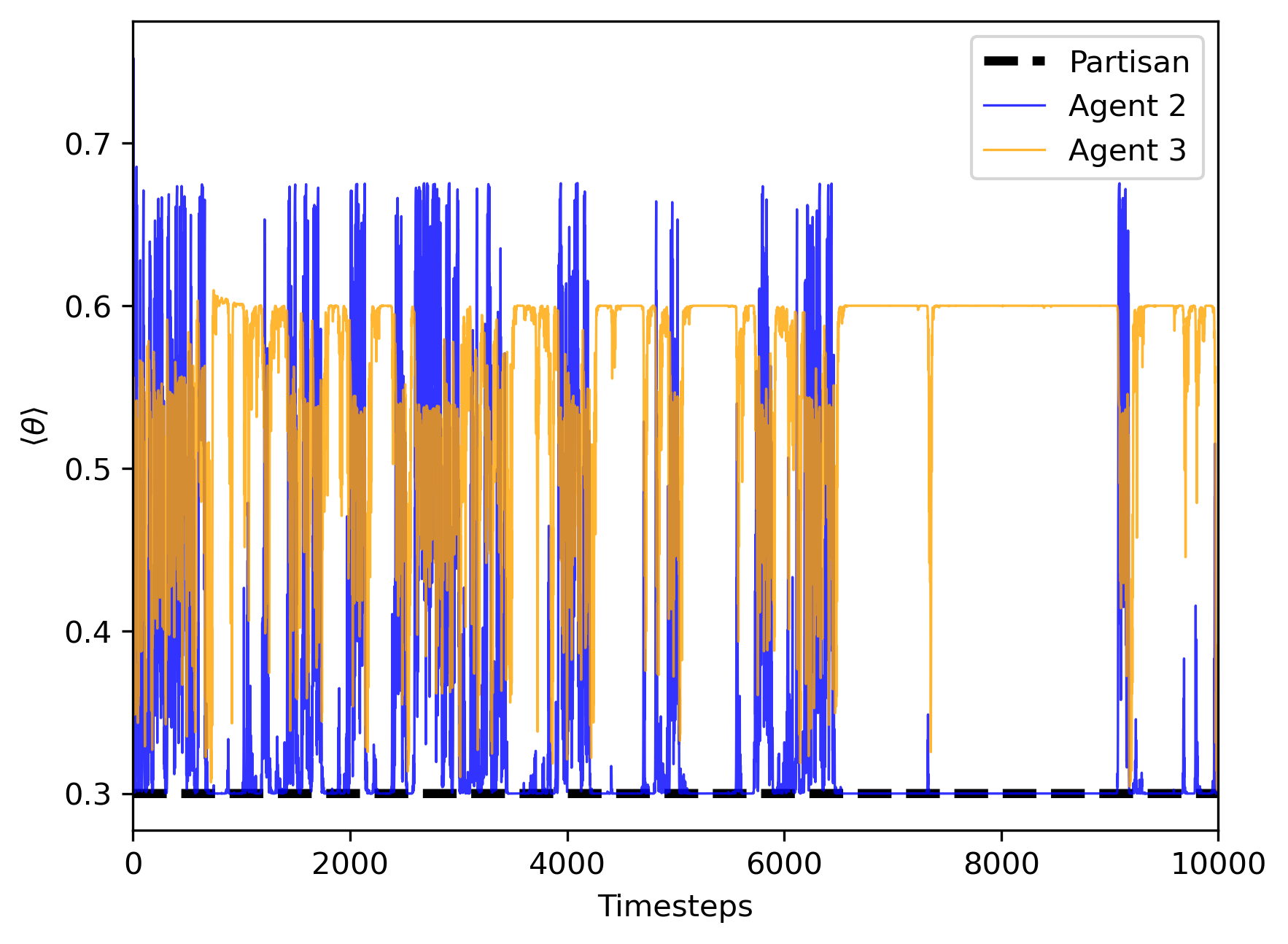}
        \caption{}\label{fig:G_2_mean}
    \end{subfigure} \\
    \begin{subfigure}{0.5\textwidth}
        \centering
        \includegraphics[width=\linewidth]{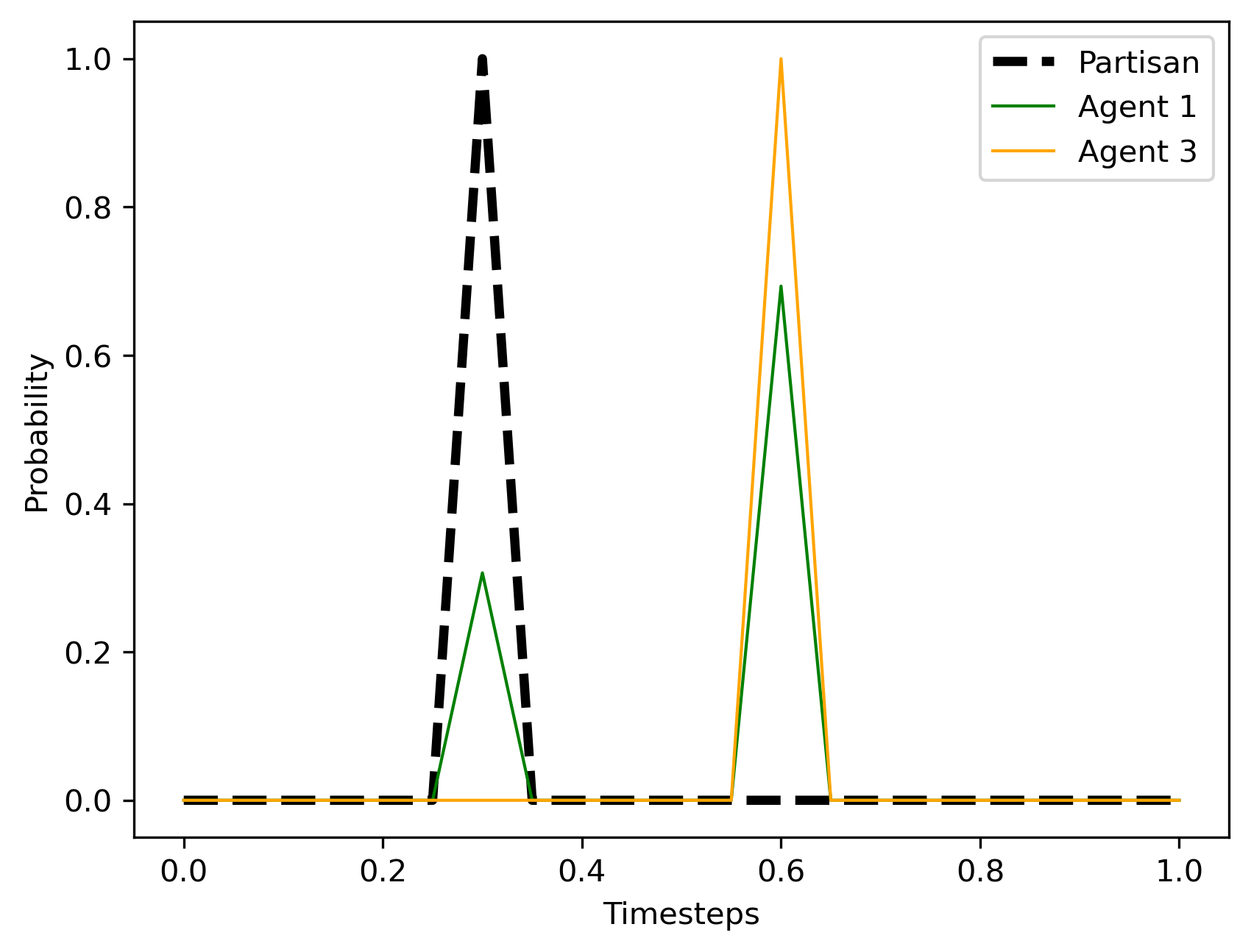}
        \caption{}\label{fig:G_2_prior_fe}
    \end{subfigure}%
    \hspace*{\fill}  
    \begin{subfigure}{0.5\textwidth}
        \centering
        \includegraphics[width=\linewidth]{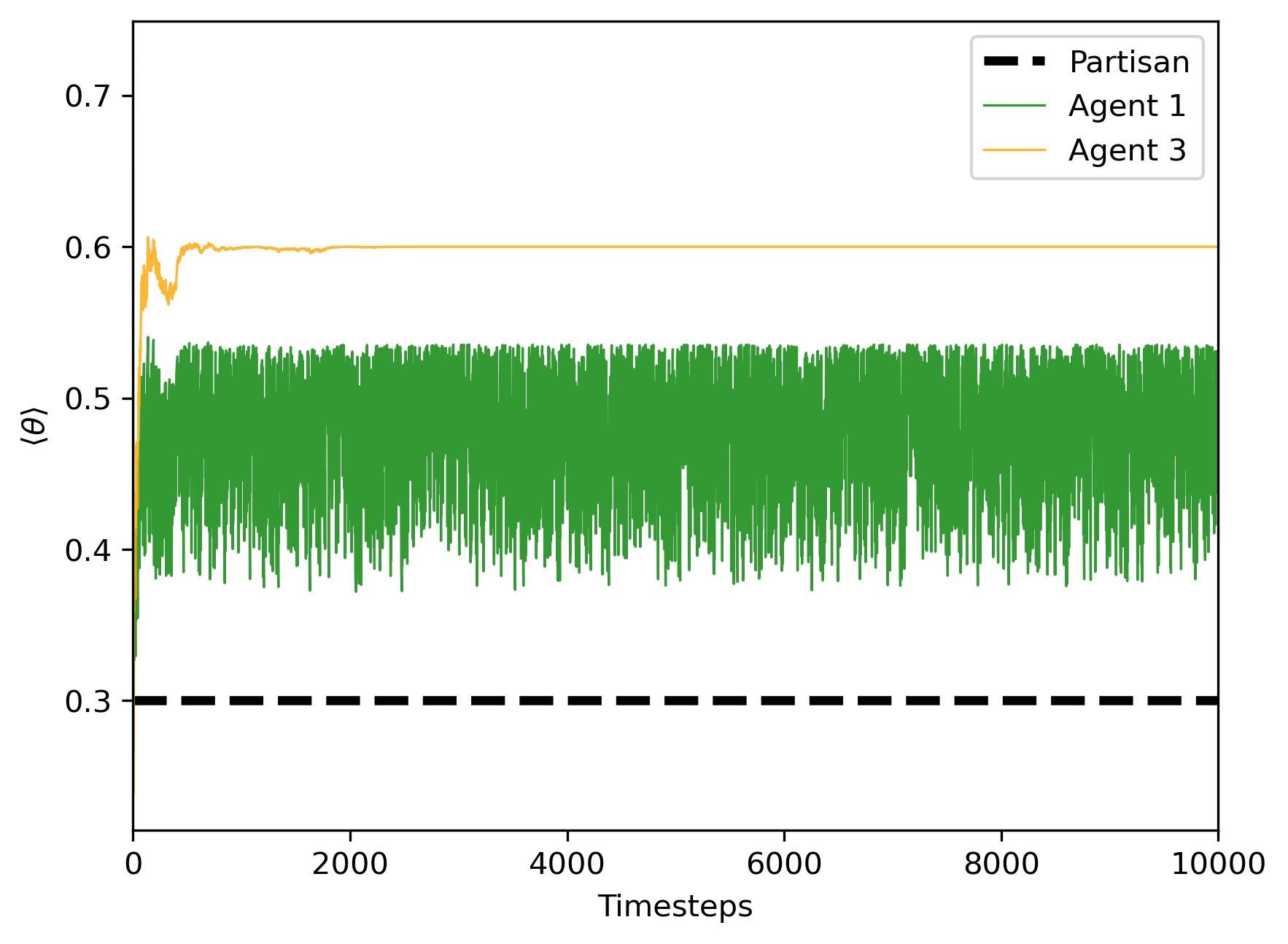}
        \caption{}\label{fig:G_2_mean_fe}
    \end{subfigure}%
    \caption{Unbalanced triad $G_2$ with internal tension, when agent 1 is the partisan [$G_{2{\rm p}1}$; panels (a) and (b)] and when agent 2 is the partisan [$G_{2{\rm p}2}$; panels (c) and (d)]. 
    (a) Snapshot of belief PDF at $t=5 \times 10^3$ for agents 1 (partisan; black dashed curve), 2 (blue curve), and 3 (orange curve).
    Agents 2 and 3 have bimodal PDFs with peaks at $\theta_{\rm p}$ (where they agree) and near $\theta_0$ (where they disagree). 
    (b) Mean belief $\langle \theta \rangle$ versus time for agent 1 (black dashed curve), agent 2 (blue curve) and agent 3 (orange curve). We observe turbulent nonconvergence for both persuadable agents. 
    (c) As for (a), but with the green curve corresponding to agent 1, who has a bimodal PDF at $\theta_{\rm p}$ and $\theta_0$, while agent 3 has an unimodal PDF peaked at $\theta_0$.  
    (d) As for (b), but with the green curve corresponding to agent 1. Agent 3 achieves asymptotic learning at $t_{\rm a1} = 636$, while agent 1 exhibits turbulent nonconvergence.
    }
    \label{fig:G_2}
\end{figure}

Let us begin with $G_{2{\rm p}1}$, which is the version of $G_2$ where the partisan (agent 1) is allied with persuadable opponents (Fig.\ \ref{fig:triad}, middle row, left graph). 
Fig.\ \ref{fig:G_2_prior} displays snapshots of $x_i(t=5 \times 10^3, \theta)$ for all three agents, while Fig.\ \ref{fig:G_2_mean} displays $\langle\theta\rangle$ versus time ($0\leq t \leq 1 \times 10^4$) for all three agents.
We observe turbulent nonconvergence for both persuadable agents, whereas only one agent (agent 1) experiences turbulent nonconvergence without a partisan in Ref.\ \cite{low_discerning_2022}. 
Agents 2 and 3 are both pulled by their alliances towards the partisan, while still observing the coin tosses. 
However, they strive to disagree with each other, so the peaks away from $\theta_{\rm p}$ in their bimodal belief PDFs occur at unequal values of $\theta$.
Dwelling still occurs, including for long intervals, e.g.\ $7374 \leq t \leq 9071$, which starts following a run of tails during $7369 \leq t \leq 7374$. 
Fig.\ \ref{fig:G_2} is a reminder that a hypothetical external observer should be cautious about inferring the truth of a specific belief by extrapolating from its popularity. In Fig.\ \ref{fig:G_2_prior}, for example, every agent believes in $\theta_{\rm p}$ to a greater or lesser extent, whereas some agents do not believe in $\theta_0$ at all, yet the partial consensus about $\theta_{\rm p} \neq \theta_0$ is misleading as a guide to the true bias.

We now turn to $G_{2{\rm p}2}$, the version of $G_2$ where agent 1 is allied with the partisan (agent 2) and agent 3, who opposes the partisan (Fig.\ \ref{fig:triad}, middle row, right graph).
Note that switching agent 1 with agent 3, or selecting agent 3 as the partisan, leads to a network with the same connections and topology.
Figs.\ \ref{fig:G_2_prior_fe} and \ref{fig:G_2_mean_fe} plot the same quantities as Figs.\ \ref{fig:G_2_prior} and \ref{fig:G_2_mean} respectively. 
They agree with the results in Ref.\ \cite{low_discerning_2022}, where agent 2 and 3 achieve asymptotic learning, 
and agent 1 experiences turbulent nonconvergence, vacillating between the beliefs of agent 2 and 3.
However, agent 3 reaches the right belief upon achieving asymptotic learning, unlike in Ref.\ \cite{low_discerning_2022}, because the partisan does not ``zero out'' agent 3's likelihood at $\theta_0$.

\begin{figure}[h!]
    \centering
    \begin{subfigure}{0.5\textwidth}
        \centering
        \includegraphics[width=\linewidth]{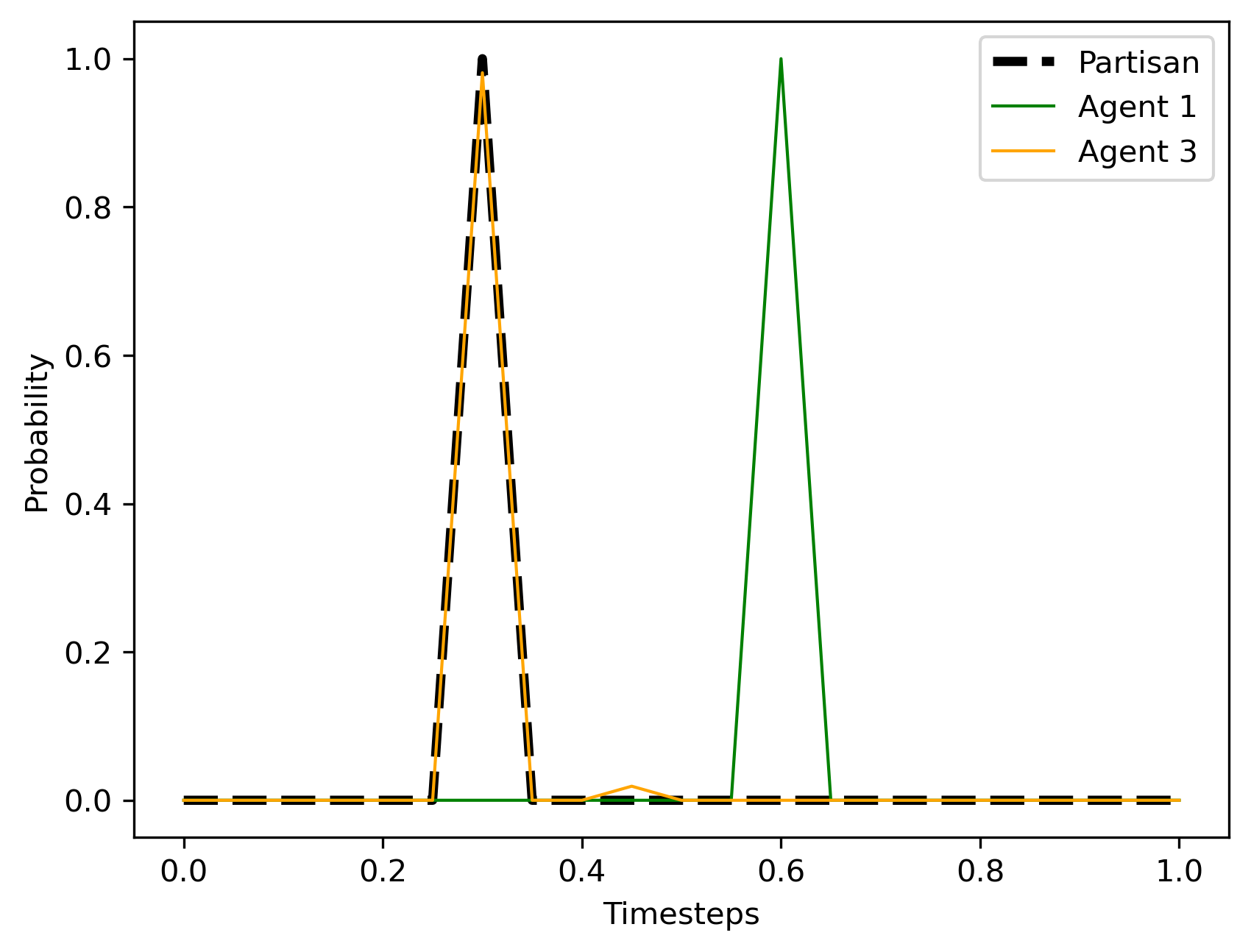}
        \caption{}\label{fig:G_3_prior}
    \end{subfigure}%
    \hspace*{\fill}  
    \begin{subfigure}{0.5\textwidth}
        \centering
        \includegraphics[width=\linewidth]{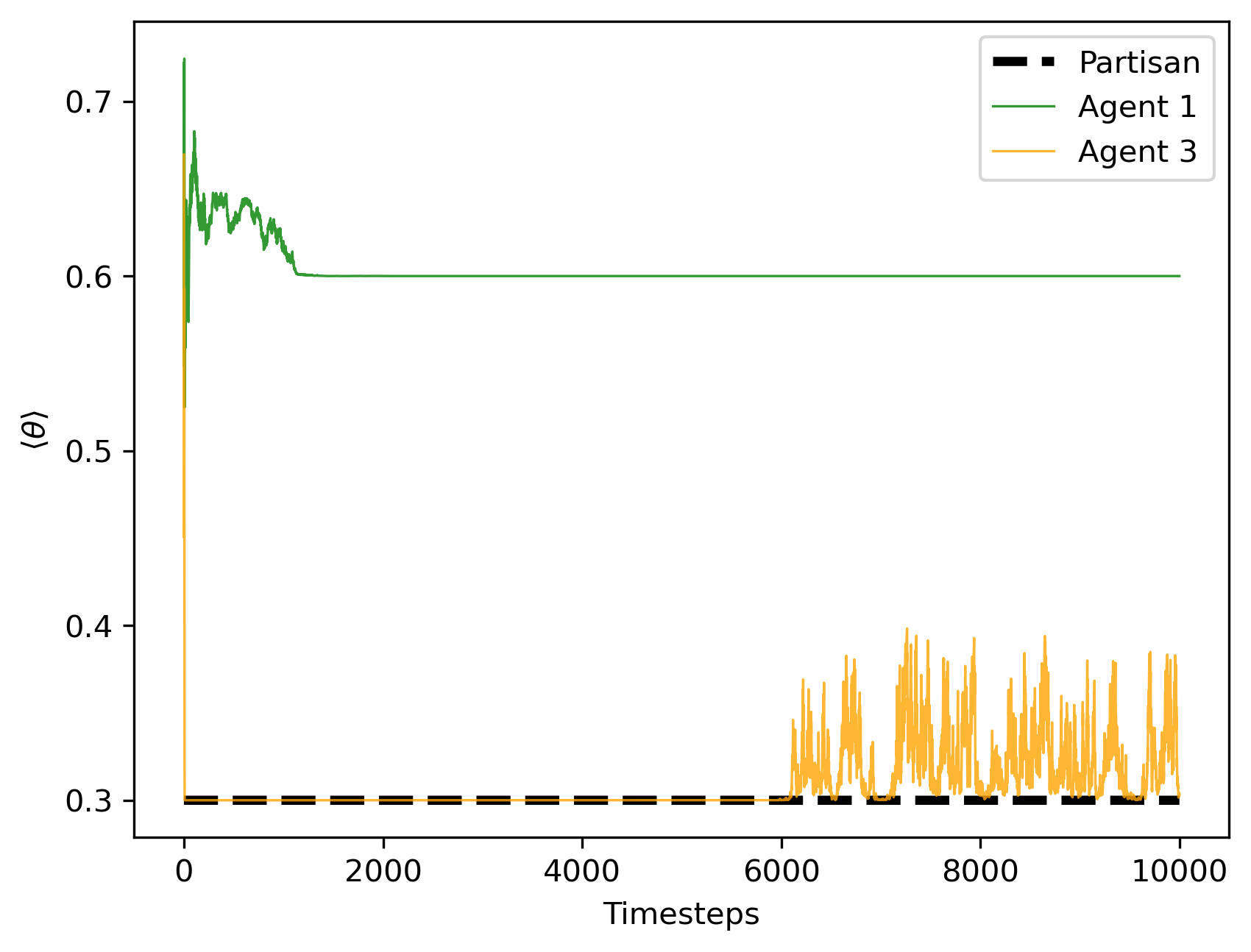}
        \caption{}\label{fig:G_3_mean}
    \end{subfigure}
    \caption{
        As for Fig.\ \ref{fig:G_2}, but for the network $G_{3{\rm p}2}$.
        Agent 1 achieves asymptotic learning at $t_{\rm a1} = 1119$. 
        Agent 3 initially agrees with the partisan then suddenly transitions to turbulent nonconvergence, starting at $t = 6100$, an example of intermittency \cite{low_vacillating_2022}. 
    }
    \label{fig:G_3}
\end{figure}

Let us now consider $G_{3{\rm p}2}$, the version of $G_3$ where agent 2 (the partisan) opposes agent 1 and allies with agent 3 (Fig.\ \ref{fig:triad}, bottom row, right graph)\footnote{Network $G_{3{\rm p}1}$ (Fig.\ \ref{fig:triad}, bottom row, left graph), in which agent 1 (the partisan) is opposed to agents 2 and 3, who form a persuadable yet allied bloc, exhibits the same dynamics essentially as an opponents-only network with $n=2$ and one partisan. Both persuadable agents achieve asymptotic learning at $\theta_0$; see Section \ref{sec:opponentonly}.}. 
The results appear in Fig.\ \ref{fig:G_3} in the same format as Fig.\ \ref{fig:G_2}.
Agent 1, opposing both the partisan and agent 3, does not experience peer pressure as the beliefs do not overlap, as discussed in Section \ref{subsec:opponent_frac_partisan}.
Hence, agent 1 only responds to the coin tosses and achieves asymptotic learning at $t_{\rm a1} = 1119$. 
Interestingly, in Fig.\ \ref{fig:G_3_mean}, agent 3 agrees quickly with the partisan (with agreement reached by $t = 7$) and maintains that belief until $t=6100$, when a long interval of turbulent nonconvergence ensues, triggered by a sequence of six heads in a row. 
This is an example of the intermittency phenomenon observed without partisans in Ref.\ \cite{low_vacillating_2022}. 
Agent 3 realizes, via Eqs.\ \eqref{eq:updatefirsthalf} and \eqref{eq:likelihood}, that six heads in a row are unlikely to be consistent with $x_3(t=6100, \theta) \approx \delta(\theta-0.3)$, and hence starts to gain confidence in higher values of $\theta$, without discounting $\theta_{\rm p}=0.3$ completely.
However, agent 1 locks agent 3 out of $\theta = 0.6$, as discussed in Section 4.2 of Ref.\ \cite{low_discerning_2022}.
Hence agent 3 is driven towards the midpoint $\approx (\theta_{\rm p} + \theta_0)/2$, so that $x_2(t,\theta)$ becomes bimodal for $t > 6100$.
We remind the reader that turbulent nonconvergence does not happen in the $G_3$ triad without a partisan, as demonstrated in Ref.\ \cite{low_discerning_2022}. 

The intermittent behavior of agent 3 in Fig.\ \ref{fig:G_3} is only one possible behavior in the $G_{3{\rm p}2}$ triad. 
Agent 3 sometimes exhibits turbulent nonconvergence from the beginning of the simulation, for example, maintaining a bimodal belief PDF with one peak at $\theta_{\rm p}$ and one at $\theta \neq \theta_0$ disagreeing with agent 1. 
The statistics of the alternative forms of intermittent behavior are complicated, as shown in Ref.\ \cite{low_vacillating_2022} even without partisans, and will be studied fully in future work.

\subsection{Larger networks}
\label{subsec:larger_network}

\begin{figure}[h!]
    \centering
    \begin{subfigure}{0.5\textwidth}
        \centering
        \includegraphics[width=\linewidth]{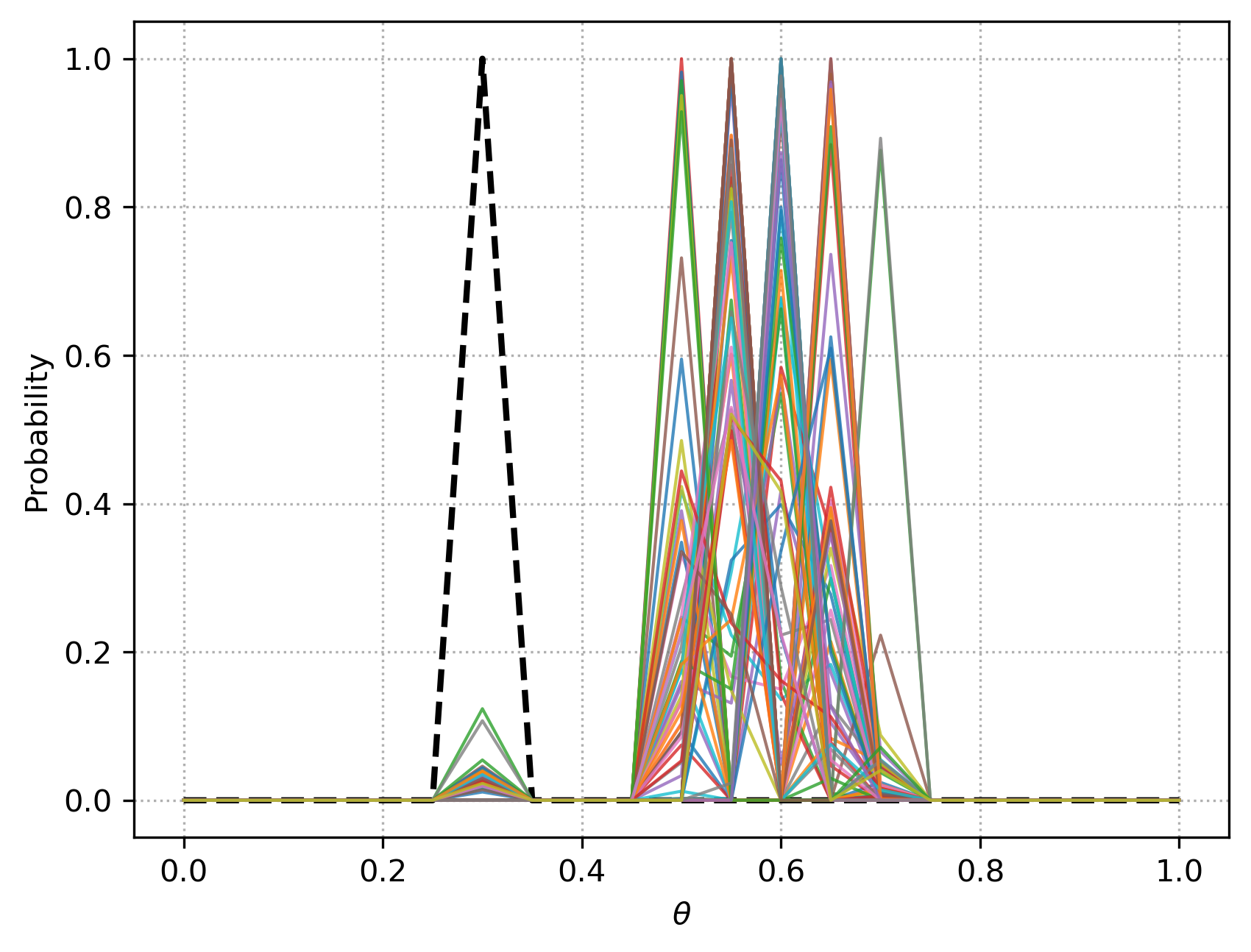}
        \caption{}\label{fig:big_mixed_final}
    \end{subfigure}%
    \hspace*{\fill}  
    \begin{subfigure}{0.5\textwidth}
        \centering
        \includegraphics[width=\linewidth]{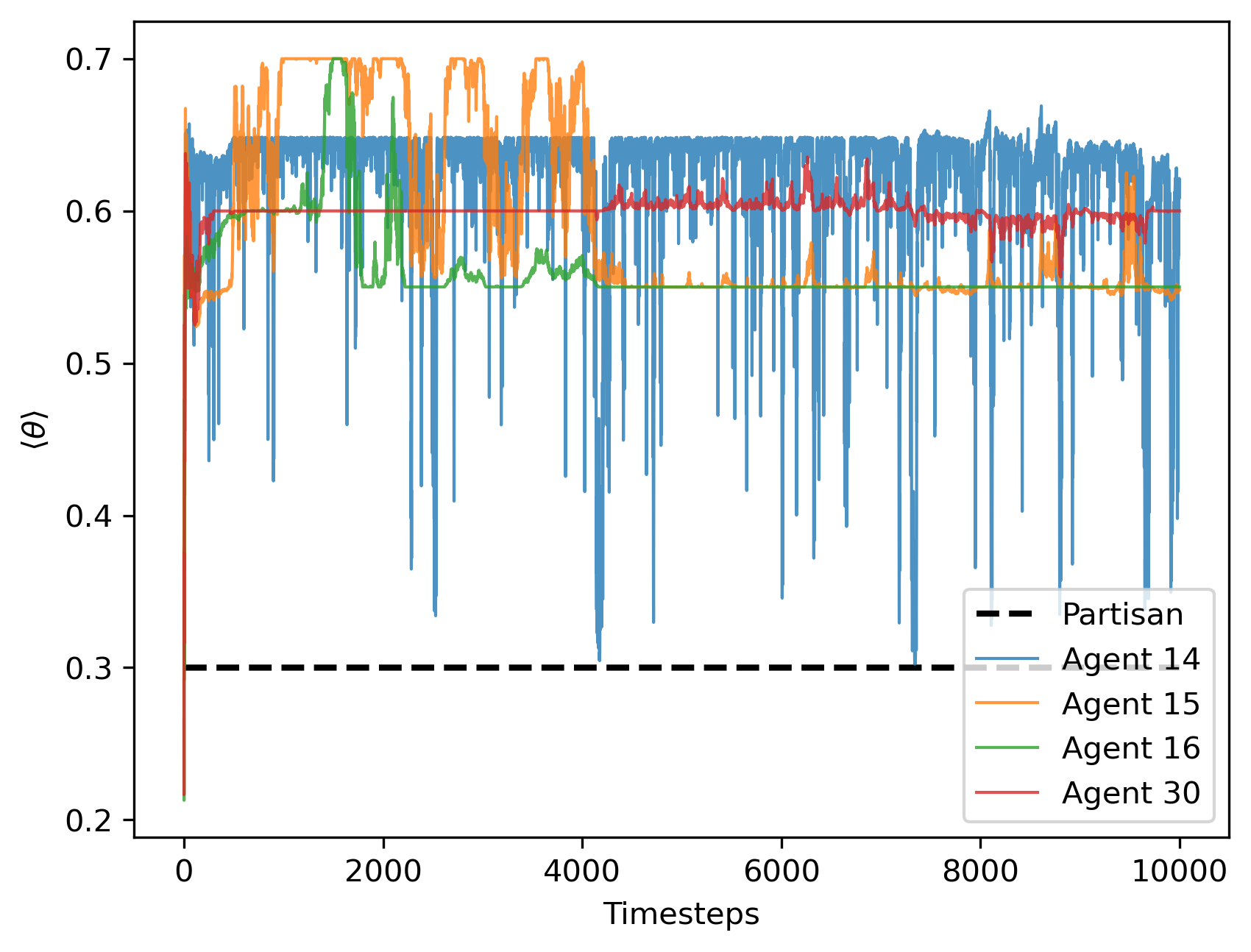}
        \caption{}\label{fig:big_mixed_mean}
    \end{subfigure} 
    \caption{
        Representative example of a mixed network with 2489 edges joining allies, 2461 edges joining opponents, one partisan, and $n=100$ (complete).
        (a) Snapshot of belief PDFs of all agents at $t = 10^4$, showing the partisan (black, dashed) and persuadable agents (variously colored, solid).
        Agents allied with the partisans have binomial PDFs with peaks at $\theta_{\rm p}$ and around $\theta_0$.
        (b) Mean belief $\langle \theta \rangle$ versus time for the partisan (black, dashed curve) and selected agents 14 (blue curve), 15 (orange curve), 16 (green curve), and 30 (red curve), illustrating examples of typical behaviors of persuadable agents. 
    }
    \label{fig:big_mixed}
\end{figure}

We now consider one representative example of a larger complete network with mixed allegiances, $n =100$, and one partisan. 
The sign of $A_{ij}$ is selected randomly with equal probability for all $i$ and $j$ and yields 2489 edges joining allies and 2461 edges joining opponents in the illustrative example analyzed here. 
Fig.\ \ref{fig:big_mixed_final} shows the belief PDF at $t = 10^4$ for each agent. 
Every PDF features a portion with $x_i(t,\theta) \neq 0$ for $| \theta - \theta_0 | \lesssim 0.1$. The PDFs are unequal for different agents and satisfy $x_i(t,\theta) \neq 0$ for multiple values of $\theta$ for some (but not all) agents. 
In addition, allies of the partisan develop a second peak at $\theta_{\rm p}$ (48 agents in this particular simulation). 
 
Fig.\ \ref{fig:big_mixed_mean} displays how $\langle\theta\rangle$ evolves for four selected agents, each displaying different but typical behavior. 
Agent 14 is allied to the partisan and exhibits turbulent nonconvergence as described in Section \ref{subsec:samethetap}. 
Agents 15, 16, and 30 oppose the partisan, satisfy $x_i(t,\theta_{\rm p})=0$ for $i=$ 15, 16, and 30, and exhibit intermittency as described in Ref.\ \cite{low_vacillating_2022}. 
Agent 15 always has a bimodal belief PDF, with the peaks moving in the range $0.5 \leq \theta \leq 0.7$ but never reaching $\theta_{\rm p} = 0.3$.
Agent 30 dwells for a long time (4238 time steps) at $\langle \theta\rangle \approx 0.6$, with PDF $\approx \delta(\theta - \theta_0)$, then suddenly transitions to turbulent nonconvergence at $t = 4239$, like agent 3 in Fig.\ \ref{fig:G_3_mean}. 
On the other hand, Agent 16 enters the long dwell interval $4165 \leq t \leq T$ with $x_{16}(t,\theta) \approx \delta(\theta-0.55)$.  
Given a longer simulation, it is possible that agent 16 will transition to turbulent nonconvergence at $t > 1\times 10^4$, like agent 30 and agent 3 in Fig.\ \ref{fig:G_3_mean}. 
A similar mixed network with $n=1000$ was also tested and returned similar results (not plotted for brevity).

In future work, we will study large networks with mixed allegiances in systematic detail, with the aim of linking an agent's behavior to their connectivity in general and their relationship with the partisan in particular (which may be null on occasion, e.g.\ in a Barab\'{a}si-Albert network).

\section{Partly connected networks}
\label{sec:BA}
It is natural to ask whether the behavior observed in Sections \ref{sec:alliesonly} -- \ref{sec:mixed} is specific to complete networks. To what extent do the results depend on the network's connectivity?
In Section \ref{subsec:wrong_conclusion_first}, for example, we find that the tendency to reach the wrong conclusion first depends on the attachment parameter in Barab\'{a}si-Albert opponents-only networks.
Moreover, it is plausible intuitively that the influence of a partisan increases, as their connections to the rest of the network increase, and that their influence reaches a maximum in a complete network, where they are connected to every other agent.

In this section, we take a first pass at generalizing the results in Sections \ref{sec:alliesonly} -- \ref{sec:mixed} to partly connected networks, as foreshadowed in Section \ref{subsec:network}. We adopt Barab\'{a}si-Albert networks as a traditional test case, motivated by previous theoretical studies \cite{low_discerning_2022,low_vacillating_2022}, and the conditions in many real social settings \cite{barabasi_emergence_1999,tang_survey_2016,kumar_structure_2016,maniu_building_2011}, and defer the study of other network topologies to future work.
Exploring the behavior for all possible values of the attachment parameter $m$ is outside the scope of this paper. Instead, we focus on networks with small $m$, i.e.\ networks that are sparsely connected, to accentuate the distinction with the complete networks studied in Sections \ref{sec:alliesonly} -- \ref{sec:mixed}.
We leave the exploration of dueling partisans within Barab\'{a}si-Albert networks to a later paper.
In Section \ref{subsec:distance}, we test how the beliefs of persuadable agents in allies-only networks depend on the minimum distance to the partisans, and investigate the implications for achieving consensus. 
In Section \ref{subsec:frac_ba}, we test how $m$ affects the trend of dwell time versus partisan fraction. 
We explore briefly the behavior of mixed allegiances in Section \ref{subsec:mixed_ba}, and the optimal placement of partisans to achieve a sociopolitical goal in Section \ref{subsec:Optimizations}. The latter two topics are subtle and multifarious and will be investigated fully in future work.

\subsection{Distance to the partisans: dissolving consensus on the path to turbulent nonconvergence}
\label{subsec:distance}

In a complete, allies-only network, even one partisan is enough to prevent asymptotic learning; the global outcome is turbulent nonconvergence, as described in Section \ref{subsec:samethetap}. However, the complete connections between persuadable agents ensure, that the persuadable agents reach a consensus promptly, which is maintained even while the persuadable agents vacillate turbulently between believing in $\theta_0$ and $\theta_{\rm p}$. In a partly connected, allies-only network, it is still true that even one partisan is enough to prevent asymptotic learning; that is, the global outcome remains unchanged.
However, the persuadable agents never reach a consensus; they follow different paths while enacting turbulent nonconvergence, because some are adjacent to the partisan and others are not. We demonstrate this behavior for smaller ($n=3$) and larger ($n=100$) Barab\`{a}si-Albert networks in this section. 

\begin{figure}[h!]
    \centering
    \begin{subfigure}{0.2\textwidth}
        \centering
        \includegraphics[width=\textwidth]{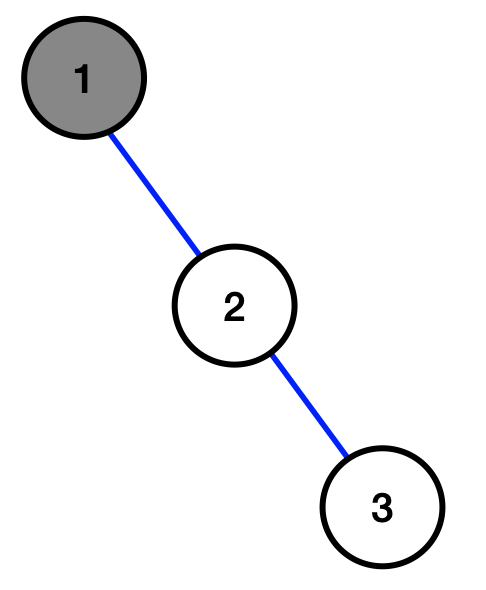}
        \caption{}\label{fig:allies_link}
    \end{subfigure} \\
    \begin{subfigure}{0.45\textwidth}
        \centering
        \includegraphics[width=\linewidth]{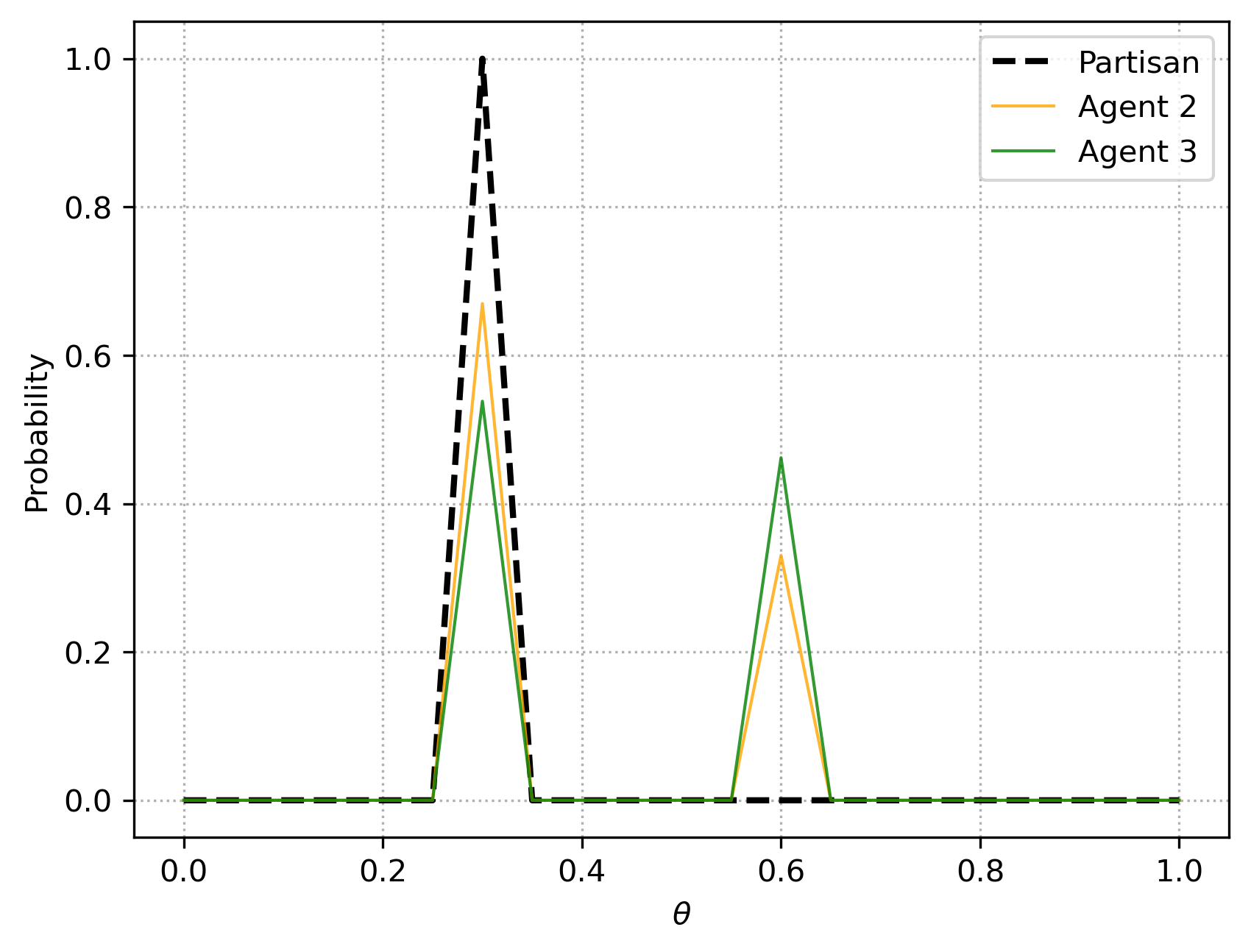}
        \caption{}\label{fig:allies_link_distr}
    \end{subfigure} 
    \hspace*{\fill}  
    \begin{subfigure}{0.45\textwidth}
        \centering
        \includegraphics[width=\linewidth]{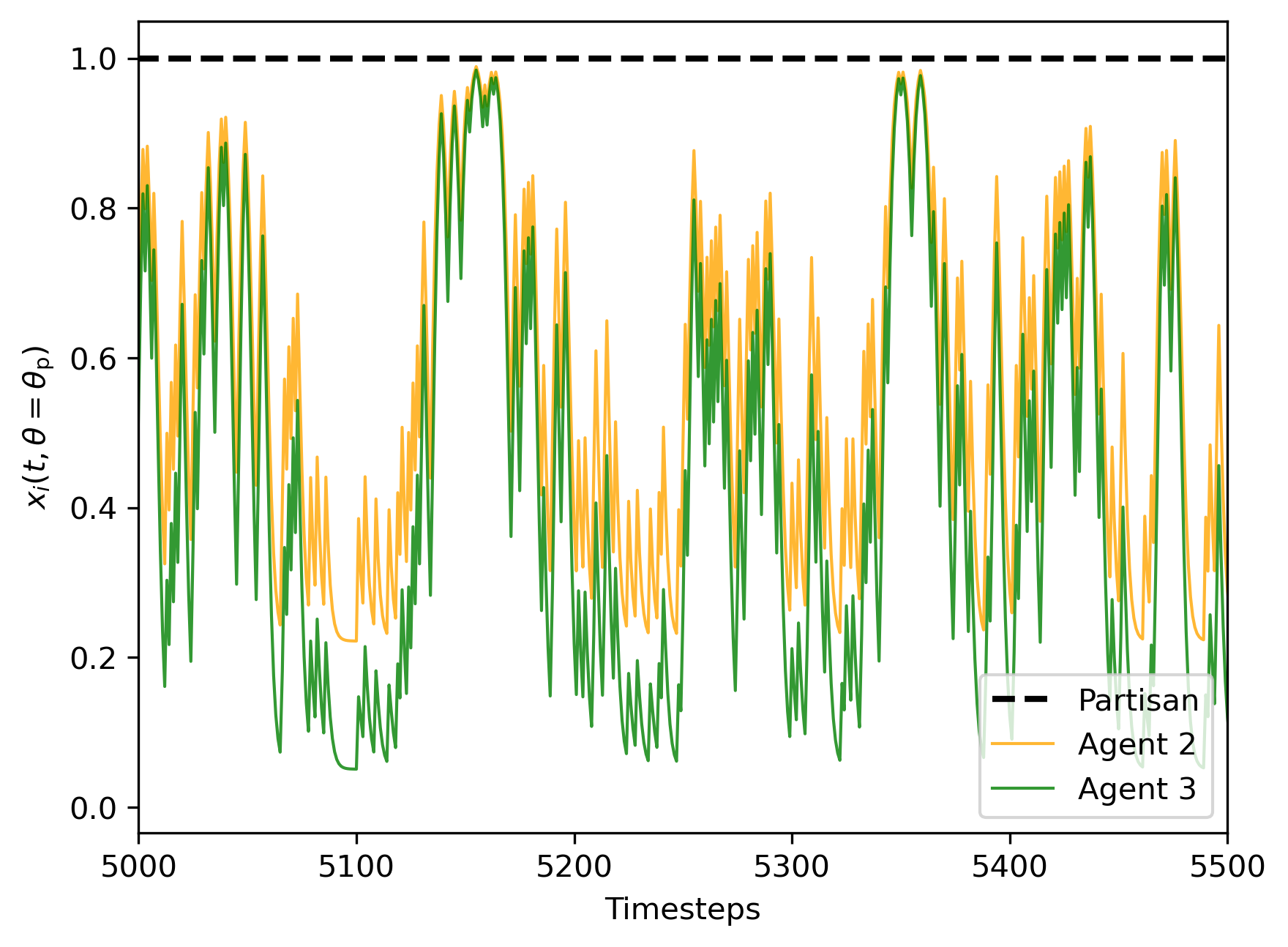}
        \caption{}\label{fig:allies_link_partisan_belief}
    \centering
    \end{subfigure}
    \caption{
        Turbulent nonconvergence without consensus in a simple partly connected network with $n=3$ and one partisan (agent 1). (a) Network topology. The blue solid edges join allies. The partisan is shaded.
        (b) Snapshot of belief PDF at $t=5 \times 10^3$ for agents 1 (partisan; black dashed curve), 2 (orange curve), and 3 (green curve).
        Agents 2 and 3 have bimodal PDFs with peaks at $\theta_{\rm p}$ and at $\theta_0$, with $ x_{2}(t, \theta = \theta_{\rm p}) > x_{3}(t, \theta = \theta_{\rm p}) $ and $ x_{2}(t, \theta = \theta_0) < x_{3}(t, \theta = \theta_0)$. 
        (c) Time evolution for $x_i(t, \theta = \theta_{\rm p})$ for $5 \times 10^3 \leq t \leq 5.5 \times 10^3$ of the partisan (black dashed line), agent 2 (orange curve), and agent 3 (green curve).
        Consensus is dissolved, with $x_2(t,\theta=\theta_{\rm p}) > x_3(t,\theta=\theta_{\rm p})$, unlike in the complete network in Fig.\ \ref{fig:meanconvergent}.
    }
\end{figure}

Consider first an allies-only network with $n=3$, illustrated in Fig.\ \ref{fig:allies_link}.
Agent 1, shaded grey, is a partisan. 
Agents 2 and 3 are persuadable. 
Agent 2 is adjacent to the partisan. 
Agent 3 is not but it is connected indirectly to the partisan via agent 2, i.e.\ one step removed.
We run a simulation for randomized priors and coin tosses and $T = 10^4$, which is analogous to the simulation in Fig.\ \ref{fig:meanconverge} but with $n=3$ instead of $n=100$.
Fig.\ \ref{fig:allies_link_distr} shows a snapshot of the belief PDF at $t = 5 \times 10^3$.  
We observe that the PDF is bimodal, as in Fig.\ \ref{fig:0.6truebias}, and that agent 2 is more confident in $\theta_{\rm p}$ than agent 3, because agent 2 is adjacent to the partisan, and agent 3 is not.
Further to the same point, Fig.\ \ref{fig:allies_link_partisan_belief} shows the evolution of $x_i(t, \theta = \theta_{\rm p})$ in the interval $5.0\times 10^3 \leq t \leq 5.5\times 10^3$. 
We find $ x_{2}(t, \theta = \theta_{\rm p}) > x_{3}(t, \theta = \theta_{\rm p})$ throughout the interval, i.e.\ persuadable agents who are closer to the partisan are more confident in $\theta_{\rm p}$ and less confident in $\theta_0$.
In a complete network, e.g.\ in Section \ref{subsec:samethetap}, we find $x_2(t,\theta_{\rm p}) = x_3(t,\theta_{\rm p})$ throughout the interval instead, i.e.\ consensus.

\begin{figure}[h!]
    \centering
    \begin{subfigure}{0.45\textwidth}
        \centering
        \includegraphics[width=\textwidth]{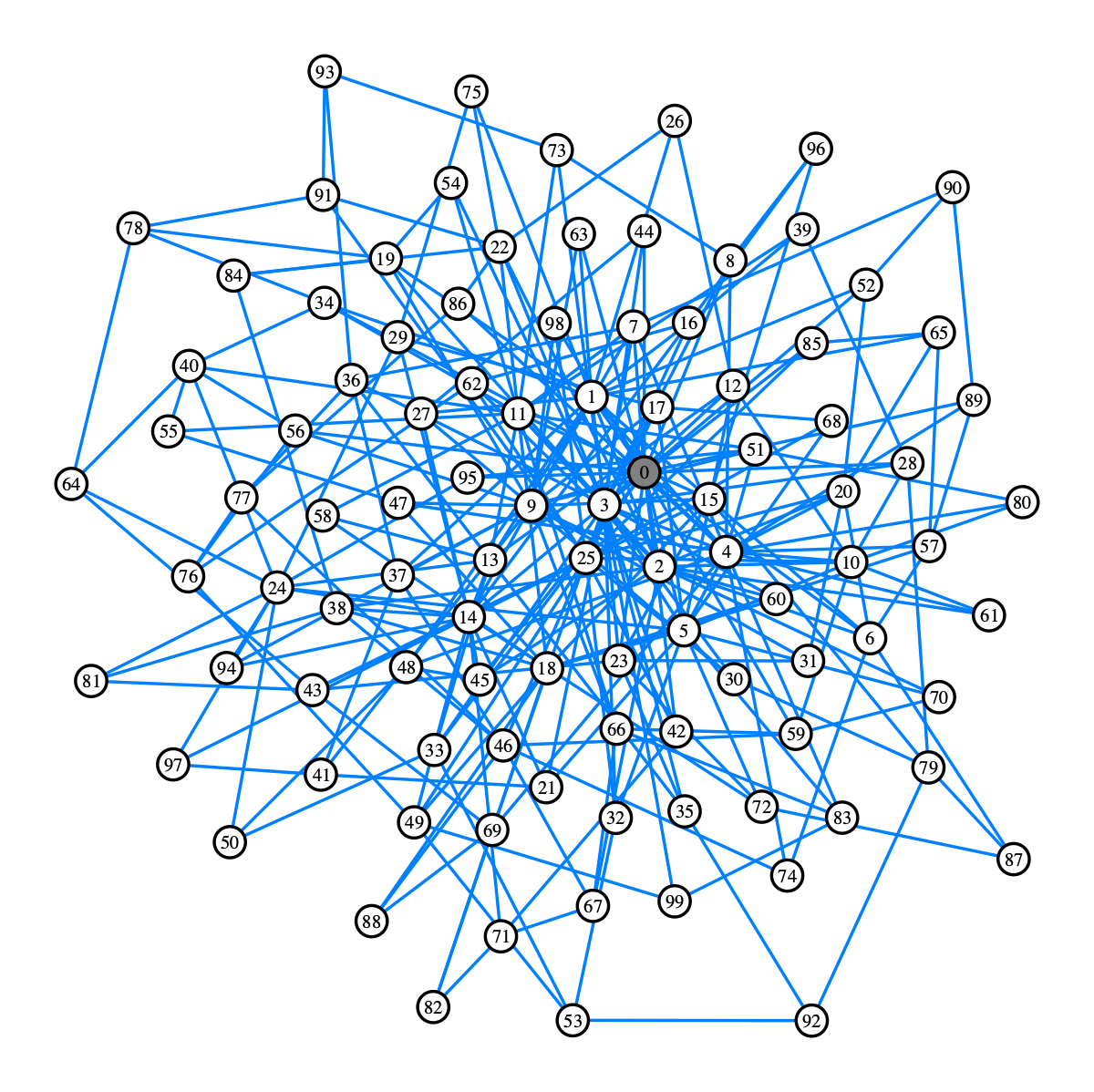}
        \caption{}\label{fig:BA_allies_m3}
    \end{subfigure} \\
    \begin{subfigure}{0.45\textwidth}
        \centering
        \includegraphics[width=\linewidth]{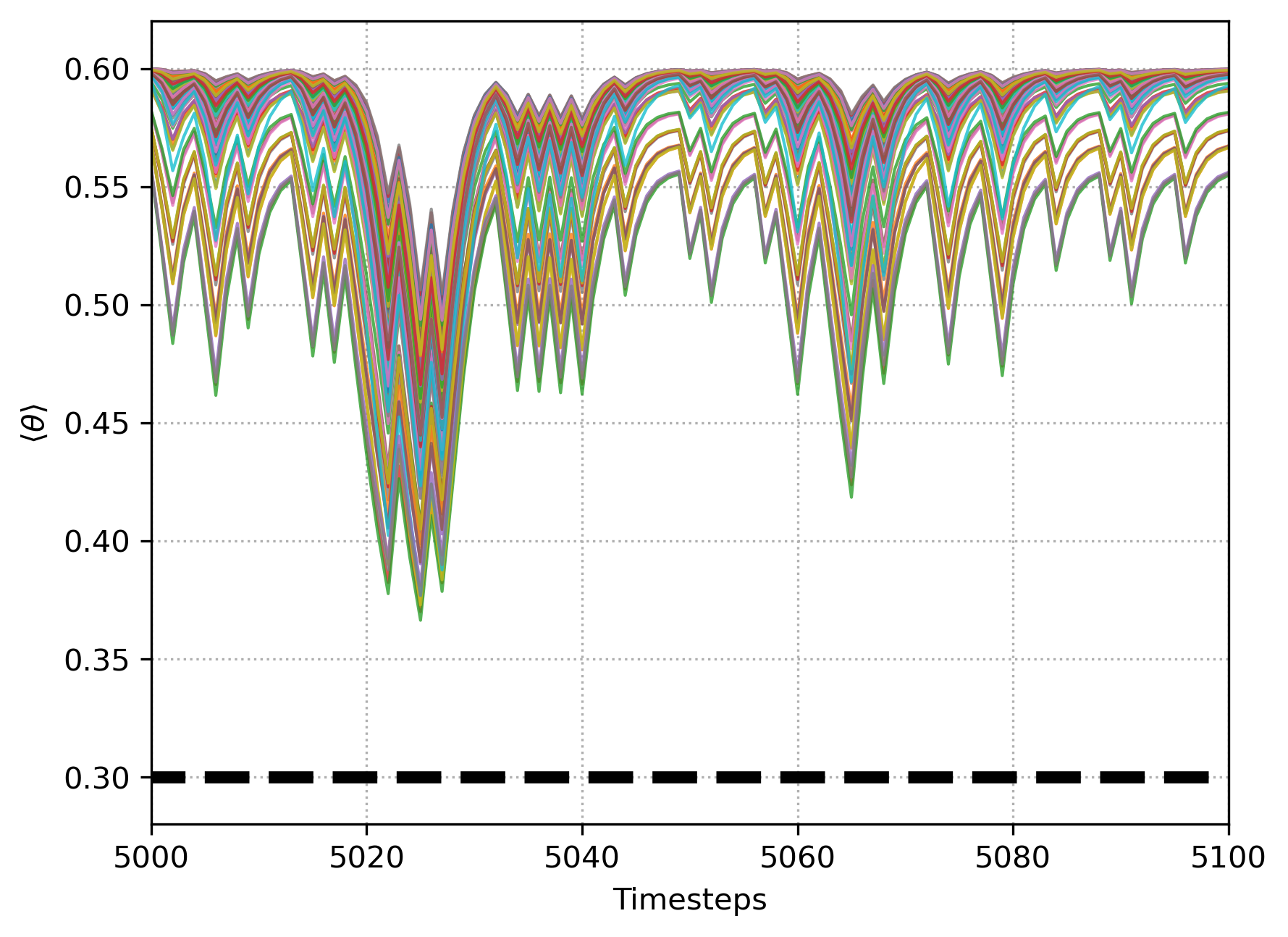}
        \caption{}\label{fig:nonconsensus}
    \end{subfigure} 
    \centering
    \begin{subfigure}{0.45\textwidth}
        \centering
        \includegraphics[width=\linewidth]{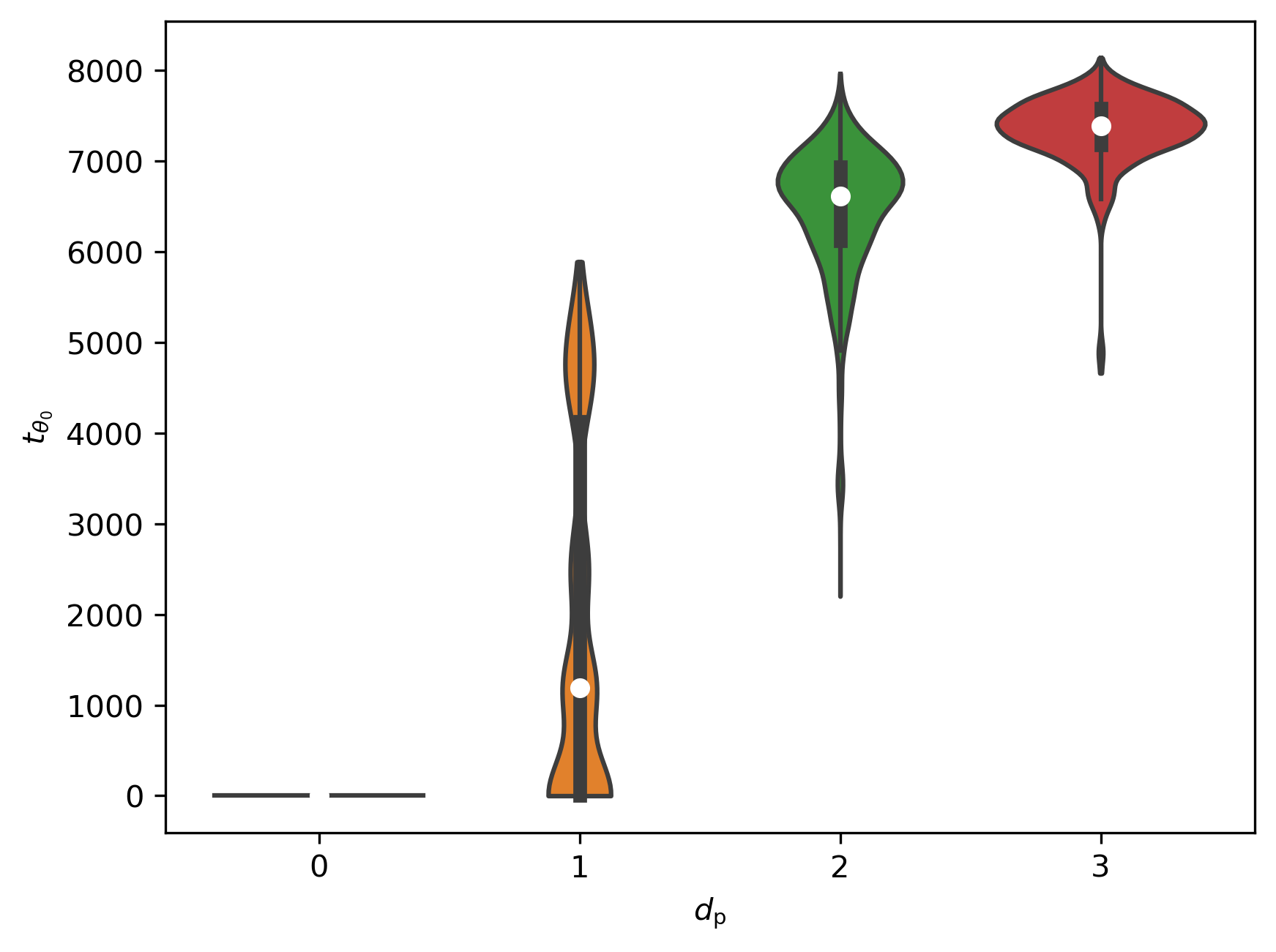}
        \caption{}\label{fig:allies_violin_samegraph}
    \end{subfigure}
    \caption{
        Turbulent nonconvergence without consensus in a larger Barab\'{a}si-Albert network with $n=100$, $m=3$, and one partisan (agent 0). 
        (a) Network topology. Blue solid edges join allies. The partisan is shaded.
        (b) Mean belief $\langle \theta \rangle$ of every agent versus time $t$ for $5 \times 10^3 \leq t \leq 5.1 \times 10^3$. The colored and black dashed curves correspond to the persuadable agents and partisan respectively. Consensus is not achieved between persuadable agents. 
        (c) Violin plot for the total number of timesteps that an agent satisfies $x_i(t, \theta = \theta_0) \geq 0.9$, denoted by $t_{\theta_0, i}$, as a function of separation from the partisan, $d_{\rm p}$, for all 100 simulations with randomized prior and coin tosses. 
        The horizontal width of the violin represents the number of agents at a given value of $t_{\theta_0, i}$. 
        The network in Fig.\ \ref{fig:BA_allies_m3} is used for all simulations.
        There are 23, 62, and 14 agents with $d_{\rm p} = 1,2,$ and 3 respectively. The single partisan has $d_{\rm p}=0$.
    }
\end{figure}

Let us now repeat the test in the previous paragraph for a larger Barab\`{a}si-Albert network with $n=100$ and $m=3$, as depicted in Fig.\ \ref{fig:BA_allies_m3}, with agent 1 being the partisan. 
Fig.\ \ref{fig:nonconsensus} displays the evolution of the mean belief $\langle\theta\rangle$ of every agent in one particular simulation for $5 \times 10^3 \leq t \leq 5.1 \times 10^3$. 
Unlike in Fig.\ \ref{fig:meanconvergent}, where the mean beliefs of persuadable agents converge mutually within $t<15$ time steps, 
the mean beliefs of persuadable agents in Fig.\ \ref{fig:nonconsensus} never converge mutually. 
That is, at any instant $t$ in the plotted range, $\langle \theta \rangle$ ranges typically from 0.35 to 0.6 for the 99 persuadable agents. 
Overall, however, $\langle \theta \rangle$ for every agent follows a similar trajectory as a function of $t$, because all persuadable agents observe the same sequence of coin tosses.
\footnote{One can quantify the degree of consensus by calculating the Kullback-Leibler divergence between agents, a topic for future work. }
This behavior differs from allies-only complete networks with partisans, discussed in Section \ref{subsec:samethetap}, and allies-only Barab\'{a}si-Albert networks without partisans, discussed in Appendix B in Ref.\ \cite{low_discerning_2022}.
Persuadable agents disagree for two reasons: 
\begin{inparaenum}[(i)]
\item they feel the influence of the partisans differently, because some are adjacent to partisans and others are not; and
\item they feel the influence of other persuadable agents differently because they are connected to different numbers of persuadable agents.
\end{inparaenum}

In order to quantify further the cause of the breakdown in consensus, we define $d_{\rm p}$ to be the length of the shortest path from a selected persuadable agent to any partisan.
We use breadth first search (BFS) \cite{moore_shortest_1959} to find the shortest path, noting that BFS only works in graphs with positive edge weights (here, allies-only networks).
Let us also define $t_{\theta_0, i}$ to be the number of total (and not necessarily consecutive) timesteps the $i$-th agent satisfies $x_i(t, \theta = \theta_0) \geq 0.9$, i.e.\ the number of timesteps when the agent is very confident in $\theta_0$,  with the threshold 0.9 having been chosen arbitrarily.  
We run an ensemble of 100 simulations with randomized priors and coin tosses on a Barab\'{a}si-Albert allies-only network with $n =100$, $m =3 $ with one partisan for $T = 10^4$.

Fig.\ \ref{fig:allies_violin_samegraph} shows violin plots of $t_{\theta_0, i}$ for each agent, as a function of $d_{\rm p}$. 
The horizontal width is a smoothed version of the histogram that shows the number of persuadable agents with a certain $t_{\theta_0, i}$, which is reflected around the vertical axis to create the shape of the violin.
The white dot indicates the mean and the thick, vertical, black bar represents the interquartile range. 
The values of $d_{\rm p}$ in this particular network are 0, 1, 2, 3, where $d_{\rm p} = 0$ refers to the single partisan; i.e.\ there are no persuadable agents with $d_{\rm p} > 3$ in this network with $n=100$.
We find that $\max(d_{\rm p})$ is related to the choice of $m$; $\max(d_{\rm p})$ is higher in more sparsely connected networks, i.e.\ smaller $m$. 
Persuadable agents with larger $d_{\rm p}$ are more confident in $\theta_0$, as shown by the fact that the white dots and thick black bars in Fig.\ \ref{fig:allies_violin_samegraph} trend higher, as $d_{\rm p}$ increases.
We also find $\max(t_{\theta_0,i}) - \min(t_{\theta_0,i}) = 5890, 5766, 3475$ for $d_{\rm p} = 1,2,3$ respectively. 
Agents with the same $d_{\rm p}$ can still have different beliefs due to their different connectivity to other persuadable agents.
In summary, the dissolving consensus among persuadable agents in Barab\'{a}si-Albert networks is attributed to the difference in connectivity and $d_{\rm p}$ values, where $t_{\theta_0, i}$ is related to $d_{\rm p}$.

\subsection{Dwell time versus $f$ and $d_{\rm p}$}
\label{subsec:frac_ba}

In the context of turbulent nonconvergence, the persistence of the beliefs of a persuadable agent is captured via the dwell time $t_{\rm d}$ defined by Eq.\ \ref{eq:dwelltime} rather than the asymptotic learning time defined by Eq.\ \ref{eq:asymlearncondition}. In Fig.\ \ref{fig:dwell_vs_frac_partisan}, we find that $\langle t_{\rm d} \rangle$ increases with the partisan fraction $f$ in a complete network. Here, we check how the trend in  Fig.\ \ref{fig:dwell_vs_frac_partisan} changes as a function of the attachment parameter, when the network is partly connected.

\begin{figure}[h!]
    \centering
    \begin{subfigure}{0.45\textwidth}
        \centering
        \includegraphics[width=\linewidth]{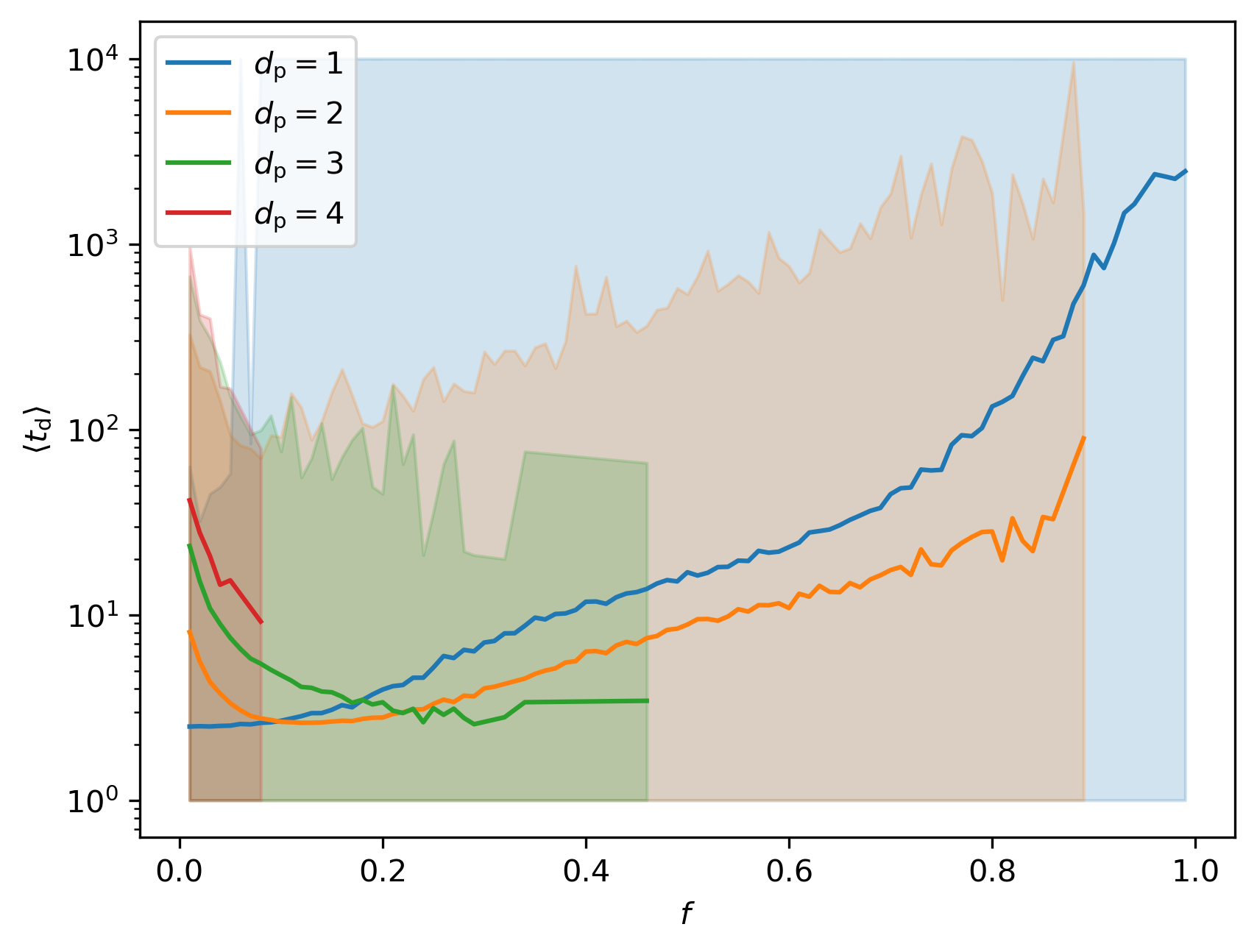}
        \caption{}\label{fig:frac_p_m3}
    \end{subfigure}%
    \hspace*{\fill}  
    \begin{subfigure}{0.45\textwidth}
        \centering
        \includegraphics[width=\linewidth]{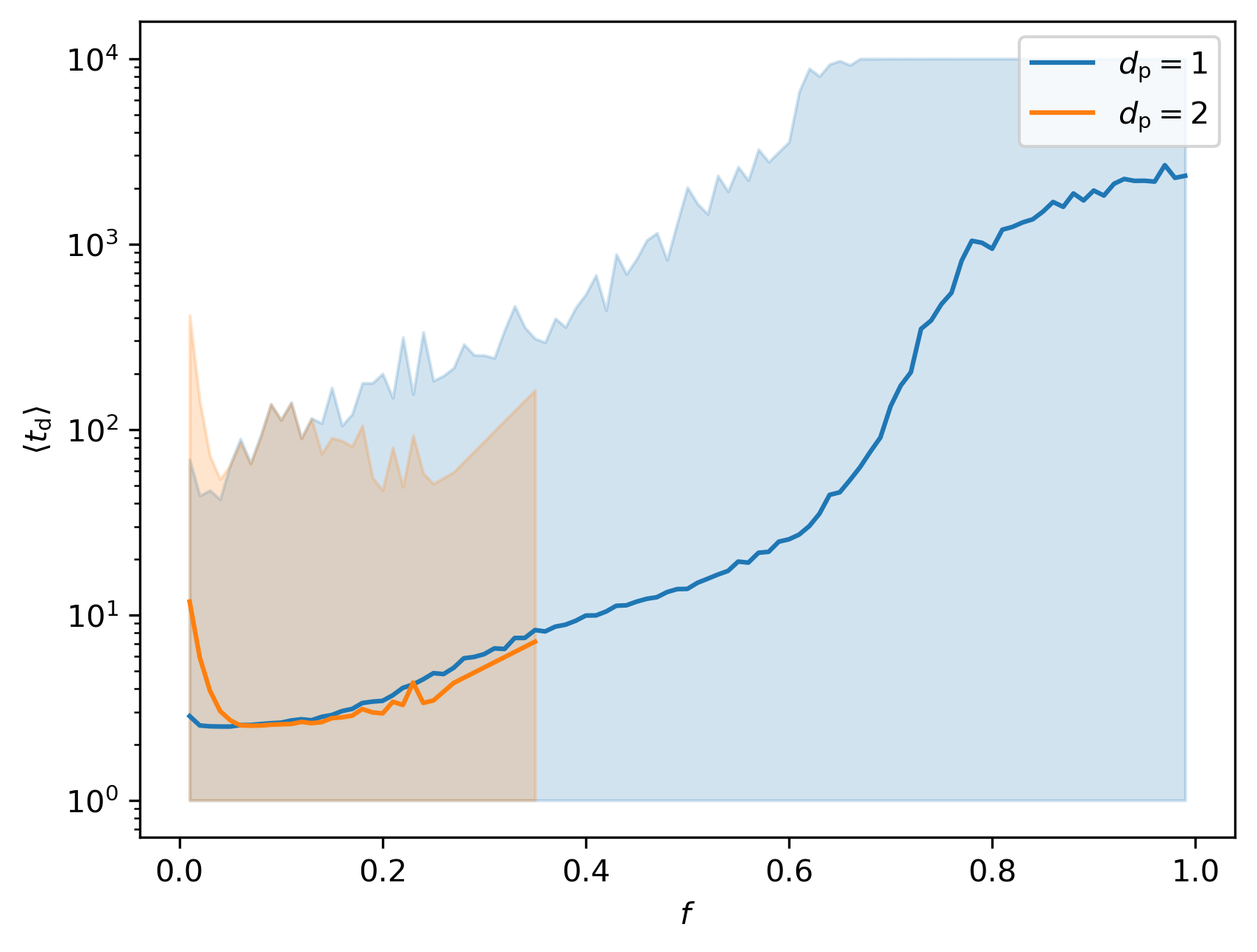}
        \caption{}\label{fig:frac_p_m20}
    \end{subfigure}%
    \hspace*{\fill}  
    \caption{
        Dwell time $t_{\rm d}$ versus partisan fraction $f$, as for Figure 13b, but for partly connected Barab\`{a}si-Albert networks with (a) $m=3$, and (b) $m=20$
        Scanning the range $0.01 \leq f \leq 0.99$ in steps of 0.01, we run an ensemble of 100 simulations with randomized priors and coin tosses, and $T=10^4$ at each value of $f$.
        Within each subpopulation, color-coded according to the legend, the persuadable agents are separated by the same $d_{\rm p}$. 
        The solid curve indicates the ensemble mean, and the shading spans the ensemble minimum and maximum. 
    }
    \label{fig:frac_BA}
\end{figure}

We consider two Barab\`{a}si-Albert networks with $n=100$, a sparse one with $m = 3$, and one of medium density with $m=20$.  
The networks with $m=3$ and $m=20$ complement the complete networks with $m = n-1 = 99$ studied in Section \ref{subsec:dwelltime_and_frac}. 
Fig.\ \ref{fig:frac_p_m3} and Fig.\ \ref{fig:frac_p_m20} display $\langle t_{\rm d} \rangle$ versus $f$ for the $m=3$ and $m=20$ networks respectively. The curves are color-coded according to $d_{\rm p}$, e.g.\ the blue curve corresponds to evaluating $t_{\rm d}$ for the subpopulation of persuadable agents adjacent to a partisan ($d_{\rm p}=1$). The aim is to test how $t_{\rm d}$ depends on the distance to the nearest partisan, and what trade-off exists between $t_{\rm d}$ and $f$.
The cut-offs for $d_{\rm p} = 2,3,4$ occur, because the number of persuadable agents with $d_{\rm p} > 1$ drops to zero, when $f$ exceeds some threshold $f_{\rm max}(d_{\rm p})$; for example, we find $f_{\rm max}(d_{\rm p} = 3) = 0.46$ in Fig.\ \ref{fig:frac_p_m3}.

Fig.\ \ref{fig:frac_BA} leads to four main conclusions. 
First, $\langle t_{\rm d} \rangle$ increases monotonically with $f$ for $d_{\rm p}=1$, with $d\langle t_{\rm d} \rangle / df$ increasing sharply for $f\gtrsim 0.6$, just like in Fig.\ \ref{fig:dwell_vs_frac_partisan}. The trend depends weakly on $m$ (compare the blue curves on the same axes in Fig.\ \ref{fig:frac_p_m3} and Fig.\ \ref{fig:frac_p_m20}). 
Second, $\langle t_{\rm d} \rangle$ decreases with $f$ at low $f \lesssim 0.2$ for $d_{\rm p} > 1$.
This is because the ``pull'' from partisans on persuadable agents with $d_{\rm p} >1 $ is weaker than the ``pull'' from the coin for low $f$, so agents dwell with belief close to $\theta_0$. 
As $f$ increases, while $d_{\rm p}$ is held fixed, the ``pull'' from the partisan increases, causing $t_{\rm d}$ at $\theta_0$ and hence $\langle t_{\rm d} \rangle$ to decrease.
Third, $\langle t_{\rm d} \rangle$ increases with $d_{\rm p}$ for $f \lesssim 0.2$ but decreases with $d_{\rm p}$ for $f \gtrsim 0.2$, as seen in Fig.\ \ref{fig:frac_p_m3}. 
This occurs because at low $f$, the ``pull'' from the partisans is weaker for agents with higher $d_{\rm p}$.
Their beliefs are dominated by the coin tosses, leading to longer $\langle t_{\rm d} \rangle$ at low $f$. 
For $f \gtrsim 0.2$, an increase in partisan population means that the ``pull'' from the partisans outweighs the ``pull'' from the coin tosses, leading to all agents dwelling at $\theta_{\rm p}$ longer.  However, persuadable agents closer to the partisans (lower $d_{\rm p}$) are more strongly influenced by the ``pull'' from the partisans, resulting in $\langle t_{\rm d} \rangle$ decreasing with $d_{\rm p}$. 
Fourth, the trends of $\langle t_{\rm d} \rangle$ versus $f$ for the $d_{\rm p}=1$ and $d_{\rm p} =2$ subpopulations resemble each other more closely for $m=20$ than for $m=3$ (compare Fig.\ \ref{fig:frac_p_m3} and Fig.\ \ref{fig:frac_p_m20}). 
This occurs because the contribution of each individual agent is weaker in a densely connected network, as the interaction between agents is averaged over all neighbors (through Eq.\ \ref{eq:xiprimed}). 
Hence the effect of $d_{\rm p}$ on $\langle t_{\rm d} \rangle$ is weaker on networks with greater $m$.  
One main difference between Fig.\ \ref{fig:dwell_vs_frac_partisan} and Fig.\ \ref{fig:frac_BA} is the maximum dwell time.
In complete networks, one obtains $t_{\rm d} \simeq 10^4$ for $f \gtrsim 0.7$, whereas in Barab\'{a}si-Albert networks with $m=3$ one obtains $t_{\rm d} \simeq 10^4$ for $f \geq 0.06$ and $d_{\rm p} = 1$. 
Agents with $d_{\rm p} = 1$ and $t_{\rm d} \simeq 10^4$ are adjacent to more than one partisan, and non-adjacent partisans also  influence their beliefs.

\subsection{Mixed allegiances}
\label{subsec:mixed_ba}

The question of mixed allegiances is subtle and multi-faceted even in complete networks, as demonstrated in Section \ref{sec:mixed}. 
A full investigation of mixed allegiances in partly connected networks, a more challenging problem, lies well outside the scope of this paper. 
Instead, as a foretaste of what can be done, we present here one representative example: a Barab\'{a}si-Albert network with $n=100$, $m=3$, a single partisan, and $A_{ij} =\pm 1$ with equal probability. The specific realization of this network studied here features 318 edges connecting allies and 270 edges connecting opponents.

When we simulate the above network for $T =10^4$, we observe that $\langle \theta \rangle$ evolves just like in a complete network, in a manner that resembles Fig.\ \ref{fig:big_mixed_mean} (the graph is omitted to avoid repetition). 
Specifically, 13 out of 99 persuadable agents reach asymptotic learning early in the simulation ($t < 10^3$), with asymptotic mean belief in the range of $0.5 \leq \langle\theta\rangle \leq 0.8 $. One persuadable agent correctly and stably infers $\theta_0$. 
The other 86 persuadable agents exhibit turbulent nonconvergence or intermittent behavior. 
In comparison, in the complete network investigated in Section \ref{subsec:larger_network}, 11 out of 99 persuadable agents reach asymptotic learning, and three persuadable agents correctly and stably infer $\theta_0$. The other 88 persuadable agents exhibit turbulent nonconvergence or intermittent behavior. 
In complete networks, we observe that only persuadable agents who are adjacent and allied to one or more partisans (i.e. $A_{i \rm{p}} = 1$) develop a peak at $\theta_{\rm p}$ in their belief PDF (see Section \ref{subsec:larger_network}). 
In contrast, in a Barab\'{a}si-Albert network, some persuadable agents who are connected but not adjacent to a partisan (e.g.\ $A_{ij} \neq 0$, $A_{j{\rm p}} \neq 0$, and $A_{i{\rm p}}=0$ for some $j \neq i$, ${\rm p}$) also develop a peak at $\theta_{\rm p}$.
Specifically, 29 agents who are connected but not adjacent to the partisan have $x(t = T, \theta = \theta_{\rm p}) \geq 0.001$. 
The 29 agents have $2 \leq d_{\rm p} \leq 4$, where $d_{\rm p}$ is defined as the shortest distance irrespective of the sign of the edges, cf. BFS in Section \ref{subsec:distance}. 
In general, the notion of distance depends on the sign of the edges; for example, it may be argued that two agents separated by three positive edges (i.e.\ three alliances) are ``closer'' than two agents separated by three negative edges (i.e.\ three adversarial links). A systematic study of this issue, with a generalized definition of $d_{\rm p}$, is deferred to future work.

\subsection{Manipulating collective opinion: optimizing the placement of partisans}
\label{subsec:Optimizations}

A challenging question with important social implications is how to optimize the placement of obdurate partisans to manipulate the beliefs of persuadable agents in the service of some social or political goal. 
Within the specific context of media bias, for example, one goal could be to camouflage the political bias of a media outlet, by convincing as many consumers as possible that the outlet is neutral ($\langle \theta \rangle \approx 0.5$, say) when in reality it is biased strongly ($\theta_0 \approx 0$ or $1$, say). 
The latter example involves persuading agents to believe something false. In other applications, both within and beyond the specific context of media bias, the goal may be to persuade agents to believe something true, e.g.\ the danger of drink driving, or the efficacy of a medical treatment. 
The obdurate partisans may be real humans or automated systems such as social network ``bots''. 
They may conduct their operations by espousing beliefs that agree with the target belief or, interestingly and counterintuitively, by espousing beliefs that disagree with the target belief, leveraging oppositional relationships in the network to shepherd persuadable agents towards the target belief. 
The long-term impact of partisanship and the optimal placement of obdurate partisans has been modeled previously in a deterministic framework (without belief PDFs) to investigate how to maximize political polarization \cite{yildiz_binary_2013,arendt_opinions_2015} or conformity with the partisans' belief \cite{abrahamsson_opinion_2019,klamser_zealotry_2017,masuda_evolution_2012}. 

\begin{figure}[h!]
    \centering
    \begin{subfigure}{0.3\textwidth}
        \centering
        \includegraphics[width=\linewidth]{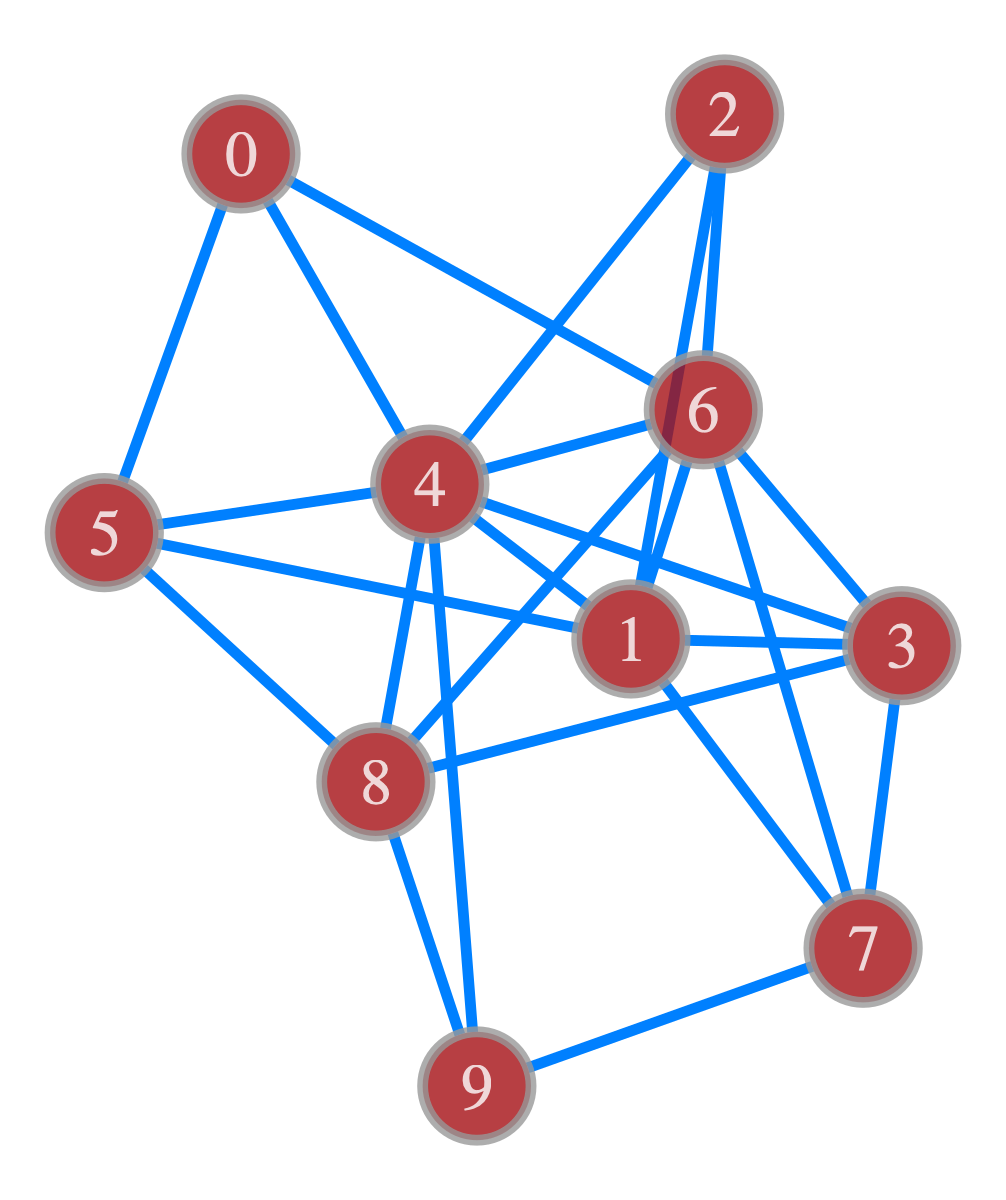}
        \caption{}\label{fig:ba_10}
    \end{subfigure}%
    \hspace*{\fill}  
    \begin{subfigure}{0.5\textwidth}
        \centering
        \includegraphics[width=\linewidth]{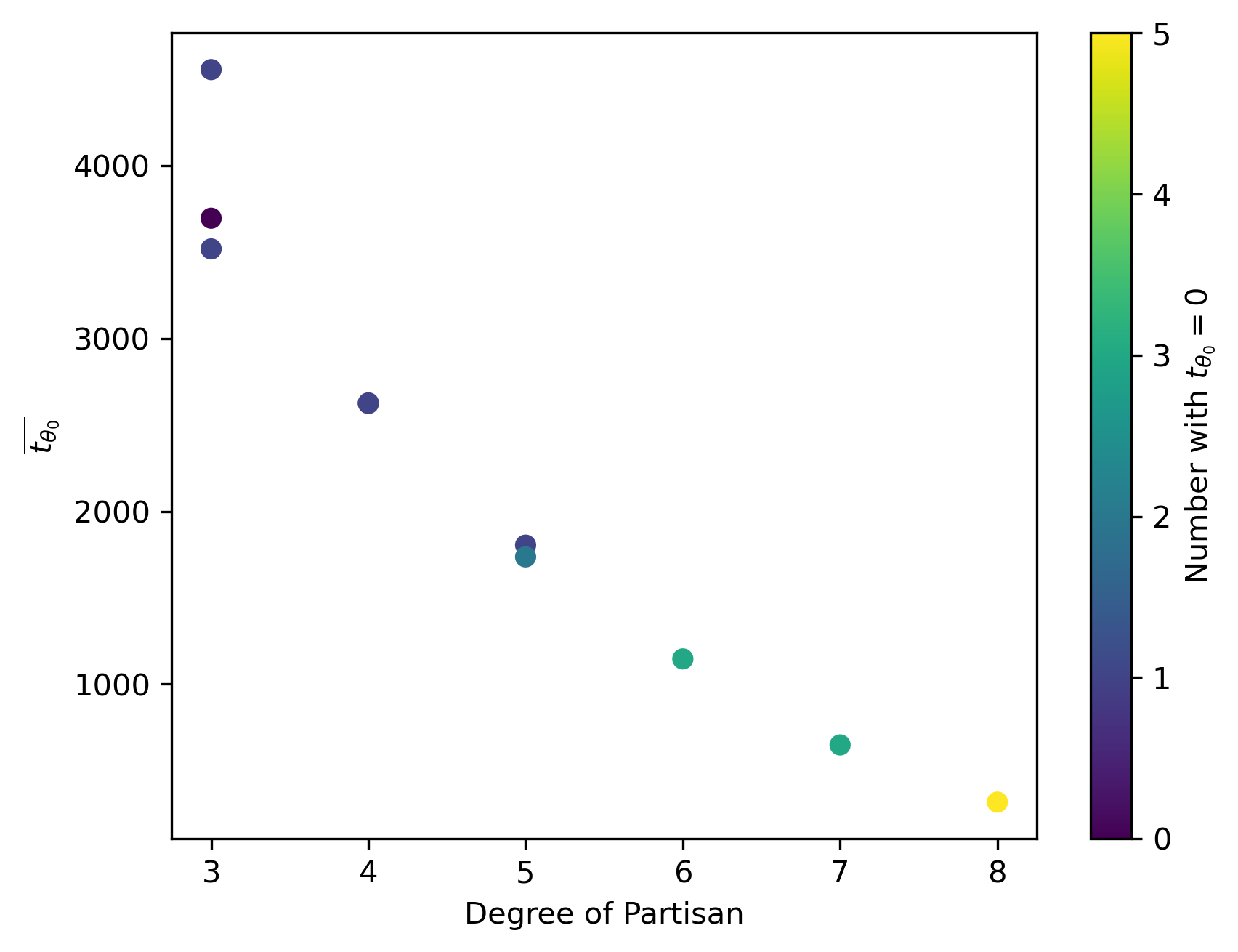}
        \caption{}\label{fig:opti_scatter}
    \end{subfigure}%
    \hspace*{\fill}  
    \caption{
        Optimizations of the placement of one partisan in an allies-only Barab\'{a}si-Albert network with $n=10$, $m = 3$.
        (a) Network topology. The degrees of the vertices are 3, 6, 3, 5, 8, 4, 7, 4, 5, 3 respectively in the order of vertex label. 
        The partisan can occupy any vertex.
        (b) Plot of $\overline{t_{\theta_0}}$ from Eq.\ \eqref{eq:bartheta0} as a function of the degree of the partisan vertex. 
        The points are color coded by the number of persuadable agents with $t_{\theta_0,i}=0$. 
    }
    \label{fig:optimization}
\end{figure}

In this section, we analyze briefly one specific, representative example of the above problem, as a foretaste of what can be investigated more broadly. The example involves a Barab\'{a}si-Albert network with $n=10$, $m=3$, and one partisan, depicted in Fig.\ \ref{fig:ba_10}. 
The degrees of the 10 vertices range from three to eight. The specific question asked is: where should the partisan be placed, to drive the beliefs of persuadable agents as far from $\theta_0$ as possible, i.e.\ to maximize $| \langle \theta \rangle - \theta_0 |$? 
To answer the question, we run 10 simulations with the same sequence of coin tosses for $T=10^4$. The simulations differ in what vertex the partisan occupies.
Let us define 
\begin{equation}\label{eq:bartheta0}
    \overline{t_{\theta_0}} = \frac{1}{(1-f) n} \sum_{i \neq {\rm p}} t_{\theta_0, i}, 
\end{equation}
where $f$ is the fraction of partisans in a network of size $n$. 
In Fig.\ \ref{fig:opti_scatter}, we observe that $\overline{t_{\theta_0}}$ is inversely related to the degree of the partisan. 
Furthermore, the number of persuadable agents with $t_{\theta_0}  = 0$ increases as the degree of the partisan increases, also as observed in Fig.\ \ref{fig:opti_scatter}. 
The partisan's degree is equal to the number of persuadable agents with $d_{\rm p} = 1$ (adjacent to the partisan). Persuadable agents with $d_{\rm p} = 1$ have shorter $t_{\theta_0, i}$, shown in Fig. \ref{fig:allies_violin_samegraph}, resulting in a shorter $\overline{t_{\theta_0}}$.
Therefore, if we wish to mislead the persuadable agents, we should place the partisan at the vertex with maximum degree, i.e.\ the most connected vertex. 
This result makes sense intuitively. However, we caution that it is not expected to apply always in networks with mixed allegiances, which contain subnetworks with internal tensions and trade-offs, such as the unbalanced triad $G_2$ in Fig.\ \ref{fig:triad}. 
An interesting avenue for future work is to optimize both the placement and beliefs of (perhaps dueling) partisans to manipulate a numerical majority of agents to hold a target belief $\theta_{\rm t}$. 
This optimization task is related to but different from minimizing the average displacement $| \langle \theta \rangle - \theta_{\rm t} |$, and may be more relevant in electoral applications \cite{bravo-marquez_opinion_2012,druckman_impact_2005}.

\section{Discussion and summary}
\label{sec:conclusion}
In this paper, we generalize previous studies of how opinions about media bias evolve in a network of politically allied or opposed Bayesian learners by inserting partisans into the network. 
Partisans, sometimes termed zealots in contexts unrelated to media bias, are obdurate agents, whose opinions do not change in response to external influences either in pursuit of a deliberate strategy or due to psychological factors. 
Individual persuadable agents, in contrast, hold partial (and changing) confidence in a spectrum of beliefs.
Their beliefs are described by a PDF, which can be multimodal and is updated via a two-stage version of Bayes's rule combining independent observations with peer pressure \cite{fang_opinion_2020,low_discerning_2022,low_vacillating_2022}.
The PDF-based analysis generalizes previous deterministic treatments of zealots, where each agent holds a unique belief \cite{mobilia_does_2003,mobilia_voting_2005,mobilia_role_2007}.

Monte Carlo simulations of the idealized, analogous system of a biased coin reveal several new behaviors, which generalize the results without partisans in Ref.\ \cite{low_discerning_2022} and \cite{low_vacillating_2022}, and the results for deterministic beliefs in previous work by Mobilia et al.\ \cite{mobilia_does_2003,mobilia_voting_2005,mobilia_role_2007} and others.
\begin{inparaenum}[(i)]
\item In allies-only networks, even a single partisan is enough to disrupt asymptotic learning, making persuadable agents vacillate in tandem between $\theta_0$ and $\theta_{\rm p}$. The dwell time is short for $f \lesssim 0.6$, with agents switching frequently from $\theta_0$ to $\theta_{\rm p}$ and vice versa, and episodes of turbulent nonconvergence in between. The dwell time rises steeply for $f\gtrsim 0.6$. 
\item Dueling partisans, who disagree either deliberately or by accident, reduce the dwell time. Even a single disagreeing partisan is enough to reduce $\langle t_{\rm d} \rangle$ by a factor $\approx 3$. The number of dwell intervals at $\theta_{\rm p1}$ and $\theta_{\rm p2}$ is proportional approximately to the number of partisans at $\theta_{\rm p1}$ and $\theta_{\rm p2}$. 
\item In opponents-only networks, multiple opponents can be ``corralled'' begrudgingly into the same belief, which may or may not equal $\theta_0$, due to repulsive peer pressure from agents with adjacent beliefs. 
\item \label{item:wrong-conclusion-first} The counterintuitive tendency to reach the wrong conclusion first in opponents-only networks, explored in Ref. \cite{low_discerning_2022}, is still observed in complete networks with $\theta_0 = \theta_{\rm p}$, although $t_{\rm a}^{\rm right} - t_{\rm a}^{\rm wrong}$ is lower. In general, for $\theta_0 \neq \theta_{\rm p}$, the tendency persists, as long as the network is sparse enough (e.g.\ Barab\'{a}si-Albert network with attachment parameter $m \lesssim 10$), as sometimes occurs in  real social settings. 
\item In mixed networks containing allies and opponents, sudden transitions occur from asymptotic learning (in the sense of Eq.\ \eqref{eq:asymlearncondition}) to turbulent nonconvergence, namely the intermittency phenomenon studied in Ref.\ \cite{low_vacillating_2022}. The strongly balanced $G_3$ triad in Fig. \ref{fig:triad} is destabilized by inserting a partisan, passing from asymptotic learning to intermittency. 
In contrast, the fate of the unbalanced triad $G_2$ depends on where the partisan is inserted.
\item Randomized tests confirm that the above results depend weakly on the priors and coin toss sequence, but there is some dependence on connectivity in opponents-only networks, as in point (\ref{item:wrong-conclusion-first}).
\end{inparaenum}

The social science theory of structural balance of Herider \cite{heider_attitudes_1946}, extended by Harary, Cartwright \cite{cartwright_structural_1956} and Davis \cite{daivs_clustering_1967} can be used to characterize networks based on clusters, as defined in Section 6 of Ref.\ \cite{low_discerning_2022}.
The stability of a network, in terms of its propensity to achieve asymptotic learning, has been linked to structural balance in the social science literature in the absence of partisans, e.g.\ 
the network of alliances of six particular countries in World War I began as an unbalanced network, and settled as a balanced network \cite{antal_social_2006}.
Networks can be categorized as: 
(i) strongly balanced, if the network can be partitioned into one or two distinct clusters, with no agents left out;
(ii) weakly balanced, if the network can be partitioned into more than two distinct clusters, with no agents left out;
(iii) unbalanced, if it is impossible to group every agent into a cluster.
The link between clusters and stability is modified in interesting ways, when partisans are introduced. 
Allies-only networks in Section \ref{sec:alliesonly} are an example of strongly balanced networks, as all agents can be grouped in the same cluster, yet they exhibit turbulent nonconvergence, when even a single partisan is introduced. 
Hence, it may be valuable to generalize the theory of structural balance by according different status to clusters including and excluding partisans, an interesting avenue for future work. 
Opponents-only networks in Section \ref{sec:opponentonly} are weakly balanced networks with each cluster containing only one agent. When partisans are introduced, we do not see major changes in the behavior of the network. 
Unbalanced networks are likely to experience turbulent nonconvergence and intermittency \cite{low_discerning_2022}, whether partisans are present or absent, but the situation is complicated from a structural balance viewpoint: the cluster that a partisan joins partly determines who in the network experiences intermittency and who does not. Clarifying the link is an interesting avenue for future work. 

We emphasize that the biased coin analyzed in this paper is a highly idealized analogy. 
In real societies, perceptions of media bias are formed through the interaction of many subtle political and psychological factors, which are not captured by the biased coin. 
These factors affect both stages of the Bayesian update rule in Section \ref{subsec:modelintro} and Algorithm \ref{alg:algo}. 
With regard to the first stage, when assessing media outputs (e.g.\ newspaper editorials), the judgments of individuals are influenced by their psychological state, which in turn depends on their beliefs, especially when those beliefs are held passionately. 
For example, an individual's preexisting perception of a newspaper's bias feeds into their reading of the text of a specific editorial in a self-reinforcing manner, not simply through the multiplicative prior $x_i(t,\theta)$ in Eq.\ \eqref{eq:updatefirsthalf} but additionally in a nonmultiplicative fashion; that is, $x_i(t,\theta)$ feeds into the individual's impressions about the tone, language, and logical arguments in the editorial and hence the likelihood $P[S(t)|\theta]$ itself. 
With regard to the second stage, the fundamental dynamics encoded by Eqs.\ \eqref{eq:undatesecondhalf} and \eqref{eq:xiprimed} --- that the beliefs of allies and opponents converge and diverge respectively --- assume that peer pressure occurs in a pairwise manner, i.e.\ the peer pressure from multiple pairs can be summed in linear superposition. 
In reality, people may not partition their response so fastidiously. Instead, they may cluster their allies and opponents using some ``fuzzy'', intuitive metric and react to clusters of different sizes (or types) nonlinearly. 
None of these effects enter the idealized model in this paper.

With the above caveats in mind, we advance with due reserve a handful of instances, where the results in this paper may enjoy some measure of social relevance in practice.
Firstly, it is striking that partisans disrupt the discovery of the truth even when they are fewer in number than persuadable agents \cite{mobilia_role_2007}. 
In this paper, even a single partisan disrupts asymptotic learning in an allies-only network; the persuadable agents are herd-like, in that they form a consensus, but the consensus switches frequently between $\theta_0$ and $\theta_{\rm p}$ at low $f$. 
These low-$f$ results represent the minority-partisan limit of the celebrated majority-partisan experiments by Asch \cite{asch_opinions_1955}, which demonstrated conformity through peer presure. 
They are relevant to debates about political bias by citizens in social networks, because it is easy for a small group of dedicated and cynical ``trolls'' to foment a climate of uncertainty and destabilize the perceptions of others by making them vacillate indefinitely, even if the ``trolls'' are too few to engineer a steady belief in something untrue. 
Secondly, the tendency to reach the wrong conclusion first in networks with significant amounts of antagonism (e.g.\ opponents-only networks as an extreme case) survives the introduction of partisans, as long as people are not connected too densely (e.g.\ attachment parameter $\lesssim 10$ in scale-free, Barab\'{a}si-Albert networks). 
Scale-free networks with modest levels of connectivity are common in society \cite{barabasi_emergence_1999,tang_survey_2016,kumar_structure_2016,maniu_building_2011}, and political opposition is patently a fact of life, so one expects a widespread tendency to reach the wrong conclusion first. 
This is related to the ``backfire effect'' in social science \cite{nyhan_when_2010}, where agents exposed to contrary beliefs become more confident in their own prior.
The effect has been modeled and studied in the scope of opinion dynamics by Chen et al.\ \cite{chen_opinion_2019}. 
Once a false conclusion becomes entrenched in a community, it can become a self-fulfilling prophecy, benefiting from its own unearned, network-bestowed, first-mover advantage. 
What is popular is not always a reliable guide to what is true; in several experiments in this paper, false beliefs championed by partisans (or even emerging accidentally from equal but opposite antagonistic interactions) are held more steadfastly for longer intervals (dwell times) by more agents than true beliefs. 
Thirdly, and finally, partisans do not always destabilize a network or deflect people from the truth, even if that is their intent. 
For example, one agent in an unbalanced triad always exhibits intermittent behavior, with or without a partisan, but a partisan prevents the non-intermittent agent from converging on the wrong belief. 

More generally, the phenomenon of intermittency in mixed networks, investigated elsewhere \cite{low_vacillating_2022,hoffman_role_2007}, survives the introduction of partisans, although the behavior depends on where exactly the partisans are located. 
This is relevant to social applications, where mixed networks are the norm, because it is a reminder that the feedback in mixed networks is complicated and unpredictable, and partisans guided by simplistic (e.g.\ pairwise-motivated) strategies to achieve certain goals may reap unintended consequences. 
In all cases, network effects are as important as individual human psychology in shaping outcomes.

The results in the paper suggest some productive directions for future work. 
\begin{inparaenum}[(i)]
\item It is interesting to investigate the behavior of persuadable agents who are indirectly connected to a partisan, using partly connected networks which more closely model real social networks (e.g.\ Barab\'{a}si-Albert). 
We make a start in this direction in Section \ref{sec:BA}, by looking briefly at outcomes including loss of consensus, controlling factors including $d_{\rm p}$, and optimizing the placement of partisans to satisfy some network-wide goal. 
\item It is also interesting to investigate the impact of one-sided relationships, such as an agent allying with another agent who opposes them, using directed networks (e.g.\ Price network \cite{price_networks_1965}).
\item One can generalize the concept of partisanship by endowing every agent, not just partisans, with some degree of stubbornness. For example, how stubborn must a ``semi-partisan'' be before they disrupt asymptotic learning in an allies-only network, as in Section \ref{sec:alliesonly}? 
\end{inparaenum}

Finally, efforts can be made to test the idealized model in this paper quantitatively against real-world data using natural language processing \cite{chowdhary_natural_2020} to infer individuals' perceptions of media bias from social media such as Twitter \cite{wang_system_2012,vilares_megaphone_2015,dragoni_unsupervised_2019} and make independent assessments of $\theta_0$ through text and image analysis of media outputs such as newspaper editorials \cite{kwon_opinion_2013}. 
There are two challenges in this sort of real-world benchmarking. The first is what question to ask. 
Is the goal to test the two-stage update rule defined by Eq.\ \eqref{eq:updatefirsthalf}--\eqref{eq:xiprimed}? 
Do we take the update rule for granted and seek to infer the existence and locations of partisans (if any)? 
Without direct access to $x_i(t,\theta)$, do we test the model for internal consistency by searching for positive or negative correlations between a quantity like $\langle \theta \rangle$ for the $i$-th agent (inferred perhaps from a proxy like social media output) and the local structure of $A_{ij}$?
Formal benchmarking frameworks of these kinds have been developed in the electrical engineering literature and elsewhere \cite{andersen_benchmarking_1995,moriarty_theory_2011}
and present an exciting avenue for future work. 
The second challenge is how to collect the relevant data under conditions that are as controlled as possible. Laboratory and field studies are both viable options \cite{jackson_social_2008}. 
Laboratory experiments offer the ability to control and fine-tune certain parameters, as is done in experimental economics \cite{weimann_experimental_2019}. 
However, some model parameters are not easily controlled. 
For instance, the learning rate $\mu$ is likely to differ from person to person, unlike in Eq.\ \eqref{eq:undatesecondhalf}. 
Field surveys, on the other hand, do not constrain parameters as strictly as laboratory experiments, but evaluate social structures in real-world settings, for example, Milgram's small world experiment \cite{milgram_milgthe_1967}. 
Data about opinions is normally limited and obtained through small sociological surveys or polls across populations and time \cite{peralta_opinion_2022}. 
The rise of computerized social experiments could be a promising means of expanding the volume of opinion data. 
It is possible to extract the topology of and agent behavior in real-world social networks from online platforms such as Facebook \cite{chang_proposed_2018,garton_studying_1997}. 
Online social networks also process large amounts of data on social network structures \cite{son_cognitive_2021,zhao_inferring_2013}, automated analysis of user-generated online content \cite{liu_sentiment_2012}, and infer agents' real-time opinions though their behavior online \cite{luceri_analyzing_2019,goel_real_2016,mu_clustering-based_2022}. 
There is some recent literature in opinion dynamics that benchmarks models against real-world data, mostly focused on detecting community structure in networks \cite{morarescu_opinion_2011,bu_graph_2020,newman_modularity_2006}. 
Ref.\ \cite{devia_framework_2022} proposes a qualitative framework through a histogram-based classification algorithm to evaluate opinion dynamics models on their accuracy of predicting the evolution of opinion in real-world values surveys.
In principle, the work in this paper can be benchmarked within such a framework.


\section*{Acknowledgements}
We thank Nicholas Kah Yean Low for sharing the automaton code in Ref.\ \cite{low_discerning_2022}, helping us to understand the model in Ref.\ \cite{low_discerning_2022}, helping to interpret the different behavior of Barab\'{a}si-Albert and complete networks in this paper, and verifying some simulation results.
We thank Yi Shuen Christine Lee and Jarra Horstman for additional discussions.
We thank Liam Saliba for providing technical programming advice. 
We thank the anonymous referees for constructive feedback and for suggesting the investigation in Section \ref{sec:BA}.
AM acknowledges funding from the Australian Research Council Centre of Excellence for Gco (OzGrav) (CE170100004).

\appendix

\section{Simulation implementation}
\label{sec:zoomzoom}
The computer code implementing Algorithm 1 is written in \texttt{Python3} \cite{van_rossum_python_1995}.
Where possible, libraries with underlying \texttt{C++} implementations are used to speed up mathematical calculations.
Networks are generated and manipulated with the graph theory library \texttt{Graph-tool} \cite{peixoto_graph-tool_2014}, which has a \texttt{C++} backend.
The \texttt{NumPy} library \cite{harris_array_2020} is used for numerical calculations, with vectorization applied to various loops to take advantage of its fast linear algebra functionality.
At each timestep, $x_i(t,\theta)$ is stored as a matrix, where each agent occupies a row, and each column holds discretized $\theta$ values. 
The likelihood broadcast to all agents by element-wise matrix multiplication synchronously via Eq.\ \eqref{eq:updatefirsthalf}. 
The interaction between agents is symmetric ($A_{ij} = A_{ji}$). 
Eq.\ \eqref{eq:xiprimed} is implemented via element-wise matrix subtraction, which calculates all interactions synchronously. 
Eq.\ \eqref{eq:undatesecondhalf} applies a mask to the $\Delta x' (t+1/2, \theta)$ matrix to obtain $x_i(t+1,\theta)$. 
Batched simulations are run in parallel on 10 threads. 
A single complete simulation with $n=100$ and $T=10^4$ typically takes 0.7 seconds on a 2021 MacBook Pro (Apple M1 Max, 32GB RAM, 3.2 GHz clock speed, 10 cores).

\section{Special case of a deterministic coin with $\theta_0=0$ (or $\theta_0=1$)}
\label{sec:twoallies}
In this appendix, for the sake of completeness, we discuss briefly the special case, where the coin returns heads or tails only. By way of illustration, we consider an allies-only network with $n = 100$ and one partisan. 
We focus on $\theta_0 = 0$, as the behavior for $\theta_0=1$ is analogous.

\begin{figure}[h!]
    \centering
    \begin{subfigure}{0.4\textwidth}
        \includegraphics[width=\linewidth]{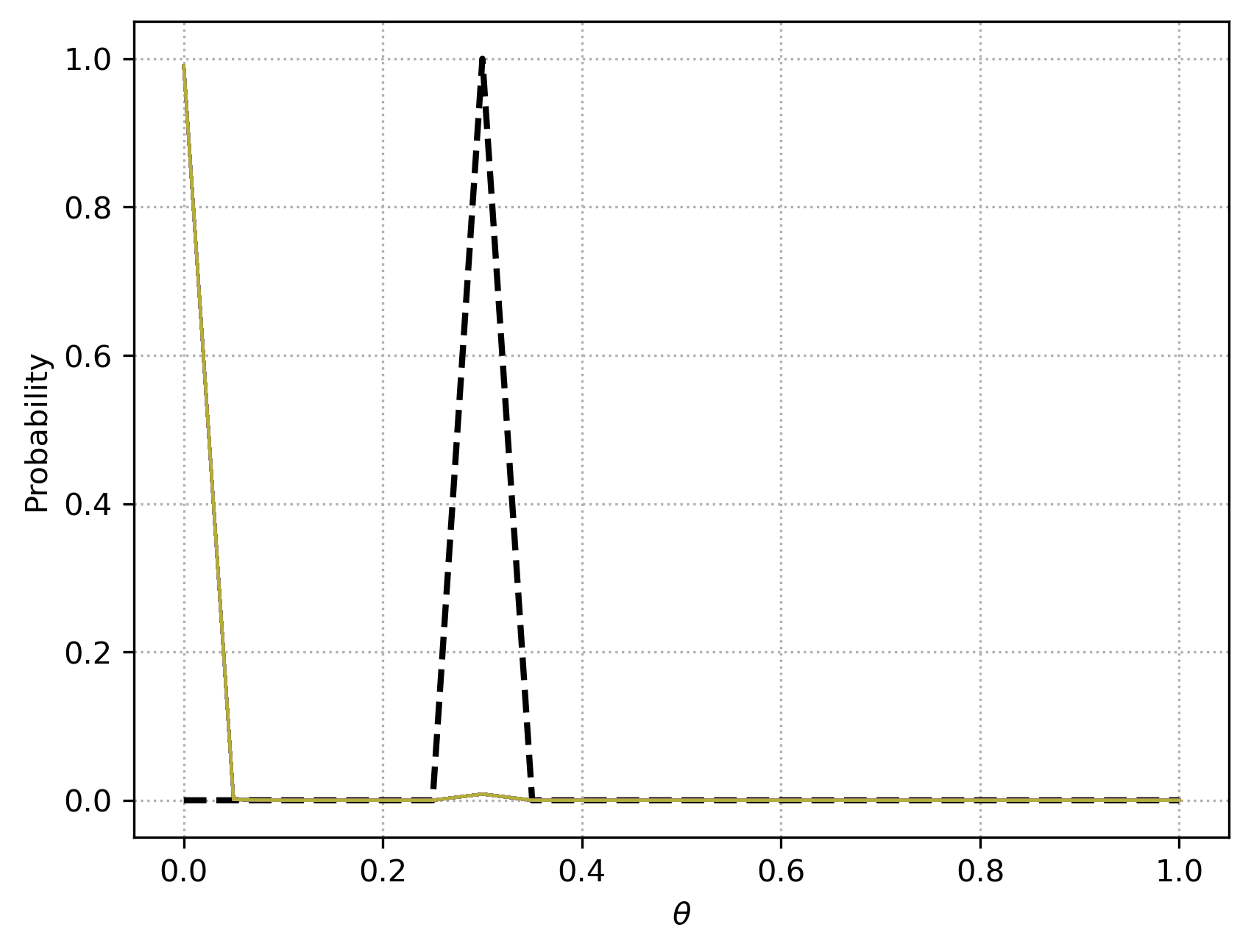}
        \caption{}\label{fig:0truebias}
    \end{subfigure}%
    \begin{subfigure}{0.4\textwidth}
        \includegraphics[width=\linewidth]{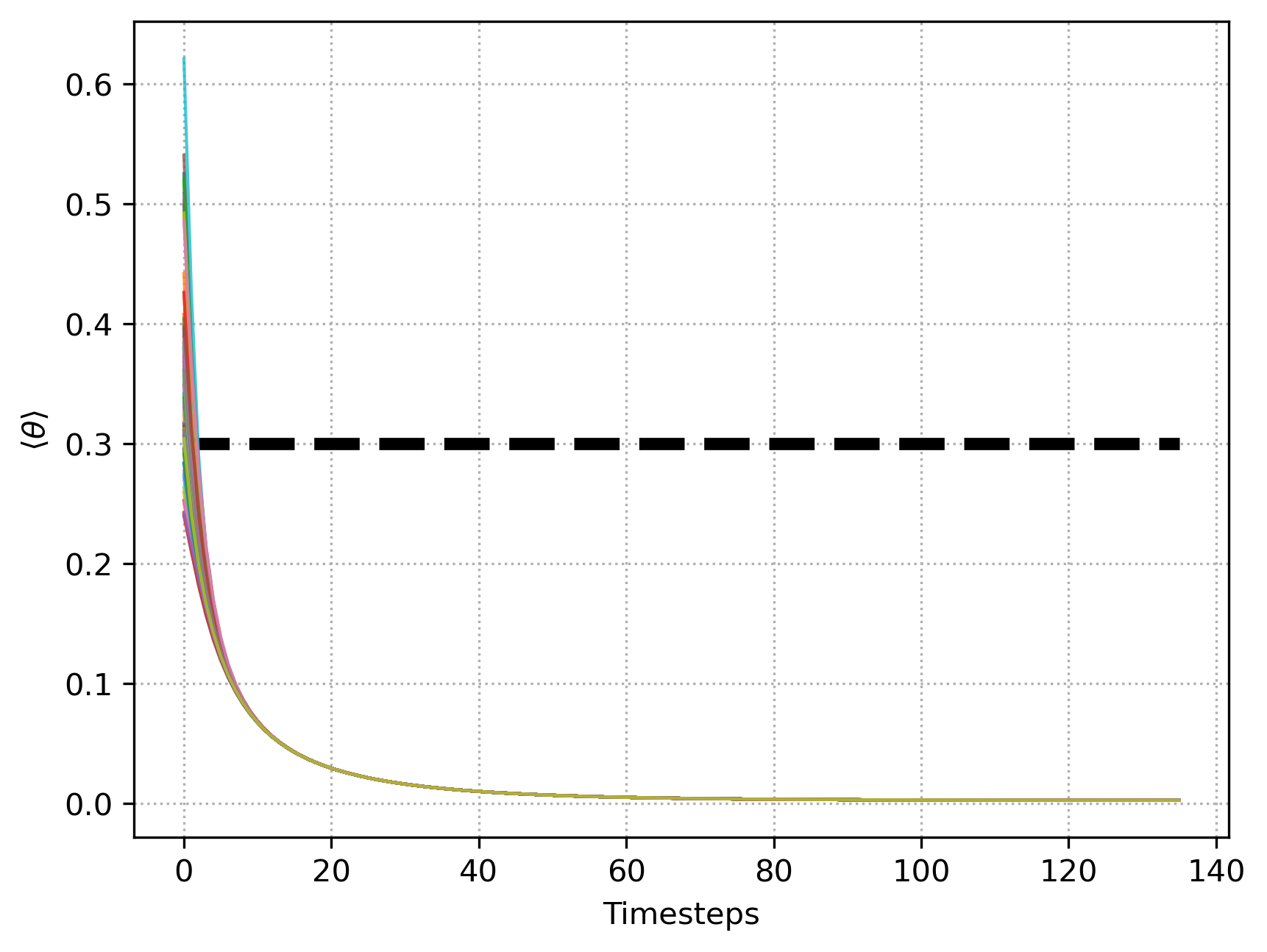}
        \caption{}\label{fig:0truebias_mean}
    \end{subfigure} \\
    \caption{
        Results for a deterministic coin: an example of an allies-only network with $n = 100$, one partisan, $\theta_0=0$, and $\theta_{\rm p} = 0.3$.
        (a) Snapshot at $t = t_{\rm a}$ of the belief PDFs of the partisan (black dashed curve) and arbitrary persuadable agents (colored curves). 
        (b) Time evolution of $\langle \theta \rangle$. 
        } 
    \label{fig:beliefatdiffercoinbias}
\end{figure}

For $\theta_0=0$, the persuadable agents achieve asymptotic learning, achieving a bimodal final distribution 
with $x_i(t \geq t_{\rm a},\theta_0) \approx 0.995$ and  $x_i(t \geq t_{\rm a},\theta_{\rm p}) \approx 0.005$, as shown in Fig.\ \ref{fig:0truebias}.  
The persuadable agents still heed the partisan but weakly. 
Unlike the turbulent nonconvergence observed in Section \ref{subsec:samethetap}, all persuadable agents' beliefs asymptotically approach $\theta_0$ and $\theta_{\rm p}$, achieving asymptotic learning at $t_{\rm a} = 136$, as shown in Figs.\ \ref{fig:0truebias_mean}. 
This is because a coin with $\theta_0 = 0$ always returns tails.
From Eq.\ \eqref{eq:likelihood}, the likelihood is given by $P[S(t)| \theta] = 1-\theta $ for all $t$, which always equals one for $\theta = 0$ and is less than one for all other values of $\theta$. 
When the likelihood is multiplied by the prior according to Eq.\ \eqref{eq:undatesecondhalf}, one obtains $x_i(t+1, \theta) < x_i(t, \theta)$ for $\theta \neq 0$, and $x_i(t,\theta) \neq 0$ decreases iteratively and monotonically to zero. 
The second peak at $\theta_{\rm p}$ is obtained though Eq.\ \eqref{eq:xiprimed} from the internal interaction between allies.  
The persuadable agents always hold some belief in $\theta_{\rm p}$, as Eq.\ \eqref{eq:undatesecondhalf} is additive rather than multiplicative.

It is interesting to ask how the persuadable agents respond, when they receive an unexpected external signal.
As a test, we run a simulation with $\theta_0=0$, returning $S(t) = {\rm tails}$ for $ 1\leq t \leq 500$, followed artificially by $S(t=501)= {\rm heads}$. 
The belief PDFs of the persuadable agents at $t = 500$ are identical to the PDF shown in Fig.\ \ref{fig:0truebias} but they change abruptly to $x_i(t=501,\theta) \approx \delta (\theta-\theta_{\rm p})$, after they observed a single heads. This readiness to agree with the partisan after a single, unexpected observation occurs for the following reason.
When persuadable agents observe $S(t=501) = {\rm heads}$, they infer $P[S(t=501)|\theta=0] = 0$ according to Eq.\ \eqref{eq:likelihood}. 
Therefore, we have $x_i(t = 500 + 1/2, \theta = 0) = 0$ via Eq.\ \eqref{eq:updatefirsthalf}, and $x_i(t=500+1/2,\theta_{\rm p}) = 1$ after renormalization.

\section{Long-term influence of initial priors and coin toss sequence}
\label{sec:coinvsprior}

In this appendix, we test how the initial priors $x_i(t=0,\theta)$ and the coin toss sequence affect the long-term evolution of $x_i(t,\theta)$ for persuadable agents.

\begin{figure}[h!]
    \begin{center}
        \begin{subfigure}{0.45\textwidth}
            \includegraphics[width=\linewidth]{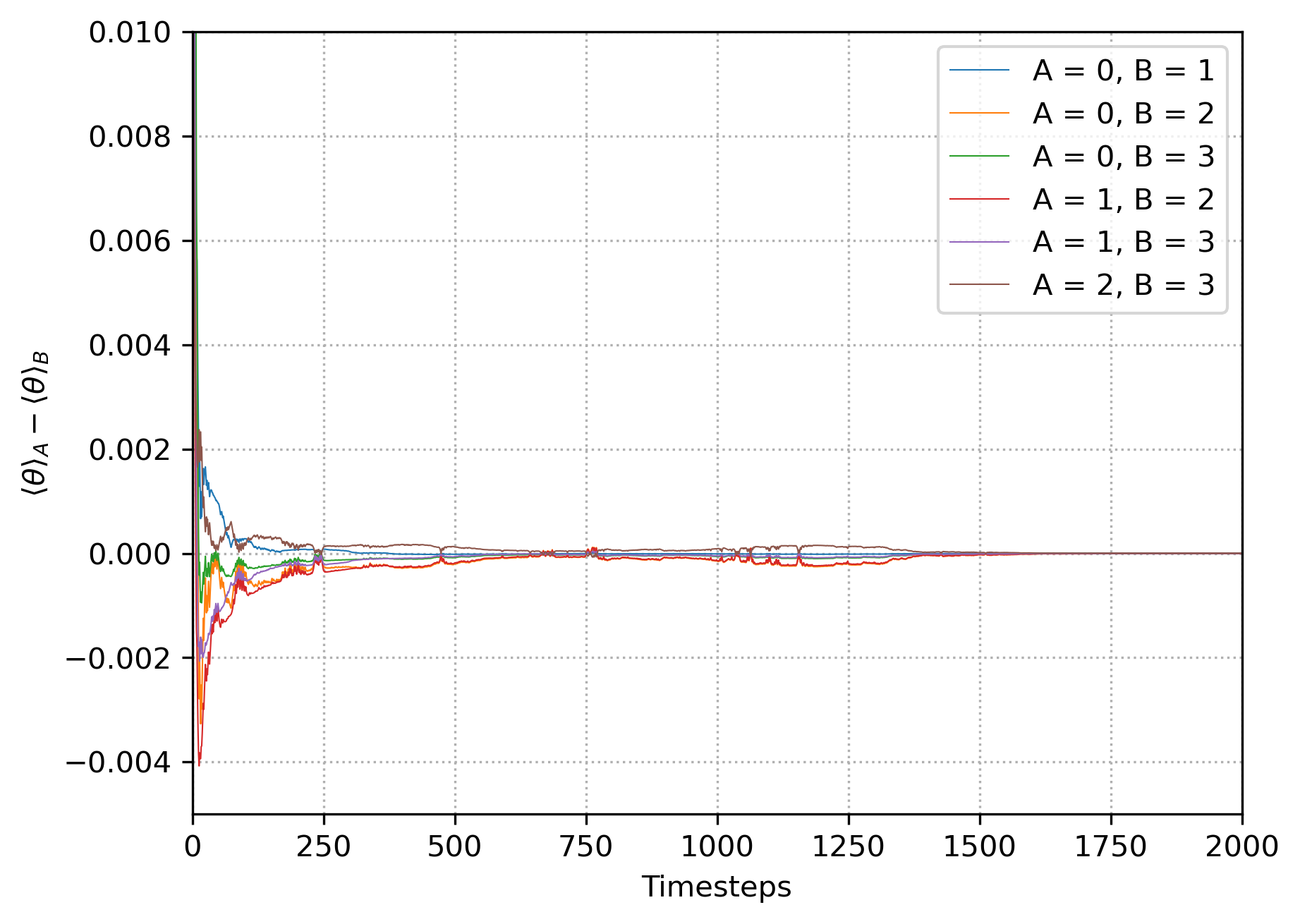}
            \caption{}\label{fig:samecoin}
        \end{subfigure}
        \begin{subfigure}{0.45\textwidth}
            \includegraphics[width=\linewidth]{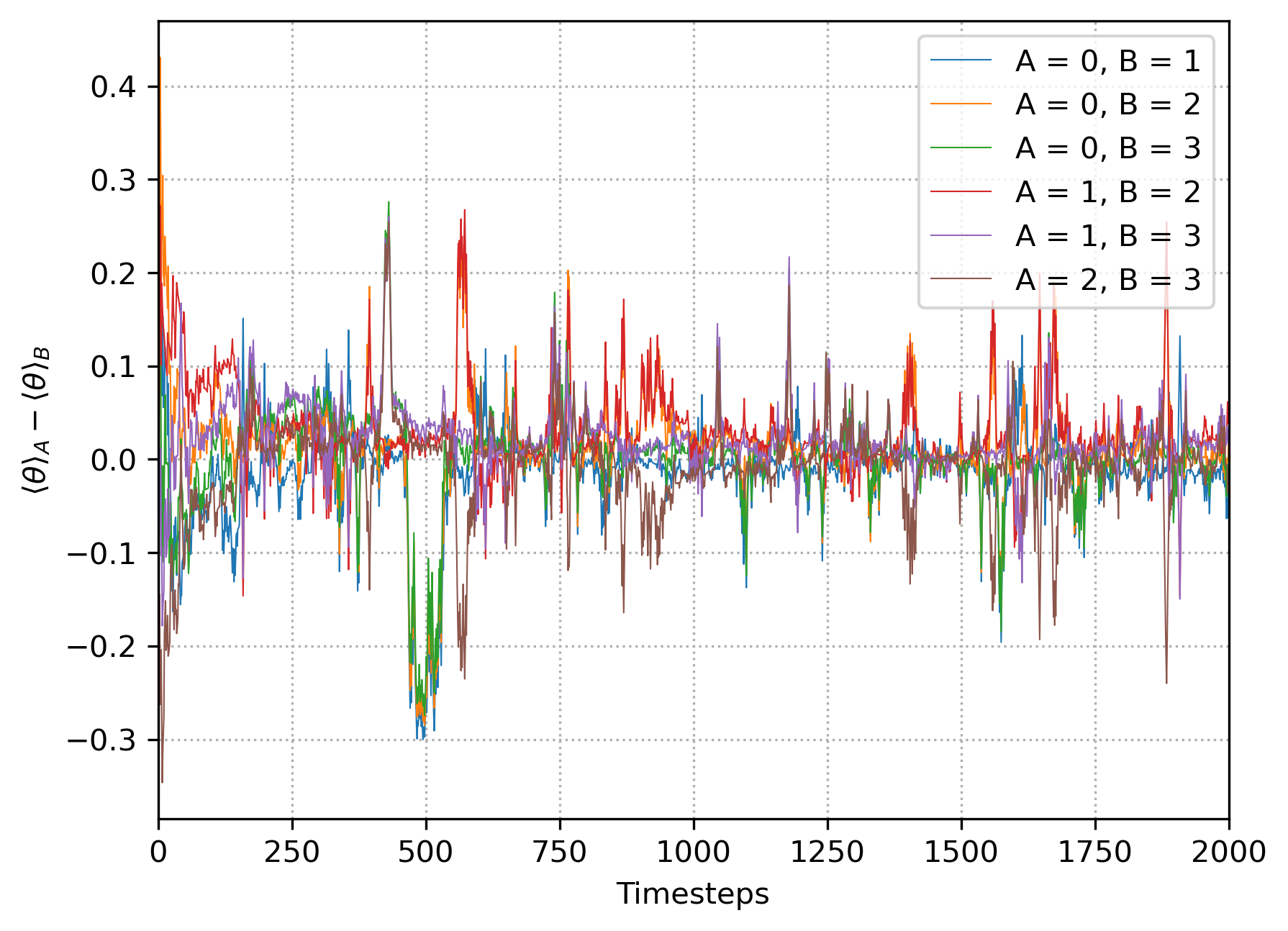}
            \caption{}\label{fig:sameprior}
        \end{subfigure}
    \end{center}
    \caption{
        Long-term impact of the initial priors and coin toss sequence in an allies-only network with $n=100$ and one partisan. 
        Six pairwise differences  $\langle\theta\rangle_A - \langle\theta\rangle_B$ involving the mean belief $\langle\theta\rangle$ of one arbitrary, persuadable agent versus time with the differences calculated between 
        (a) four random priors indexed by $A$ and $B$, and 
        (b) four random coin toss sequences indexed by $A$ and $B$.} 
    \label{fig:coinvsprior}
\end{figure}

We start by simulating $10^4$ copies of the same network, which are identical, except that $x_i(t=0,\theta)$ for all the persuadable agents $i$ is drawn afresh in each copy from the truncated Gaussian distribution defined in Section \ref{subsec:automaton}.
All copies of the network witness the same coin toss sequence. Fig.\ \ref{fig:samecoin} plots the six independent pair-wise differences $\langle\theta\rangle_A - \langle\theta\rangle_B$ for four network copies (indexed by $A$ and $B$) and a single, arbitrary persuadable agent as functions of time.
The differences in mean belief evolve towards zero shortly after the simulation starts, with $| \langle \theta \rangle_A - \langle \theta \rangle_B | \leq 10^{-5}$ for $t \geq 5\times 10^2$.
This behavior is typical: the long-term belief PDF does not depend strongly on $x_i(t=0,\theta)$. It is also observed in networks without partisans \cite{acemoglu_opinion_2013}. 

Next we test the sensitivity to the coin toss sequence. In Fig.\ \ref{fig:sameprior}, we plot again the six pair-wise differences $\langle\theta\rangle_A - \langle\theta\rangle_B$ against time for a single, arbitrary, persuadable agent and four random, independent coin toss sequences, indexed by $A$ and $B$. 
The differences do not decay to zero, unlike in Fig.\ \ref{fig:samecoin}. 
Instead, they fluctuate steadily for all $0\leq t \leq 2\times 10^3$, with standard deviation $\approx 0.05$ throughout the interval. 
This behavior matches Section \ref{subsec:samethetap}: the partisan disrupts the system so that it never reaches equilibrium, and the coin tosses are an ongoing factor, competing with the pull of the partisan at every time step. 


\newpage
%
\bibliographystyle{elsarticle-num} 
\bibliography{paper1}

\end{document}